\documentclass{aa}  

\usepackage{graphicx}
\usepackage{xcolor}
\usepackage{placeins}
\usepackage{lscape}
\usepackage{ulem}
\definecolor{notecolor}{rgb}{0.8,0,0}

\usepackage{txfonts}
\newcommand{\HI}{\mbox{H\,{\sc i}}}

\newcommand{\CII}{\mbox{C\,{\sc ii}}}

\newcommand{\CIV}{\mbox{C\,{\sc iv}}}
\newcommand{\NV}{\mbox{N\,{\sc v}}}
\newcommand{\OI}{\mbox{O\,{\sc i}}}

\newcommand{\MgII}{\mbox{Mg\,{\sc ii}}}
\newcommand{\AlII}{\mbox{Al\,{\sc ii}}}
\newcommand{\AlIII}{\mbox{Al\,{\sc iii}}}
\newcommand{\SiII}{\mbox{Si\,{\sc ii}}}

\newcommand{\SiIV}{\mbox{Si\,{\sc iv}}}
\newcommand{\SII}{\mbox{S\,{\sc ii}}}
\newcommand{\FeII}{\mbox{Fe\,{\sc ii}}}

\newcommand{\MnII}{\mbox{Mn\,{\sc ii}}}

\newcommand{\Lya}{Ly$\alpha$}
\newcommand{\Lyb}{Ly$\beta$}
\newcommand{\kms}{km s$^{-1}$}

\begin{document}

%   \title{Low-ionization systems in the E-XQR-30 sample: evidence of Pop~III stars' chemical signatures at $z\sim 6$?}

    \title{Evidence of Pop~III stars' chemical signature in neutral gas at $z\sim6$}
    \subtitle{A study based on the E-XQR-30 spectroscopic sample}
    
    \titlerunning{}
    \authorrunning{Sodini et al.}

\author{
Alessio Sodini\inst{1}, Valentina D'Odorico\inst{1,2,3}\fnmsep\thanks{E-mail: valentina.dodorico@inaf.it}, Stefania Salvadori\inst{4,5}, Irene Vanni\inst{4,5}, Manuela Bischetti\inst{1,6}, Guido Cupani\inst{1,3}, 
Rebecca Davies\inst{7,8}, George D. Becker\inst{9}, Eduardo Ba\~nados\inst{10}, Sarah Bosman\inst{10,11}, 
Frederick Davies\inst{10}, Emanuele Paolo Farina\inst{12}, Andrea Ferrara\inst{2}, 
Laura Keating\inst{13}, Girish Kulkarni\inst{14}, Samuel Lai\inst{15}, Emma Ryan-Weber\inst{7,8}, Alma Maria Sebastian\inst{7,8}, 
Fabian Walter\inst{10}}
\institute{
INAF - Osservatorio Astronomico di Trieste, Via G. Tiepolo 11, 34143, Trieste, Italy
\and
Scuola Normale Superiore, Piazza dei Cavalieri 7, 56126, Pisa, Italy
\and
IFPU - Institute for Fundamental Physics of the Universe, via Beirut 2, I-34151 Trieste, Italy
\and
Dipartimento di Fisica e Astronomia, Università degli Studi di Firenze, Via G. Sansone 1, Italy
\and
INAF - Osservatorio Astrofisico di Arcetri, Largo E. Fermi 5, I-50125 Firenze, Italy
\and
Dipartimento di Fisica, Sezione di Astronomia, Universit\'a di Trieste, via Tiepolo 11, 34143 Trieste, Italy
\and
Centre for Astrophysics and Supercomputing, Swinburne University of Technology, Hawthorn, Victoria 3122, Australia
\and
ARC Centre of Excellence for All-Sky Astrophysics in 3 Dimensions (ASTRO 3D), Australia
\and
Department of Physics \& Astronomy, University of California, Riverside, CA 92521, USA
\and
Max Planck Institut f\"ur Astronomie, K\"onigstuhl 17, D-69117, Heidelberg, Germany
\and
Institute for Theoretical Physics, University of Heidelberg, Philosophenweg 16, 69120 Heidelberg, Germany
\and
Gemini Observatory, NSF’s NOIRLab, 670 N A’ohoku Place, Hilo, Hawai'i 96720, USA
\and
Institute for Astronomy, University of Edinburgh, Blackford Hill, Edinburgh, EH9 3HJ, UK
\and
Tata Institute of Fundamental Research, Homi Bhabha Road, Mumbai 400005, India
\and
Research School of Astronomy and Astrophysics, Australian National University, Canberra, ACT 2611, Australia
}

   \date{Received ; accepted }

% \abstract{}{}{}{}{} 
% 5 {} token are mandatory
 
  \abstract
   {}
   {This study explores the metal enrichment signatures attributed to the first generation of stars (Pop~III) in the Universe, focusing on the E-XQR-30 sample -- a collection of 42 high signal-to-noise ratio spectra of quasi-stellar objects (QSOs) with emission redshifts ranging from 5.8 to 6.6. We aim to identify traces of Pop~III metal enrichment by analyzing neutral gas in the interstellar medium of primordial galaxies and their satellite clumps, detected in absorption.}
   {To chase the chemical signature of Pop~III stars, we studied metal absorption systems in the E-XQR-30 sample, selected through the detection of the neutral oxygen absorption line at 1302 \AA. The \OI\ line  is a reliable tracer of neutral hydrogen and allowed us to overcome the challenges posed by the Lyman-$\alpha$ forest's increasing saturation at redshifts above $\sim5$ to identify damped Lyman-$\alpha$ systems (DLAs). We detected and analyzed 29 \OI\ systems at $z \geq 5.4$, differentiating between proximate DLAs (PDLAs) and intervening DLAs. Voigt function fits were applied to obtain ionic column densities, and relative chemical abundances were determined for 28 systems. These were then compared with the predictions of theoretical models.}
   {Our findings expand the study of \OI\ systems at $z \geq 5.4$ fourfold. No systematic differences were observed in the average chemical abundances between PDLAs and intervening DLAs. The chemical abundances in our sample align with literature systems at $z > 4.5$, suggesting a similar enrichment pattern for this class of absorption systems. A comparison between these DLA-analogs at $4.5 < z < 6.5$ with a sample of very metal-poor DLAs at $2 < z < 4.5$ shows in general similar average values for the relative abundances, with the exception of [C/O], [Si/Fe] and [Si/O] which are significantly larger for the high-$z$ sample.
   Furthermore, the dispersion of the measurements significantly increases in the high-redshift bin. This increase is predicted by the theoretical models and indicates a potential retention of Pop III signatures in the probed gas.}
   {This work represents a significant advancement in the study of the chemical properties of highly neutral gas at $z \geq 5.4$, shedding light on its potential association with the metal enrichment from Pop~III stars. Future advancements in observational capabilities, specifically high-resolution spectrographs, are crucial for refining measurements and addressing current limitations in the study of these distant absorption systems.}


   \maketitle
%
%________________________________________________________________

\section{Introduction}
\label{sec:Introduction}
Measuring the abundances of chemical elements in galactic and intergalactic gas provides some of the most useful observables for investigating stellar formation and evolution in the Universe. 
In fact, the comparison between the observed stellar chemical abundances and those predicted by theoretical models can be used to place constraints on stellar evolution models. 
In turn, these models are a key ingredient in constraining the history of galaxy formation and evolution. \citep[e.g.][]{Maiolino2019}.

Of particular interest is the first generation of stars in the Universe, the so-called Population III (Pop III) stars \citep[for a review, see][and references therein]{Klessen2023}. These stars formed from gas produced by the Big Bang nucleosynthesis, rich in hydrogen and helium but almost metal free. 
Models predict that they produced abundant quantities of UV photons which started the process of \HI\ reionization in the Universe; when they exploded as supernovae (SNe) they enriched the interstellar medium (ISM) and the intergalactic medium (IGM) with the first heavy elements. On the other hand, if, at the end of their life, they turned into black holes (BHs) they could provide the seeds of the super-massive BH observed at $z\sim 6-7$.

In the conventional framework in which Pop III stars are predominantly massive, i.e., $\approx 10 - 1000\, M_{\sun}$ \citep[e.g.][]{Hirano2014}, those with $140 M_{\sun}\leq M_{\rm PopIII} \leq 260 M_{\sun}$ will explode as pair instability supernovae (PISNe, \citealt{Heger2002,Takahashi2018}), that leave no remnant and distribute the first metals in the gas in and out of galaxies \citep[e.g.][]{Bromm2004}. At intermediate masses, $10 M_{\sun} \leq M_{\rm PopIII} \leq 100 M_{\sun}$, first stars can also evolve as SNe but they can have a variety of explosion energies, thus yielding very different chemical elements ratios depending upon the mass of the progenitor star and the SN explosion energy \citep[e.g.][]{Heger2010,Nomoto2013,Limongi2018}. The nucleosynthetic signature of Pop~III stars may therefore be very different from that of subsequent generations of more metal-rich stars, i.e., Pop~II/I stars \citep[e.g.][]{Salvadori2019,Vanni2023}.

A technique commonly used for testing the predicted existence of Pop~III stars is to identify their chemical traces by analyzing the abundance pattern of their direct descendants, which can be identified among ancient, metal-poor stars in the Local Group. 
Indeed, if massive Pop III stars quickly ended their lives as energetic SNe, their nucleosynthetic products were efficiently dispersed into the surrounding gas where subsequent generations of stars were able to form. 

Local observations have reported the existence of many carbon enhanced metal-poor stars (CEMP stars, [C/Fe]$>0.7$ \citealt{BeersChristlieb2005}), both in the galactic halo \citep[e.g.][]{Bonifacio2015,Yoon2019} and in ultra-faint dwarf galaxies \citep[e.g.][]{Spite2018,simon2019}. These stars appear predominantly at [Fe/H]$<-3$ and the most iron-poor among them have chemical abundance patterns in agreement with those produced by Pop III stars which exploded as low-energy SNe \citep[e.g.][]{Iwamoto2005a, Marassi2015a}.
Stars imprinted by more energetic hypernovae or PISN are more difficult to be identified since they have normal [C/Fe]$\leq 0.0$ \citep[e.g.][]{Koutsouridou2023, Pagnini2023}. However, thanks to other peculiar chemical abundances, two of them have been recently discovered \citep[][]{Skuladottir21,Placco21}. %such as [Mg/Ca]$<-0.5$  
Furthermore, a rare halo star imprinted by a massive and energetic PISN has been found among alpha-poor stars in the LAMOST survey \citep{Xing2023}, with strong implications for the initial mass function of Pop~III stars \citep{Koutsouridou2024}.

The advent of the James Webb Space Telescope (JWST) has opened a new route to study Pop~III stars through, in particular: possible direct detection \citep[e.g.][]{Zackrisson2023} and the identification of chemical enrichment from the first sources in $z\gtrsim 10$ galaxies \citep{Katz2023,Deugenio2023}. 

An additional method to find the traces of Pop III stars at high-redshift is by measuring the chemical abundances in gas associated with primordial galaxies and detected in absorption in bright source spectra \citep[e.g.][but see also the partial revision of the latter paper results by \citealt{Vanni2024}]{Welsh2021,saccardi2023,christensen2023}. 
The main advantages of this observational technique are that it can be applied using the same diagnostics over a broad range of redshifts and that the conversion from the properties of the observed ionic absorption lines to the total abundance of elements is often simple. 
This is especially true for systems with a high neutral hydrogen (\HI) column density, known as damped Lyman-$\alpha$ systems (DLAs; $\log ({N(\mathrm{\HI})/ {\rm cm}^{-2}}) \gtrsim 20.3$), which trace galactic environments  along QSO lines of sight \citep[see][for a review]{Wolfe2005}. 

In DLAs, the large amount of neutral hydrogen shields the innermost gas from the ionizing radiation. Consequently, oxygen, which has a first ionization potential similar to that of hydrogen\footnote{In fact, oxygen is in charge-exchange equilibrium with \HI\ because the two ionization potentials are almost identical.}, will be present in the gas predominantly in the neutral state. While other elements such as carbon, silicon, magnesium and iron, which have a first ionization potential significantly lower than $13.6\,\rm eV$, will be mainly singly ionized. 
As a result, in the simplest case, the abundance for a given element can be calculated directly from the measured column density of ions in a single electronic state \citep[e.g.][]{Rafelski2012}. Complications arise at lower \HI\ column densities, whereby the ionization effects start to become significant or in the case of metal depletion onto dust grains \citep[e.g.][]{Peroux2020}. 

Thanks to the discovery of metal-enriched absorbers up to $z \sim 6$ \citep{Becker2006,RyanWeber2006,Simcoe2006,Dodorico2013,Dodorico2022,RDavies2023a}, it is now possible to investigate chemical abundances in the gas when the Universe was about one billion years old. 
At that time, the observed metals could only have been produced by massive stars, which have a sufficiently short lifespan and which explode as SNe in their final phase. 

At redshifts $z \ge 5$, however, the large density of absorption lines in the \HI\ Lyman-$\alpha$ (\Lya) forest makes it very difficult to measure the \HI\ column density of single absorption systems. Luckily, many chemical elements (in particular: C, N, O, Mg, Al, Si, S, Cr, Mn, Fe, Zn) have ionic transitions that fall outside the \Lya\ forest, in the region of the spectrum redward of the \Lya\ emission of the QSO. 
As a consequence, although it is rarely possible to estimate absolute chemical abundances (e.g., relative to hydrogen) many relative abundances can be determined, which provide unique constraints on nucleosynthesis in massive (and presumably metal-poor) stars. 

\citet[][]{Becker2012} analyzed the relative abundances of C, O, Si and Fe in nine low ionization QSO absorption systems at $4.7 <z <6.3$, considered to be DLAs or sub-DLAs (the latter are defined to have $19.0 \lesssim \log{N_{\rm HI}}\lesssim 20.3$) thanks to the presence of the neutral oxygen absorption line at $1302$ \AA. 
The authors compared the relative abundances of the $z>4.7$ systems with those measured in very metal-poor\footnote{\citet{Cooke2011b} defines VMP-DLAs as systems with [Fe/H]~$< -2$ following the definition used for stars.} (VMP) DLA systems at lower redshifts ($2<z<4.5$) observed by \citet{Cooke2011b}. The two samples show minimal dispersion in the ratio between the column densities of any two ions (\CII, \OI, \SiII\ and \FeII) and no apparent evolution with redshift. The lack of dispersion suggests that the relative abundances of the elements are correctly estimated from the ratios of the corresponding ions, with minimal ionization or depletion effects due to dust formation (in particular, for the high-$z$ systems). It also suggests that heavy element enrichment in VMP DLAs at $2<z<6$ are dominated by the integrated returns of SNe of type II and I. This is confirmed also in later studies \citep[e.g.][]{Poudel2018,Poudel2020}, in which the lack of exotic abundances at $z\sim 5-6$ suggests that ordinary Pop II stars (rather than massive Pop III stars) likely enriched the interstellar gas with metals even approaching the epoch of hydrogen reionization.

In this work, we exploit the new XQR-30 sample of high-quality spectra of QSOs at $z\sim 6$ \citep{Dodorico2023} to significantly increase the number of \OI\ absorption systems at $z\gtrsim 5.4$. Our aim is to  study the relative abundance of elements present in primordial galaxies and to obtain information on the nature of the stars that contributed to the enrichment of the gas. 

The paper structure is the following. 
In Sect.~\ref{sec:Sample}, the spectroscopic sample used in this work and the data reduction are briefly described. Section \ref{sec:Analysis} reports how we performed the analysis of the spectra, the search and identification of absorption lines, the fit of the velocity profiles and the estimate of the \HI\ column densities. 
In Sect.~\ref{sec:Results}, we present the relative abundances of the observed systems and discuss the results obtained by comparing them with those of similar systems present in the literature at low and high redshift. Section \ref{sec:discussion} is dedicated to the discussion of our results and the comparison with the predictions of theoretical models of enrichment. Finally in Sect.~ \ref{sec:Conclusion}, we conclude summarizing the obtained results.

%__________________________________________________________________

\section{Sample and data reduction}
\label{sec:Sample}

The data used for this work (summarized in Tab.~\ref{tab:QSO}) include optical and NIR spectra of 42 QSOs in the redshift range $5.8\lesssim z\lesssim 6.6$. Thirty of these QSOs are part of the "XQR-30 survey", an ESO Large Programme  (ID. 1103.A-0817, P.I. V. D'Odorico) which obtained 248 hours of observation with the X-SHOOTER spectrograph \citep{Vernet2011} at the Very Large Telescope (VLT). For these QSOs, optical spectra (VIS, $\lambda=5500-10,250$ \AA) were obtained using a slit width of $0.9''$ corresponding to  a nominal resolving power $R_{\rm VIS} \simeq8900$, and near-infrared spectra (NIR, $\lambda=9800-24,800$ \AA) using a slit width of $0.6''$ corresponding to $R_{\rm NIR} \simeq 8100$. The ultraviolet spectrum of the UVB arm was not used, since the spectrum of high-redshift QSOs is completely absorbed in this spectral region. 

The other 12 QSOs which are part of the analyzed sample, have been retrieved from the X-SHOOTER archive and have similar characteristics to those of XQR-30 (see Tab.~\ref{tab:QSO}). The total sample, dubbed the "enlarged" XQR-30 sample (E-XQR-30), is thoroughly described in \citet{Dodorico2023}. 

Here, we briefly review the steps of the spectroscopic data reduction procedure. The reduction was performed with a proprietary software \citep[see][]{Becker2019}, which has demonstrated to perform better than the default ESO pipeline, in particular for what concerns the sky subtraction \citep[see][]{Lopez2016}. 
First, the individual frames were corrected for the bias and the dark current and then divided by the flat field frames. 
A subtraction of the sky spectrum on each frame was then performed following  \citet{Kelson2003}. The frames were calibrated in  wavelength and flux calibrated\footnote{Note that this is a relative flux calibration and not an absolute calibration.} based on a standard star. 
Corrections for telluric absorptions were performed using a model based on the ESO SKYCALC Cerro Paranal Advanced Sky Model \citep{Noll2012,Jones2013}. Finally, the frames of the individual exposures thus obtained were added together and a one-dimensional final spectrum was extracted.

As previously mentioned, the nominal resolving power of the VIS and NIR arm of X-SHOOTER depend on the choice of the slit width. However, if the seeing disk at the time of observation is smaller than the slit, the resolving power of the observed spectrum will be larger than the nominal one. The determination of the correct resolving power of the spectrum is particularly relevant when fitting absorption lines. The "true" resolution of the E-XQR-30 spectra was estimated using an empirical relation between the FWHM of the order spatial profile and the FWHM of the model fitting the telluric lines in the single frames \citep[see][for further details]{Dodorico2023}. 
In summary, we find that the resolving power of the final spectra is always larger than  the nominal one (see the values reported in  Tab.~\ref{tab:QSO}) with medians of $R_{\rm VIS} \simeq 11,400$ and $R_{\rm NIR} \simeq 9800$. 

Finally, the systemic redshifts of the QSOs, listed in Tab.~\ref{tab:QSO}, are based on CO or [\CII] emission, when available, otherwise we considered the maximum\footnote{We adopt the maximum value because the \MgII\ line is generally blueshifted by $\sim500$ \kms\ with respect to systemic redshift \citep[e.g.][]{venemans2016,mazzucchelli2017,schindler2020}. } between the redshift measured from the \MgII\ emission line \citep{Bischetti2022,Dodorico2023} and that determined from the first \Lya\ absorption line observed on the peak of the \Lya\ emission \citep[following][]{Zhu2021, Becker2019}. All the spectra of the E-XQR-30 sample are shown in \citet{Dodorico2023}.

\section{Data analysis}
\label{sec:Analysis}
The analysis of the spectra of the QSO sample was performed with the \textsc{Astrocook} software \citep{Cupani2020}. \textsc{Astrocook} is a Python 3 \citep{VanRossumDrake2009} package for analyzing QSO spectra, featuring a graphical user interface with numerous analysis methods.
Below is a description of the analysis procedure that was used for the E-XQR-30 spectra. 

We note that the procedure of detection and identification of all the metal absorption lines in the 42 spectra of the E-XQR-30 QSO sample is reported in the general catalog paper \citep{RDavies2023a}, which is the official reference for the metal lines of the survey. However, we decided to report here the procedure adopted for the search of the low ionization systems which was carried out before and independently from the work of \citet{RDavies2023a} for the master thesis of A. Sodini. The results of the Voigt fitting adopted in this work are in agreement with those in \citet{RDavies2023a} within $1\sigma$.

\subsection{Detection of the absorption lines and estimate of the intrinsic emission spectrum}
\label{sec:Cont}

In order to prepare the spectra for our analysis, first we have extracted from the VIS spectrum the wavelength region that goes from the \Lya\ emission of the QSO to $10,100$ \AA, to exclude the \Lya\ forest.
We then detected the absorption lines present in this part of the spectrum. The procedure in  \textsc{Astrocook} looks for the local minima of flux density in the spectrum, after smoothing with a Gaussian profile to reduce the noise. Different values of the Gaussian variance are used in descending order to detect lines of different widths. We used values of the Gaussian variance from $100$ \kms\ to $15$ \kms, while we set a threshold to define the prominence of the minima, expressed as a multiple of the variance of the local flux density, of about $2- 3\,\sigma$. The regions between the local maxima adjacent to each line are then masked.

At this point, the intrinsic emission spectrum is fitted with \textsc{Astrocook} by interpolating nodes in the unmasked spectrum with a spline function of chosen degree. The nodes are placed at fixed velocity intervals (we used $500-750$ \kms) considering the median flux density of the surrounding region, from which outliers are excluded. 

We then analyzed the NIR region of the spectrum, selecting the wavelength range from $10,100$ to $22,000$ \AA. The VIS and NIR regions of the spectrum were equalized, rescaling the flux density in the NIR part; for this operation we have selected a wavelength range that goes from $10,000$ to $10,200$ \AA\ common to both regions, in which \textsc{Astrocook} calculates the rescaling factors from the ratios of the median flux densities.
Subsequently, with the same procedure used for the VIS part of the spectrum, we detected the absorption lines and we fitted the emission spectrum also for the NIR part, setting the nodes for the interpolation at intervals of $1000- 1500$ \kms. Finally, the two parts were stitched together at 10,150 to obtain a single spectrum, and we revised the fit of the emission spectrum by manually eliminating some nodes and adding new ones directly on the main spectrum. 

\subsection{Identification and fit of the metal absorption lines}
\label{sec:Abs_line}

The absorption systems were first automatically identified with \textsc{Astrocook}, which divides the wavelengths of the detected lines by the laboratory wavelengths of a list of transitions commonly observed in QSO spectra, and finds coincidences between the possible redshift values resulting from the cross-comparison. Identified lines were then fitted with Voigt profiles using \textsc{Astrocook}, which provides for each line the central redshift, the column density and the Doppler width.

In the spectra of the QSOs analyzed in this work, we first identified the lines due to the most common ionic doublets, such as  \CIV\ $\lambda\lambda\  1548,\,1550$ \AA, \SiIV\ $\lambda\lambda\ 1393,\,1402$ \AA\ and \MgII\ $\lambda\lambda\ 2796,\,2803$ \AA. Identifications proposed by \textsc{Astrocook} were visually inspected to check the correspondence of the line velocity profiles and then fitted. 

Then, we proceeded with the identification of the systems showing absorption due to \OI\ (dubbed "DLA-analogs"), which are the focus of this work. The transition due to \OI\ $\lambda\ 1302$ \AA\ was identified by requiring the simultaneous identification of the transition due to \CII\ $\lambda\ 1334$ \AA\ and, possibly, also that due to  \SiII\ $\lambda\ 1304$ \AA\ at the same redshift.
Subsequently, other lines at low ionization associated with the identified systems were searched for. In particular, we considered:  \SiII\  $\lambda\ 1260$, $1526$ \AA, \AlII\ $\lambda\ 1670$ \AA\ and the transitions of the \FeII\ multiplet with rest frame wavelengths $\lambda\ 1608$, $2344$, $2374$, $2382$, $2586$ and $2600$ \AA. %All the identified lines were fitted with Voigt profiles. 

We found in most cases a redshift correspondence also with the \MgII\ lines previously found in the NIR part of the spectrum. 
However, for systems in the redshift range $5.5<z_{\rm abs}<5.9$, the doublet of the \MgII\ falls in the spectral region strongly affected by telluric absorptions between the photometric $H$ and $K$ bands and it is not always detectable (see details in Appendix~\ref{app:LIS}).
Finally, we verified if these low ionization systems were also associated with previously identified high ionization lines such as those of \CIV\ and \SiIV. 

When at the redshift of a given system, we did not detect significant absorption due to some of the previously mentioned  ions, or the absorption was below the 3$\sigma$ detection threshold, we estimated a $3\sigma$ upper limit to their column density. This limit was estimated starting from the average signal-to-noise ratio (SNR) per pixel of the spectrum in a wavelength interval around the expected occurrence of the absorption. 
This ratio is linked to the equivalent width of the line, possibly present at that point, by the relation \citep{Herbert2006}:
\begin{equation}
    {\rm SNR}\simeq \frac{3\,\lambda_{0,\rm X}}{c\,w_{0,\rm X}}\sqrt{4.24264\,b\,\Delta v}
    \label{eq:s_n}
\end{equation}
where $\lambda_{0,\rm X}$ and $w_{0,\rm X}$ are respectively the wavelength and the equivalent width in the rest frame of the transition X, $b$ is the Doppler parameter, and $\Delta v$ is the size of the pixels which, for the analyzed spectra, is equal to 10 \kms. The $b$ parameter for weak low ionization lines is assumed to be equal to that of the other ions detected in the system, while for the upper limits on \SiIV\ and \CIV\ we assumed $b=26$ \kms, which is the average value of the Doppler parameter for the detected systems. 
The corresponding upper limit on the  column density is estimated from the relation\footnote{This relation assumes that the line is on the linear part of the curve of growth, i.e., that the relation between column density and equivalent width does not depend on the Doppler parameter.}:
\begin{equation}
    N_{\rm X}=1.13\cdot 10^{20} \frac{w_{0,\rm X}}{f_{\rm X}\,\lambda_{0,\rm X}^2}\ [\rm cm^{-2}]
    \label{eq:upperlimit}
\end{equation}
where $f_{\rm X}$ is the oscillator strength of transition X.

To conclude the identification process, we associated, when possible, the other lines present in the spectrum with the  \CIV, \SiIV\ and \MgII\ systems found initially and which did not match with the low ionization systems. In some systems containing \CIV, the \SiIV\ doublet and the \MgII\ doublet are also present; in some of these systems we have also detected lines of \SiII, \CII, \FeII\ and \AlII. 
Some of these low ionization systems could also be DLA-analogs but they are at $z_{\rm abs}<\lambda_{\rm Ly\alpha}(1+z_{\rm em})/\lambda_{\rm OI}-1$, and the \OI\ line falls into the \Lya\ forest. For this reason they were not taken into consideration for the subsequent analysis.

The Doppler $b$ parameter is given by the quadrature sum of the thermal and turbulent components. 
In the analysis of low ionization systems, we adopted the following procedure: when the fit of  a velocity component returned comparable $b$ values for the different ionic transitions, we re-performed the fit of the component by linking their $b$ parameters, assuming that the turbulent motions are dominant over thermal agitation. 
Furthermore, to take into account the instrumental resolution, when the fit of a line returned a value lower than $b_{\rm ins}/3$\footnote{Where $b_{\rm ins}=(c / R) / (2 \sqrt{\ln 2})$ is the Doppler parameter corresponding to the instrumental broadening.}, we re-performed the fit with the column density and the redshift as free parameters, but fixing $b$ to this minimum value.

\begin{figure}
\centering
	\includegraphics[width=\columnwidth]{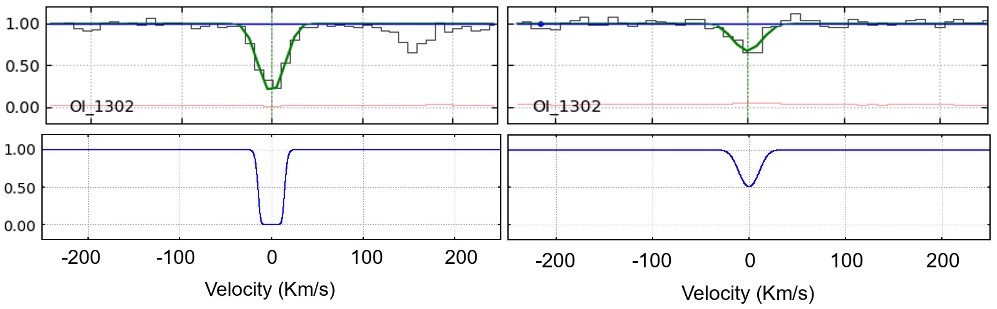}
    \caption{Comparison between two absorption lines fitted with Voigt profiles in XQR-30 (upper panels; black line: observed spectrum; red line: observed error; blue line: continuum level; green line: Voigt profile fit) and their theoretical profiles without the convolution with the instrumental broadening (lower panels). 
    The theoretical profile of the line on the left is saturated,  while the theoretical profile of the line on the right is not.}
    \label{fig:systsatnosat}
\end{figure}

Some of the detected lines could also be affected by saturation, which however at this resolution can be masked by the instrumental broadening of observed feature. To determine which lines were saturated, we therefore simulated the theoretical Voigt profile of the lines using the parameters obtained from the Voigt fitting, without the enlargement due to the resolution of the instrument, as shown in Fig.~\ref{fig:systsatnosat}. 
When the theoretical profile was saturated, we considered the value of the column density obtained from the fit as a lower limit.
The velocity plots and the parameters obtained from the fit of the studied DLA-analogs are reported in Appendix \ref{app:LIS}.

\subsection{Analysis of the DLA-analogs}
\label{sec:LIS}

In the spectra of the 42 analyzed QSOs of the E-XQR-30 sample, we detected 29 DLA-analogs (i.e., low ionization systems traced by the presence of \OI), along the line of sight to 19 QSOs. We note that in our sample we have two more \OI\ absorbers with respect the official catalog by \citet{RDavies2023a}. This is due to the fact that the \OI\ $\lambda\,1302$ lines of the systems at $z=5.6993$ in PSO J060+24 and at $z=5.7974$ in SDSS J0100+2802 are falling at the beginning of the \Lya\ forest and thus they were excluded from the official E-XQR-30 sample of metal absorption lines, while we decided to consider them because the \OI\ column density can be reliably determined. On the other hand, the proximate system at $z=5.8441$ along the line of sight to PSO J023-02 shows absorption lines due to fine-structure transitions of \CII\ and \SiII, high-ionization transitions due to \NV, and also a possible signature of partial coverage in the \HI\ \Lya\ profile implying that it could be very close to the ionizing source, possibly arising in a gas outflow powered by the AGN. For these peculiar characteristics, we decided to exclude it from the subsequent chemical analysis.

For those systems that have more than one velocity component, the total column density for each ion was computed as the sum of the column densities of the single components and the error was obtained propagating the individual errors.
The total column densities of each ion detected in each \OI\ system are reported in Tab.~\ref{tab:LIS}. This table also shows the redshift of the system, for which we have considered the absorption redshift of the \OI. For systems with multiple velocity components, the average redshift of the system was calculated as the column-density weighted average of the \OI\ redshifts of the various components:
\begin{equation}
    z_{\rm abs}=\frac{\sum_i{z_iN_i}}{\sum_i{N_i}}. 
    \label{eq:zion}
\end{equation}
Finally, for a system for which the \OI\ absorption could not be fitted ($z=5.8990$ in DELS J1535+1943, see Appendix \ref{app:LIS}), we considered the redshift of \CII.

Depending on the velocity separation of the systems from the QSO emission redshift, we distinguish two types of absorption systems: the ``proximate'' DLAs (or PDLAs), with $v_{\rm abs} \leq 5000$ \kms, and the ``intervening'' DLAs at larger velocity separations. 
PDLAs are systems which are thought to be ``associated'' with the QSO itself, since they are close enough to be affected by the ionizing flux of the QSO or even to be part of high velocity outflows powered by the AGN. Intervening DLAs, on the other hand, are generally independent structures such as galaxies or neutral gas condensations close to galaxies along the QSO line of sight \citep[e.g.][]{Perrotta2016}.

The velocity separation of absorption systems from the QSO emission redshift is determined with the relation \citep{Peterson1997}:
\begin{equation}
    v_{\rm abs} =\frac{(1+z_{\rm em})^2-(1+z_{\rm abs})^2}{(1+z_{\rm em})^2+(1+z_{\rm abs})^2}c.
    \label{eq:vsep}
\end{equation}

In our sample of DLA-analogs, 10 are PDLAs, while the other 19 are intervening systems (see Tab.~\ref{tab:LIS}). 

\subsection{Estimate of the \HI\ column density}
\label{sec:lya}

\begin{figure*}
\sidecaption
	\includegraphics[width=1.4\columnwidth]{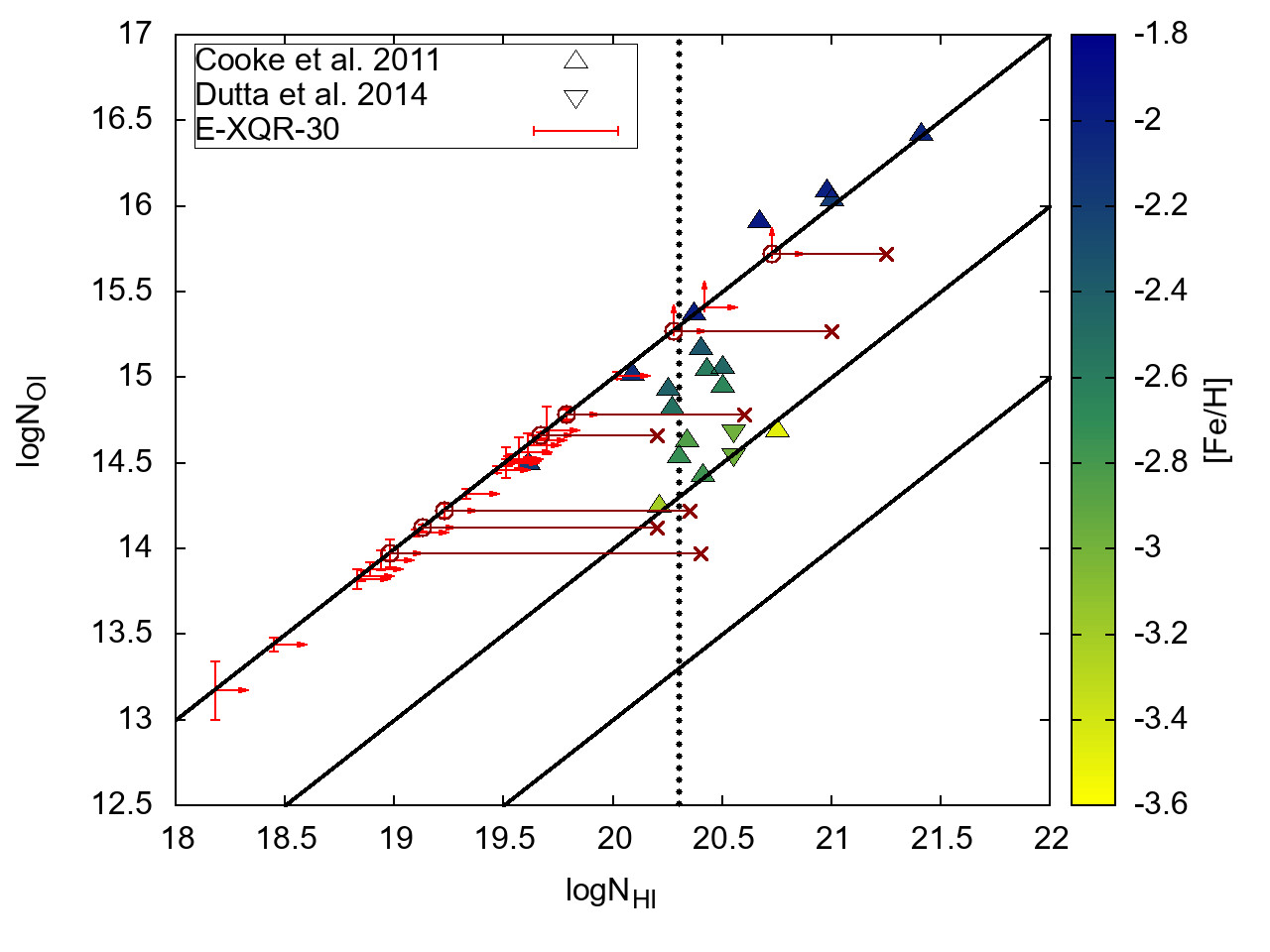}
    \caption{Log $N({\rm \OI})$ as a function of $\log N({\rm \HI})$ for the VMP DLAs of \citet[][triangles]{Cooke2011b}. The color of the triangles refers to their metallicity measured as [Fe/H] and reported in the side band. The solid black lines are approximately marking metallicities of [Fe/H]~$\simeq -2$, $-3$ and $-4$ (from top to bottom). The dotted black line indicates the column density threshold defining DLAs. The red lower limits on $\log N({\rm \HI})$ are the \OI\ absorbers in our sample, red circles indicate PDLA systems. Dark-red crosses mark estimated $\log N({\rm \HI})$ for the latter systems, obtained from the fit of the red damping wing of the absorption.  }
   \label{fig:NHI_OI}
\end{figure*}

\begin{figure}
\centering
	\includegraphics[width=\columnwidth]{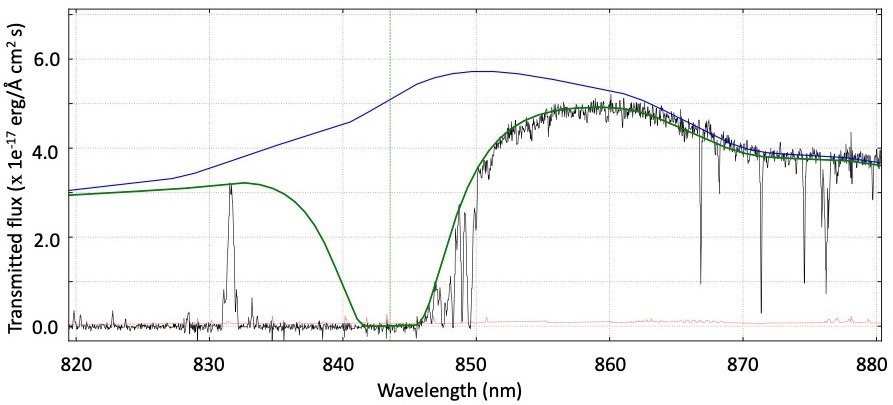}
    \caption{Region of the spectrum of the QSO SDSS J2310+1855 corresponding to the \Lya\ emission. The black line is the observed spectrum and the red one is the observed error. The blue line is the intrinsic emission spectrum modeled with a template (see text), while the green line is the fit of the \HI\ \Lya\ absorption line for the PDLA at $z=5.9388$.}
    \label{fig:lyafit}
\end{figure}

The measurement of the \HI\ column density in the observed systems is complicated by the increasing saturation of the \Lya\ forest observed at such high redshifts. 
Nonetheless, we have tried to estimate these column densities to understand whether the systems we analyzed are predominantly neutral (as we assume) or partially ionized.   

To this aim, we considered the relation between \OI\ and \HI\ column densities in the sample of VMP-DLAs at $2 < z < 4.5$ by \citet{Cooke2011b}. As shown in Fig.~\ref{fig:NHI_OI}, we found that there is a clear relation that links these column densities, which depends upon the metallicity of the system, measured as [Fe/H]. 
We verified that VMP-DLA systems with similar [Fe/H] are approximately aligned on straight lines defined by the relation:
\begin{equation}
\log N(\mathrm{\OI}) = \log N(\mathrm{\HI}) + \log (n_{\rm O} / n_{\rm H})_{\sun}+0.3 + [{\rm Fe}/{\rm H}], 
\label{eq:relOH}
\end{equation}
implying that the column density ratio of \OI\ to \HI\ for these systems is consistent with the solar oxygen abundance rescaled at their metallicity and increased by the $\alpha$ over iron excess, [O/Fe]~$\approx0.3$, produced by Type II SNe \citep[see e.g.][]{Rafelski2012}.
Solving eq.~\ref{eq:relOH} for $\log N({\rm \HI})$ and assuming that the \OI\ absorbers in our sample have a metallicity [Fe/H]~$\leq -2$, we derived lower limits on their \HI\ column densities from the measured \OI\ ones. We note that the majority of our DLA-analogs have [O/Fe] in the range $\sim 0.0-0.5$ consistent with stars and DLAs at [Fe/H]~$\le-2$, and there are also 5 cases with [O/Fe]~$> 0.55$ which hint at metallicities [Fe/H]~$<-3$ \citep{Cooke2011b,Welsh2022}. 
The obtained column density limits are reported in Tab.~\ref{tab:LIS}: 3 systems have  $\log N({\rm \HI}) \geq 20.3$, while 21 have $\log N({\rm \HI}) \geq 19$.

For the PDLAs, we estimated the \HI\ column density from the direct fit of the red damping wing of the typical Lorentzian profile of DLAs which appears in the region of the \Lya\ emission of the QSO \citep[see e.g.][and also Fig.~\ref{fig:lyafit}]{Dodorico2018}.   
In order to carry out the fit, it is necessary to model the QSO intrinsic emission spectrum in the region of the \Lya\ emission line. To this aim, we have used the template spectra built as in \citet{Bischetti2023} to determine the presence of BAL systems. Briefly, starting from the total catalog of 11\,800 SDSS QSOs in the redshift range $2.13<z_{\rm em}<3.20$ \citep{Shen2011}, the composite template spectrum is built as the median of a hundred randomly selected, non-BAL \citep{Gibson2009,Shen2011} QSO spectra matching within $\pm$ 20\% the slope of the rest-frame UV continuum and the equivalent width of the \CIV\ emission line of the considered E-XQR-30 QSO. The composite template is normalised to the median flux value of the QSO spectrum  in the rest-frame  1650-1750 \AA\ spectral interval, avoiding prominent emission lines and strong telluric absorption for the redshift interval covered by our sample. Further details can be found in \citet{Bischetti2022,Bischetti2023}. 

The composite spectra created in the rest frame were redshifted to the emission redshift of the considered QSO, rescaled to make the \CIV\ peak emissions coincide and then used as the continuum level in the context of \textsc{Astrocook}. In particular, to perform the fit of the \Lya\ line, we fixed the absorption redshift at the value determined by the low ionization metal lines of the system, and the Doppler parameter considering a value of $20$ \kms\ for each \OI\ component present in the velocity profile\footnote{Note that the shape of the Lorentzian profile does not significantly depend on the chosen Doppler parameter value}. In this way, we determined the column density of \HI\ that best fit the damping wing observed in the spectrum, with an estimated error of approximately $\pm 0.1$ dex. This error does not take into account the uncertainty on the true shape of the spectrum in the \Lya\ emission region, which is very difficult to estimate. For some cases, it was possible to use also the \Lyb\ absorption to constrain, in particular, the width of the \HI\ line (see figures in Appendix~\ref{app:LIS}).

Figure~\ref{fig:NHI_OI} shows that in all the cases in which it was possible to measure directly $\log N({\rm \HI})$, the obtained value was consistent with the system being a DLA, even if the value obtained from eq.~\ref{eq:relOH} was much lower than $\log N({\rm \HI})=20.3$. Based on this result, we proceeded with the calculation of the relative chemical abundances for the considered systems assuming they are VMP-DLAs. 

\section{Observational results}
\label{sec:Results}

\subsection{Determination of chemical abundances}
\label{sec:RelAbb}

\begin{table*}
\tiny
    \centering
    \caption{Chemical abundances relative to iron of the elements detected in the observed \OI\ systems.}
    \label{tab:RelAbbFe}
    \begin{tabular}{lcccccc}
        \hline
        QSO & $z_{\rm abs}$ & $\rm [C/Fe]$ & $\rm [O/Fe]$ & $\rm [Mg/Fe]$ & $\rm [Al/Fe]$ & $\rm [Si/Fe]$\\
        \hline
        PSO J308-27 & $5.4400$ & $0.18\pm 0.07$ & $0.33\pm 0.06$ & $0.39\pm 0.08$ & $<0.34$ & $0.56\pm 0.09$\\
        PSO J308-27 & $5.6268$ & $0.21\pm 0.04$ & $0.26\pm 0.04$ & $0.31\pm 0.06$ & $0.17\pm 0.13$ & $0.33\pm 0.03$\\
        PSO J023-02 & $5.4869$ & $0.70\pm 0.18$ & $0.49\pm 0.10$ & $0.44\pm 0.15$ & $<0.74$ & $<0.78$\\
        PSO J025-11$^*$ & $5.7763$ & $0.46\pm 0.06$ & $-0.07\pm 0.06$ & $0.50\pm 0.12$ & $0.53\pm 0.17$ & $0.59\pm 0.06$\\
        PSO J025-11$^*$ & $5.8385$ & $>0.03$ & $>0.03$ & $>-0.02$ & $-0.02\pm 0.20$ & $0.28\pm 0.10$\\
        PSO J108+08 & $5.5624$ & $>0.28$ & $>0.74$ & $\cdots$ & $\cdots$ & $\cdots$\\
        SDSS J0818+1722 & $5.7912$ & $0.25\pm 0.05$ & $0.36\pm 0.05$ & $\cdots$ & $\cdots$ & $0.37\pm 0.05$\\
        SDSS J0818+1722 & $5.8767$ & $0.35\pm 0.09$ & $0.44\pm 0.08$ & $0.37\pm 0.12$ & $<0.46$ & $0.46\pm 0.08$\\
        PSO J007+04$^*$ & $5.9917$ & $-0.05\pm 0.21$ & $0.21\pm 0.22$ & $<0.17$ & $<0.58$ & $0.11\pm 0.20$\\
        SDSS J2310+1855$^*$ & $5.9388$ & $>0.08$ & $>0.76$ & $>0.07$ & $-0.10\pm 0.39$ & $0.39\pm 0.10$\\
        PSO J158-14 & $5.8986$ & $0.24\pm 0.04$ & $0.27\pm 0.08$ & $\cdots$ & $0.17\pm 0.21$ & $0.50\pm 0.05$\\
        PSO J239-07$^*$ & $5.9918$ & $>0.11$ & $>0.49$ & $>0.23$ & $\cdots$ & $>0.45$\\
        ULAS J1319+0950$^*$ & $6.0172$ & $-0.21\pm 0.19$ & $0.29\pm 0.16$ & $0.06\pm 0.22$ & $0.44\pm 0.27$ & $0.04\pm 0.16$\\
        PSO J060+24 & $5.6993$ & $0.33\pm 0.19$ & $0.35\pm 0.11$ & $\cdots$ & $0.39\pm 0.24$ & $0.67\pm 0.08$\\
        PSO J065-26 & $5.8677$ & $0.18\pm 0.04$ & $0.40\pm 0.03$ & $0.12\pm 0.06$ & $0.10\pm 0.16$ & $0.44\pm 0.04$\\
        PSO J065-26$^*$ & $6.1208$ & $>0.34$ & $>0.25$ & $>0.35$ & $0.29\pm 0.04$ & $0.56\pm 0.03$\\
        PSO J065-26$^*$ & $6.1263$ & $>0.60$ & $-0.06\pm 0.04$ & $>0.87$ & $0.78\pm 0.06$ & $0.65\pm 0.04$\\
        SDSS J0100+2802 & $5.7974$ & $0.04\pm 0.03$ & $0.32\pm 0.04$ & $\cdots$ & $-0.01\pm 0.08$ & $0.43\pm 0.04$\\
        SDSS J0100+2802 & $5.9450$ & $0.39\pm 0.05$ & $0.00\pm 0.06$ & $0.29\pm 0.06$ & $0.25\pm 0.15$ & $0.51\pm 0.09$\\
        SDSS J0100+2802 & $6.1114$ & $0.33\pm 0.07$ & $0.85\pm 0.14$ & $0.35\pm 0.05$ & $-0.04\pm 0.10$ & $0.35\pm 0.04$\\
        SDSS J0100+2802 & $6.1434$ & $0.26\pm 0.03$ & $0.55\pm 0.03$ & $0.24\pm 0.04$ & $-0.13\pm 0.11$ & $0.34\pm 0.03$\\
        DELS J1535+1943 & $5.8990$ & $0.54\pm 0.16$ & $\cdots$ & $0.40\pm 0.20$ & $\cdots$ & $0.52\pm 0.08$\\
        PSO J183+05 & $6.0642$ & $0.23\pm 0.08$ & $0.18\pm 0.05$ & $0.59\pm 0.20$ & $0.45\pm 0.08$ & $0.54\pm 0.05$\\
        PSO J183+05$^*$ & $6.4041$ & $0.22\pm 0.07$ & $0.22\pm 0.05$ & $>0.60$ & $0.07\pm 0.08$ & $0.58\pm 0.04$\\
        WISEA J0439+1634 & $6.2743$ & $>0.50$ & $>0.38$ & $\cdots$ & $\cdots$ & $>0.58$\\
        VDES J0224-4711 & $6.1228$ & $0.21\pm 0.14$ & $0.41\pm 0.06$ & $0.18\pm 0.07$ & $<0.13$ & $0.50\pm 0.08$\\
        PSO J036+03 & $6.0611$ & $>0.41$ & $>0.56$ & $>0.66$ & $>0.73$ & $>0.95$\\
        DELS J0923+0402 & $6.3784$ & $>0.38$ & $>0.84$ & $>0.23$ & $\cdots$ & $>0.57$\\
        \hline
    \end{tabular}
    \tablefoot{
    \tablefoottext{*}{These systems are PDLA.}
    }
\end{table*}

We convert directly ionic column densities into total abundances of a given element in the hypothesis that our systems are predominantly neutral - thus ionic corrections are not needed - and very metal poor - implying a negligible, if present, dust depletion \citep[see e.g.][]{Becker2012}. 

The relative abundance between two elements X and Y was then calculated using the relationship:
\begin{equation}
    \bigg[\frac{\rm X}{\rm Y}\bigg]=\log\frac{N_{\rm X}}{N_{\rm Y}}-\log\frac{n_{\rm X_{\sun}}}{n_{\rm Y_{\sun}}}
    \label{eq:RelAbb}
\end{equation}
where $N_{\rm X}$ and $N_{\rm Y}$ are the column densities of the ions of element X and Y detected in the systems, and $n_{\rm X_{\sun}}$ and $n_{\rm Y_{\sun}}$ are the solar abundances of X and Y in number. All quantities are calculated with respect to the solar photospheric values presented in \citet{Asplund2009}. The error on relative abundances has been propagated neglecting the error on solar abundances as the uncertainties on the measured column densities are dominant. Lower and upper limits on column densities are reflected in the calculated relative abundances. When both ionic column densities of X and Y were limits, the relative abundance was not computed. Note that the chemical abundances of carbon and silicon are determined from the column densities of \CII\ and \SiII, implying that \CIV\ and \SiIV, if present, arise in a different gas phase from the low ionization lines.

\begin{figure*}
\centering
	\includegraphics[width=8cm]{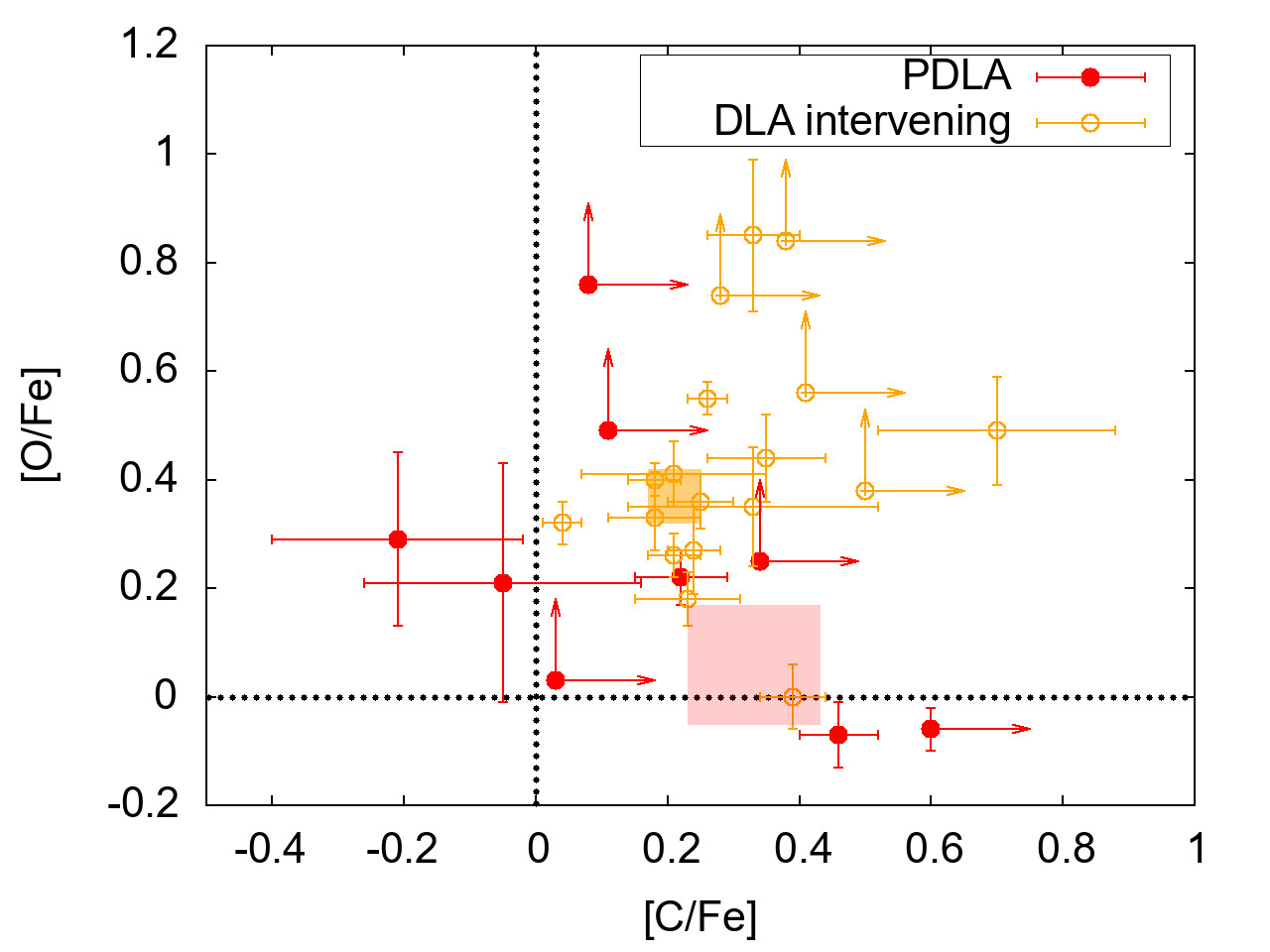}
	\includegraphics[width=8cm]{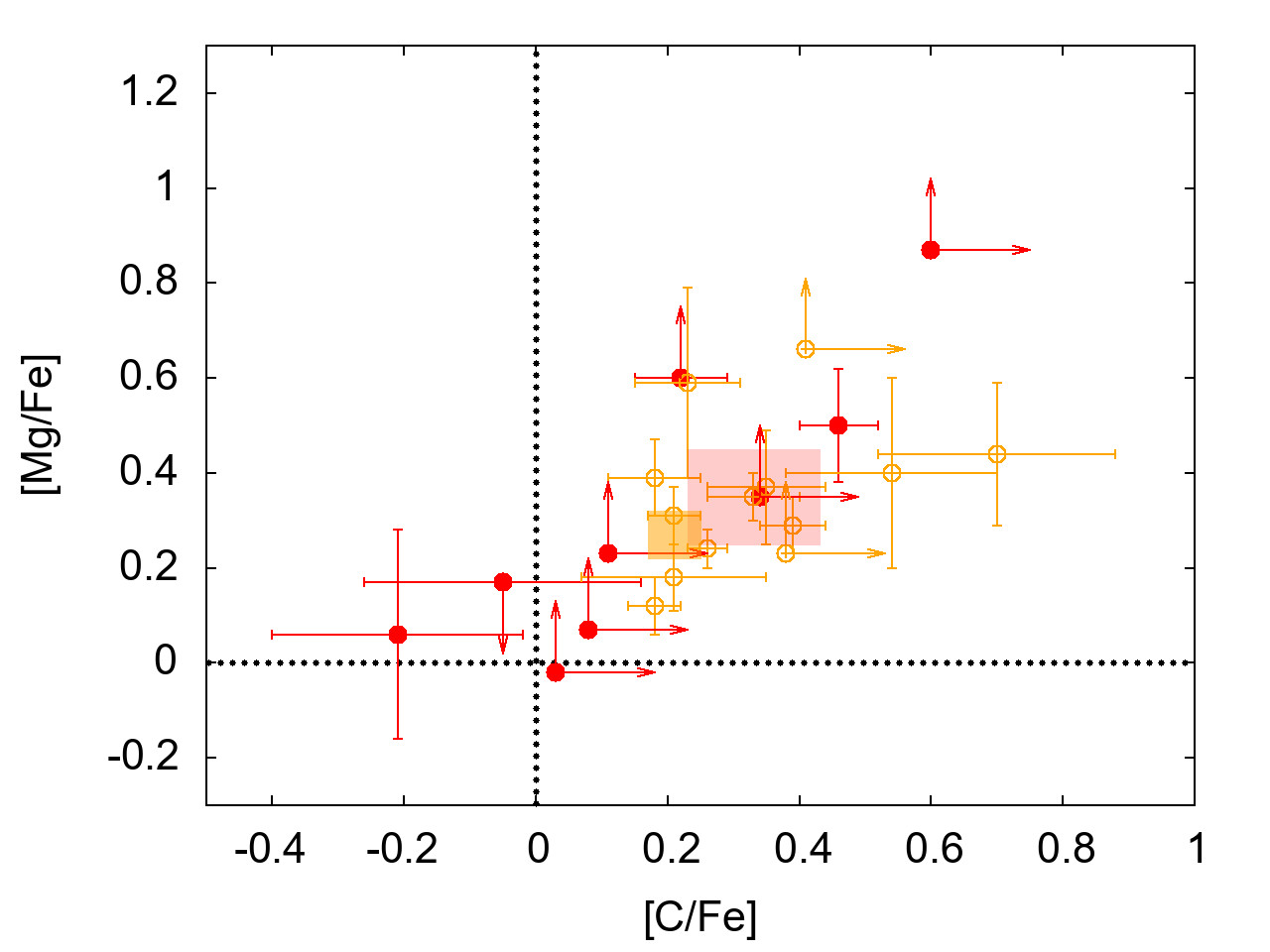}
	\includegraphics[width=8cm]{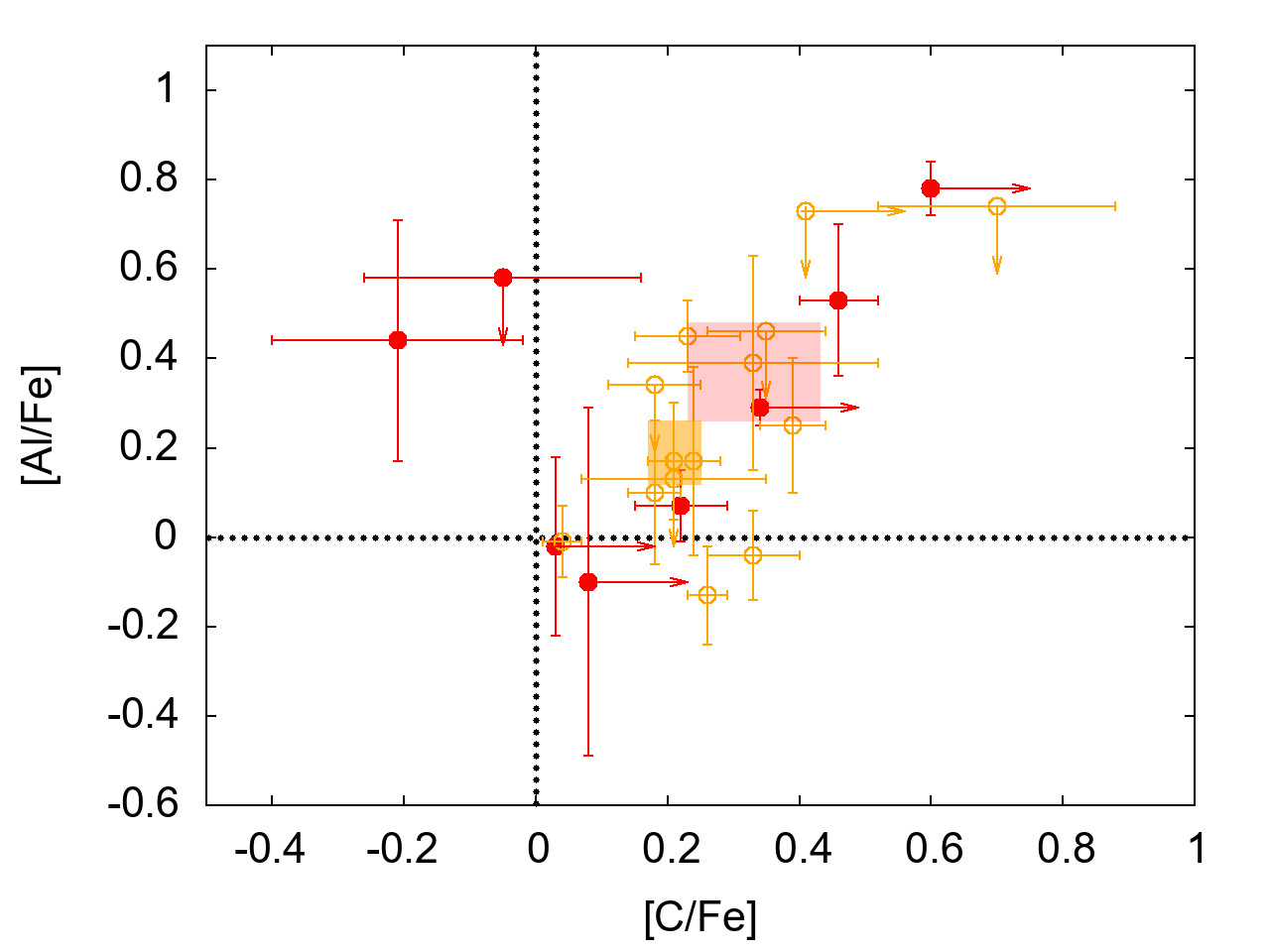}
	\includegraphics[width=8cm]{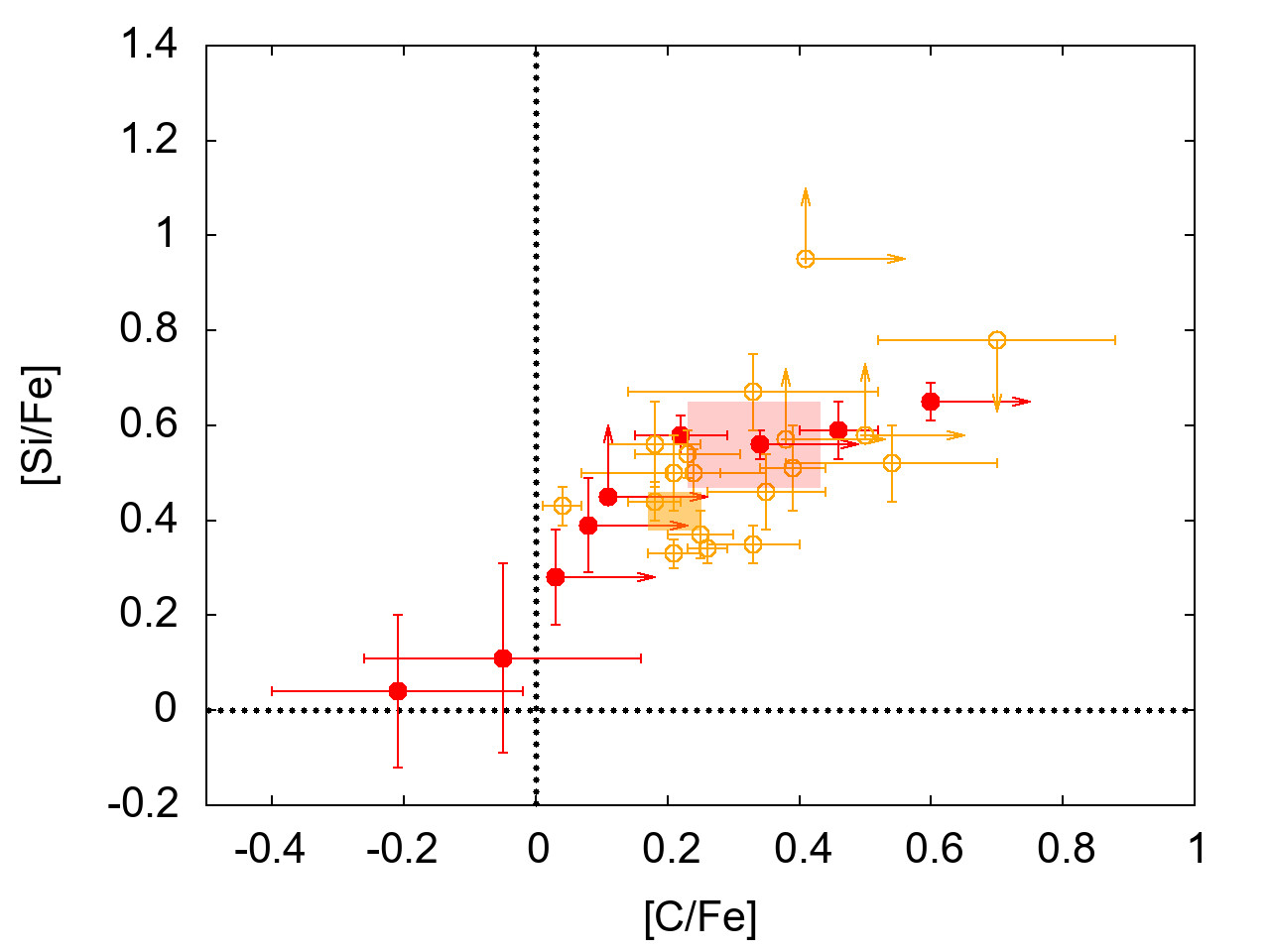}
    \caption{Abundances relative to iron of the elements detected in our \OI\ systems. In the panels, PDLAs (red filled circles) are distinguished from intervening DLAs (orange empty circles). The orange shaded box (red shaded box) represents the weighted average value of the DLA (PDLA) abundances with the relative dispersion, calculated as the standard deviation of the sample mean. }
    \label{fig:DLAconf}
\end{figure*}

Table \ref{tab:RelAbb} reports the chemical abundances for the DLA-analogs for which it was possible to estimate the column density of \HI, while Tabs.~\ref{tab:RelAbbFe} and \ref{tab:RelAbbO} show the relative abundances, with respect to iron and oxygen, of all analyzed systems. 
Figure \ref{fig:DLAconf} shows the abundances of O, Mg, Al and Si relative to iron as a function of $[{\rm C/Fe}]$ for all the systems in our sample. In these plots, we distinguish between PDLAs (red squares) and intervening DLAs (orange circles) to check if there are systematic differences between these two classes of absorbers (although the definition is based empirically only on the velocity separation from the QSO emission redshift). 
The colored boxes represent the weighted average values of the two categories of absorbers, with their dispersion calculated as the standard deviation of the sample mean. The weighted averages are computed treating lower and upper limits as measurements with an error of 0.1 dex. 

The plots show that, in general, the average relative abundances of these two types of absorbers are in agreement within errors. This suggests that the ionizing flux of the QSO does not have a clear influence on the nature of the associated systems.
 
We note that PDLAs present a larger spread in the abundance values with respect to the intervening ones, this is probably due to the smaller size of the sample and to the large uncertainties characterizing, in particular, the systems along the sightline to PSO J007+04 and to J1319+0950.
These two systems also present a subsolar abundance of C/Fe (see Fig.~\ref{fig:DLAconf}).
However, as we will see in Sect.~\ref{sec:Conf_lett}, these abundances are not peculiar, there are other intervening DLAs at $z>4.5$, detected by \citet{Becker2012} and \citet{Poudel2020}, which present $[{\rm C/Fe}]<0$ as shown in Fig.~\ref{fig:literatureconf}.
In the following analysis we will consider the full sample comprising PDLAs and intervening DLAs.

\subsection{Comparison with the literature}
\label{sec:Conf_lett}

\begin{figure*}
\centering
	\includegraphics[width=8.5cm]{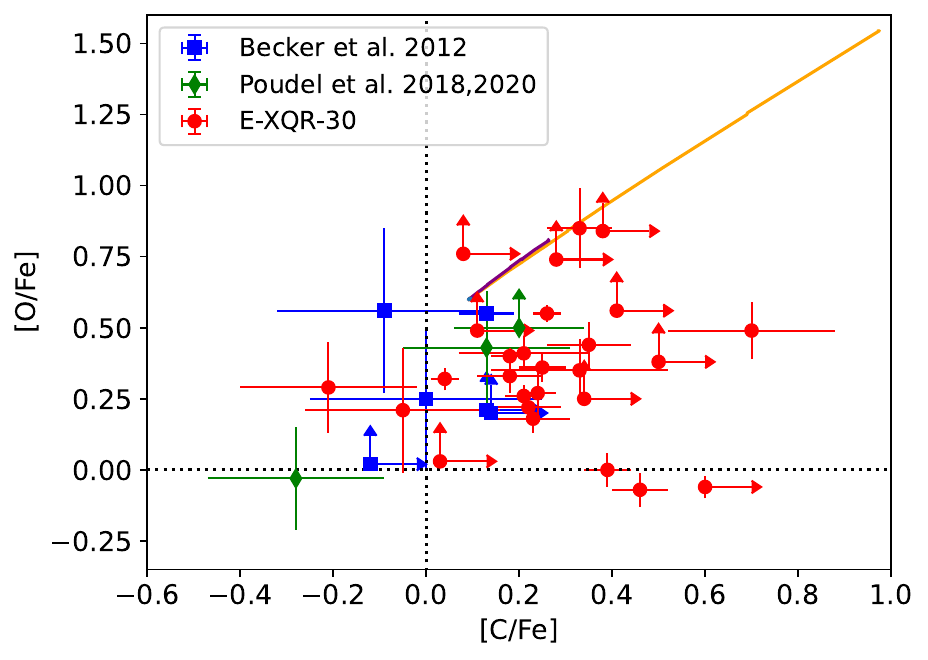}
	\includegraphics[width=8.5cm]{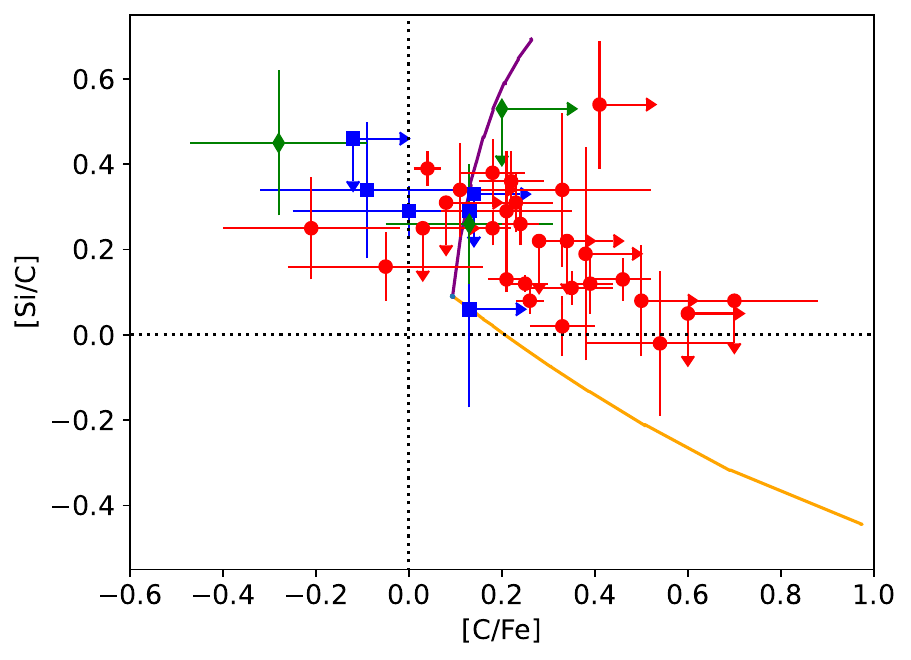}
	\includegraphics[width=8.5cm]{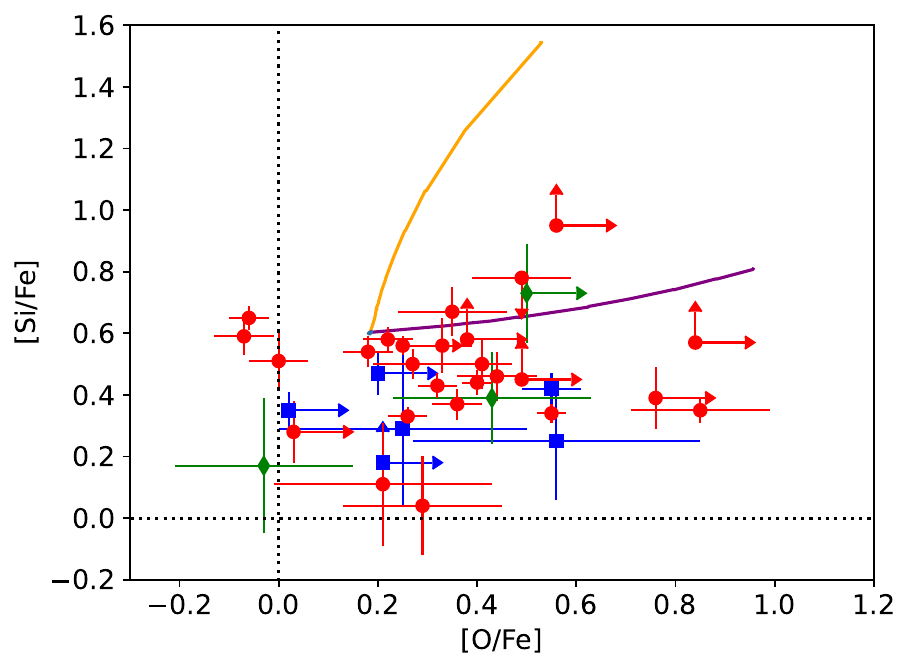}
	\includegraphics[width=8.5cm]{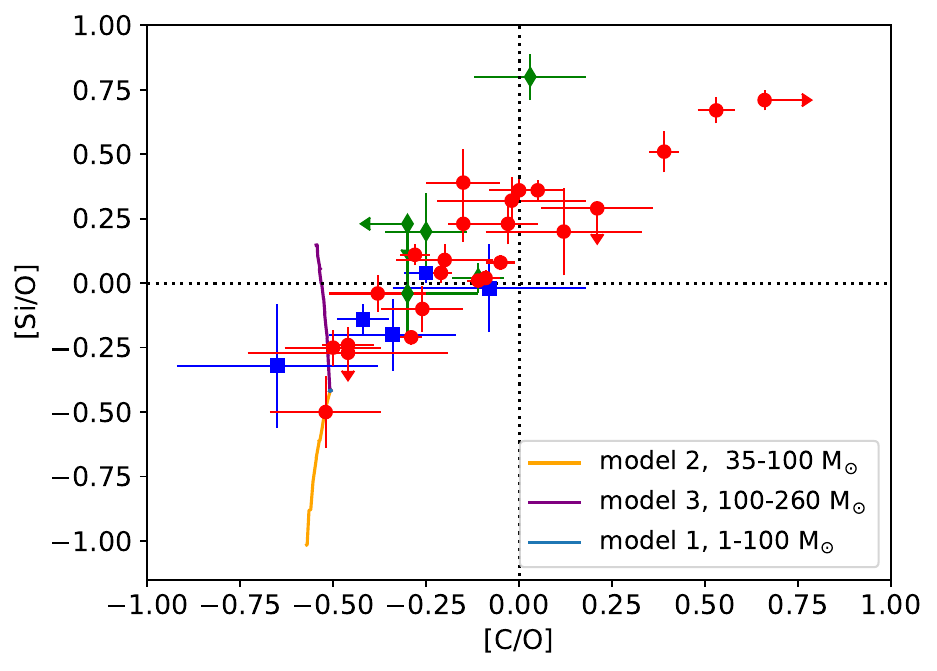}
    \caption{Comparison of some of the relative abundances of the \OI\ systems/DLAs at high redshift ($z>4.5$) found in this work (red dots), in \citet{Becker2012} (blue squares), and \citet{Poudel2018,Poudel2020} (green diamonds). Dotted lines mark the solar abundance values. We report also the predictions of the simulations by \citet{kulkarni2013} at $z =6$, the three models are described in greater detail the text.  Model~1 assigns a 1--100~M$_\odot$ Salpeter IMF to Pop~III stars.  This is 35--100~M$_\odot$ Salpeter in case of model~2 and 100--260~M$_\odot$ Salpeter for model~3.  The prediction of model 1 is the light blue dot at the junction of the lines of model 2 and 3.}
    \label{fig:literatureconf}
\end{figure*}

We compared the relative abundances of the elements in the DLA-analogs identified in this work with those of low ionization systems at similar redshifts and selected in the same way presented in \citet{Becker2012}. For the systems detected in SDSS J0818+1722, which are in common between the two samples, we adopted our chemical abundances.  We considered also the systems in the redshift range $4.5 \lesssim z \lesssim 5.5$ studied by   \citet{Poudel2018,Poudel2020} selecting only those with $\log N_{\rm HI} \ge 20.10$ and metallicity [Fe/H]~$\le -2.0$ (or [O/H]~$\le -2.0$ when iron is not measured), for a total of 5 systems. 
Figure \ref{fig:literatureconf} reports 4 combinations of relative abundances computed with the available chemical elements, it shows that our measurements are in general agreement with previous ones. It is important to note that due to the simultaneous saturation of the carbon and oxygen lines in 
2 systems of \citet{Becker2012} and 3 of our systems, the sample for which the [C/O] abundance could be calculated is limited.

\begin{table}
    \centering
    \caption{Weighted average of the chemical abundances of the VMP-DLAs at $2 \le z \le 4.5$ and DLA-analogs at $4.5 < z \le 6.5$. }
    \label{tab:meanAbb}
    \begin{tabular}{ccccc}
        \hline
        $z$ & N & Mean & $\sigma_{\rm sample}$ & $\sigma_{\rm mean}$\\ 
        \hline
        \multicolumn{5}{c}{$[{\rm C/Si}]$}\\
        \hline
        $2 \le z \le 4.5$ & 11 & $-0.21$ & 0.08 & 0.02 \\
        $4.5 < z \le 6.5$ & 40 & $-0.19$ & 0.12 & 0.02 \\
        \hline
        \multicolumn{5}{c}{$[{\rm C/O}]$}\\
        \hline
        $2 \le z \le 4.5$ & 11 & $-0.38$ & 0.12 & 0.04 \\
        $4.5 < z \le 6.5$ & 34 & $-0.11$ & 0.24 & 0.04 \\
        \hline
        \multicolumn{5}{c}{$[{\rm C/Fe}]$}\\
        \hline
        $2 \le z \le 4.5$ & 9 & 0.14 & 0.08 & 0.03 \\
        $4.5 < z \le 6.5$ & 37 & 0.22 & 0.14 & 0.02 \\
        \hline
        \multicolumn{5}{c}{$[{\rm O/Fe}]$}\\
        \hline
        $2 \le z \le 4.5$ & 20 & 0.42 & 0.05 & 0.01 \\ 
        $4.5 < z \le 6.5$ & 36 & 0.32 & 0.21 & 0.04 \\
        \hline
        \multicolumn{5}{c}{$[{\rm Si/Fe}]$}\\
        \hline
        $2 \le z \le 4.5$ & 20 & 0.32 & 0.04 & 0.01 \\
        $4.5 < z \le 6.5$ & 36 & 0.46 & 0.13 & 0.02 \\
        \hline
        \multicolumn{5}{c}{$[{\rm Si/O}]$}\\
        \hline
        $2 \le z \le 4.5$ & 22 & $-0.11$ & 0.08 & 0.02 \\
        $4.5 < z \le 6.5$ & 39 & $0.11$   & 0.25 & 0.04 \\
        \hline
    \end{tabular}
    \tablefoot{The columns report: the redshift interval, the number of systems, the weighted average, the standard deviation of the sample and the standard deviation of the mean.  }
\end{table}

\begin{figure*}
\centering
	\includegraphics[width=9cm]{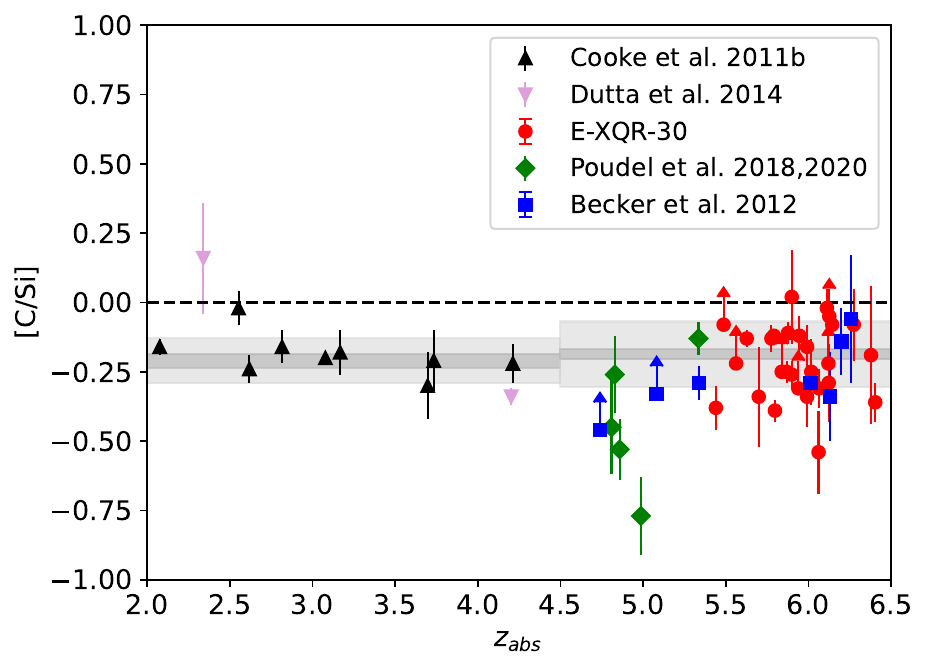}
	\includegraphics[width=9cm]{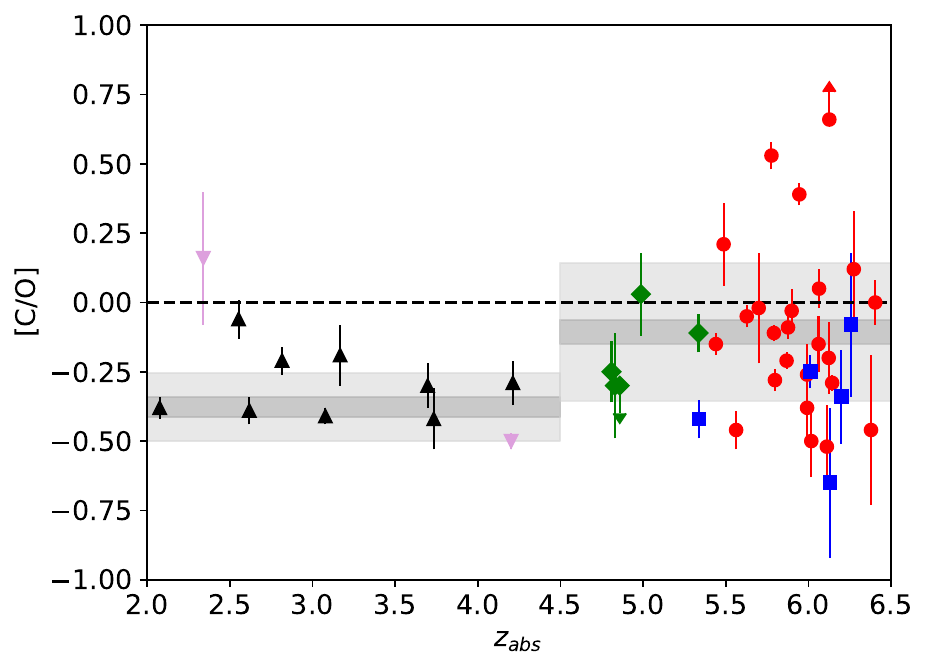}
	\includegraphics[width=9cm]{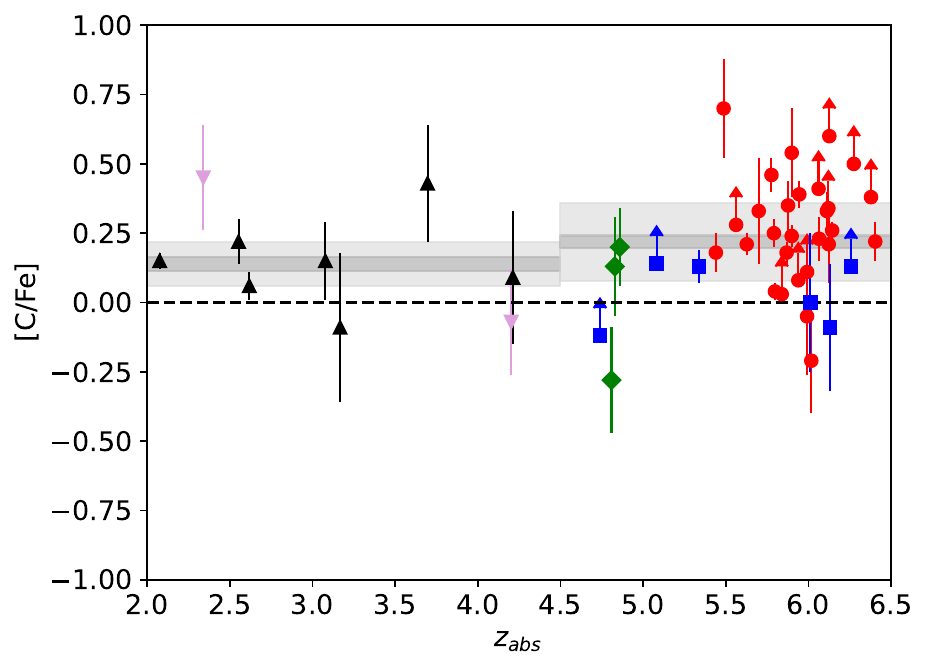}
	\includegraphics[width=9cm]{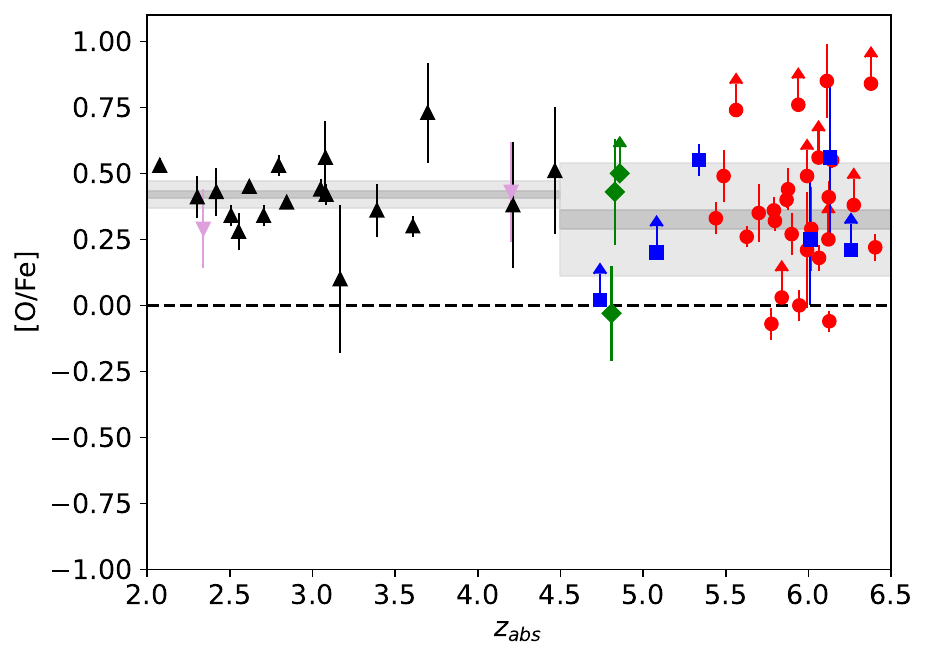}
	\includegraphics[width=9cm]{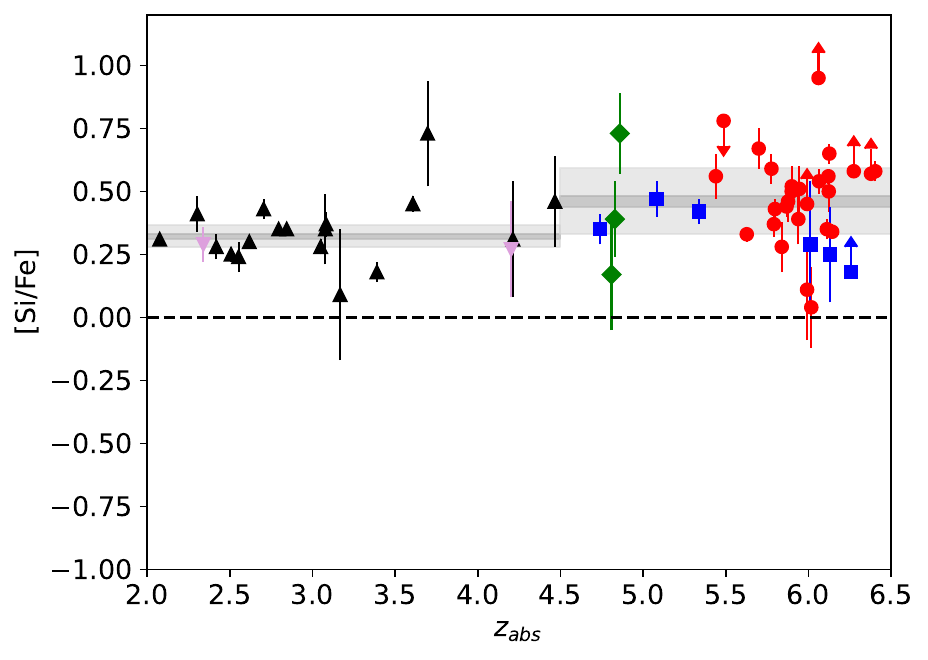}
	\includegraphics[width=9cm]{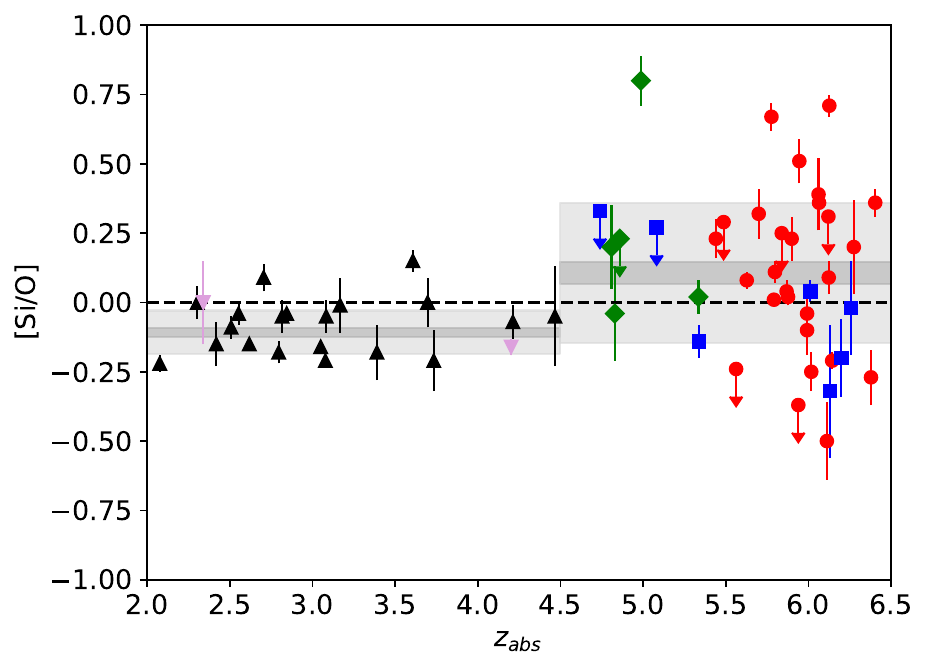}
    \caption{Relative abundances of DLA systems at $z>2$, as a function of redshift. The figure shows the DLA analogs studied in this work  (solid red dots), in \citet[][solid blue squares]{Becker2012} and in \citet[][solid green diamonds]{Poudel2018,Poudel2020}, and the VMP-DLAs analyzed in \citet[][solid black triangles]{Cooke2011b} with the exception of the systems identified in the QSOs J0035-0918 and J1558-0031, and those in \citet[][solid plum inverted triangles]{Dutta2014} in the QSOs J0035-0918 and J0953-0504. The shaded regions are centred on the weighted average values and represent the standard deviation of the sample (light gray band) and of the sample mean (dark gray band) for the systems in the two redshift bins: $2 \le z \le 4.5$ and $4.5 < z \le 6.5$.}
    \label{fig:redshift}
\end{figure*}

Following the analysis in \citet{Becker2012}, we compared the relative abundances of the identified high redshift systems ($z>4.5$), with the abundances of the VMP-DLAs at lower redshift ($2<z<4.5$) presented in \citet{Cooke2011b} and \citet{Dutta2014}, to evaluate the evolution of the chemical composition of the gas with redshift. 
In the VMP-DLA sample of \citet{Cooke2011b}, we omitted the system at $z=2.34010$ found in QSO J0035-0918 and the system at $z=2.70262$ in J1558-0031 due to possible large uncertainties in column densities. In the sample of \citet{Dutta2014}, we have instead considered only the system at $z=2.340$ in J0035-0918 (in which the values of \citealt{Cooke2011b} are corrected) and the one at $z=4.202$ in J0953-0504. 

Figure~\ref{fig:redshift} shows several relative chemical abundances computed for the low and high-redshift samples as a function of the redshift of the absorbers. The total number of systems at $z>4.5$ presented in this study has increased by a factor of $\sim4$ the sample studied in previous works, allowing us to carry out the comparison in a more quantitative way. In order to establish if the distributions of values in the low and high-redshift samples are significantly different, we determined the weighted averages of the two groups of measurements (considering lower and upper limits as measurements). 
The results are reported in Tab.~\ref{tab:meanAbb} and in Fig.~\ref{fig:redshift}, where the standard deviation of each sample, $\sigma_{\rm sample}$, is represented by the light grey shaded regions, while the standard deviation of the mean, $\sigma_{\rm mean} = \sigma_{\rm sample} / \sqrt{N}$, is represented by the dark grey regions, both centred on the mean values. 

\citet{Becker2012} and \citet{Poudel2018,Poudel2020} had found very good agreement between the mean values and the dispersions of the VMP-DLAs at $z<4.5$ and their $z>4.5$ samples.  This result suggested that the sources enriching the high redshift systems had the same nature as those that enriched the low redshift ones, which most likely were type II SNe from Pop II stars.  Differently, the mean values of the measured relative chemical abundances in our sample are not all in agreement with those of the lower $z$ sample of VMP-DLAs, in particular if we consider the error of the mean, the [C/O] and [Si/Fe] differ by $\sim 7\, \sigma_{\rm mean}$, while [Si/O] differs by $\sim 5.5\, \sigma_{\rm mean}$.  Furthermore, we observe an increase in the dispersion of the measurements of the high-$z$ sample.

We have also carried out the Epps-Singleton test, which is useful to compare two samples and understand if they are extracted from the same distribution, for the cases in which both the low and high-redshift samples have a relatively large number of measurements ([Si/Fe], [Si/O] and [O/Fe]). In all three cases, we obtain a significance level lower than the generally adopted threshold $p = 0.05$, implying that the null hypothesis (samples are drawn from the same distribution) is rejected.   
The implications of these results are discussed in Sec.~\ref{sec:TheoreticalAbb}-\ref{sec:discussion}. 

%%%%-----------------------------------
\section{Are we observing the imprint of the first stars?}
\label{sec:TheoreticalAbb}
In order to understand if our high-redshift absorbers have been imprinted by the first stellar generations, we compare the measured chemical abundance ratios with the results of chemical evolution models in the literature and, in particular, with: the semi-analytic code by \citet{kulkarni2013} and the general parametric study by \citet{Vanni2024}.

\subsection{Comparison with Kulkarni et al. 2013}
In Fig.~\ref{fig:literatureconf}, we plot, together with our observational results, the predictions of the semi-analytic chemical evolution model by \citet{kulkarni2013}.  This  model incorporates global effects like reionization and photoionization feedback along with a range of possible Pop~III initial mass functions (IMFs).  The model is consistent with a variety of observational constraints on galaxy and IGM evolution. \citet{kulkarni2013} calculate the minimum mass of star-forming galaxies self-consistently and produce galaxies that lie on observational curves such as the stellar-to-halo mass relation at low redshift and the mass–metallicity relation. They then explore the influence of Pop~III stars on the predictions of this model, by varying the Pop~III stellar IMF.  They consider haloes of different masses that grow according to growth rates calibrated on the Millennium simulation \citep{fakhouri2010}.  A semi-analytic model of galaxy formation and chemical evolution is implemented on these haloes.

Pop~I/II stars are formed following a Salpeter IMF with masses in the range $m_*=0.1-100\,M_{\odot}$, while Pop~III stars can have three different IMFs: model 3 covers the PISN range ($m_*=100-260\,M_{\odot}$ Salpeter), whereas models 1 and 2 choose Salpeter IMFs with $m_*=1-100\,M_{\odot}$ and $m_*=35-100\,M_{\odot}$, to explore the role of asymptotic giant branch (AGB) stars and core-collapse SNe, respectively. Chemical yields of Pop~III stars are taken from the calculations of \citet{Heger2002}. 
%Stellar spectra of Pop~I/II stars are calculated using starburst99 \citep{leitherer1999,vazquez2005} with respective metallicities. 
The transition from Pop~III to Pop~II star formation in any halo is implemented via a critical metallicity threshold, $Z_{\rm crit} = 10^{-4}\,Z_{\odot}$ \citep{bromm2001,schneider2003}.  

In Fig.~\ref{fig:literatureconf}, we show the simulated abundances at $z=6$; moving to lower redshift, the spread of the values decreases, and relative abundances tend to converge to the average abundance of the PopII stars, represented by the predictions for model 1 \citep[see Fig.~8 of][]{kulkarni2013}. Given the four combinations of relative abundances that we have considered for this comparison, Fig.~\ref{fig:literatureconf} suggests that the spread of the observed data is not in agreement with either of the three models considered by \citet{kulkarni2013}.  There does appear to be a weak preference for model 3, in which the Pop~III IMF is 100--260~M$_\odot$ Salpeter, however this does not appear conclusive. 

The poor agreement between models and data is perhaps not surprising, as the three Pop~III IMFs considered by \citet{kulkarni2013} were chosen by these authors for illustrative purposes only, and recent studies based on stellar archaeology showed that the Pop III IMF have different mass ranges and shapes \cite{Rossi2021, Pagnini2023, Koutsouridou2024}. A future study could investigate whether there is a Pop~III IMF that fits the data shown in Fig.~\ref{fig:literatureconf} far better. Nonetheless, the general conclusion by \citet{kulkarni2013} was that the distribution of metal absorbers in the relative abundance plane at high redshifts is sensitive to the Pop~III IMF, and a generic signature of Pop~III enrichment would be a correlation between relative abundances along a particular direction.  While, with enrichment only from a universal IMF, one would generally expect the relative abundances to be clustered around a point in the relative abundance plane.  This behavior is clearly borne out by the measurements shown in Fig.~\ref{fig:literatureconf}.  

A similar approach, with three different IMFs for Pop~III stars but applied to a full 3D numerical chemistry simulation, was proposed in \citet{Ma2017}. Comparing the distribution of our measurements in Fig.~\ref{fig:literatureconf} with their results, we can exclude for our absorbers an enrichment from very massive Pop~III stars (PISNe, left column of their Fig. 5), while we observe a better agreement in all three abundance planes reported by \citet{Ma2017} considering intermediate enrichment from a population of massive Pop~III stars (Salpeter IMF  with $m_*=10-100\,M_{\odot}$, central column of their Fig. 5).  

\subsection{Comparison with Vanni et al. 2024} 
\begin{figure*}
    \centering
    \includegraphics[width=\linewidth]{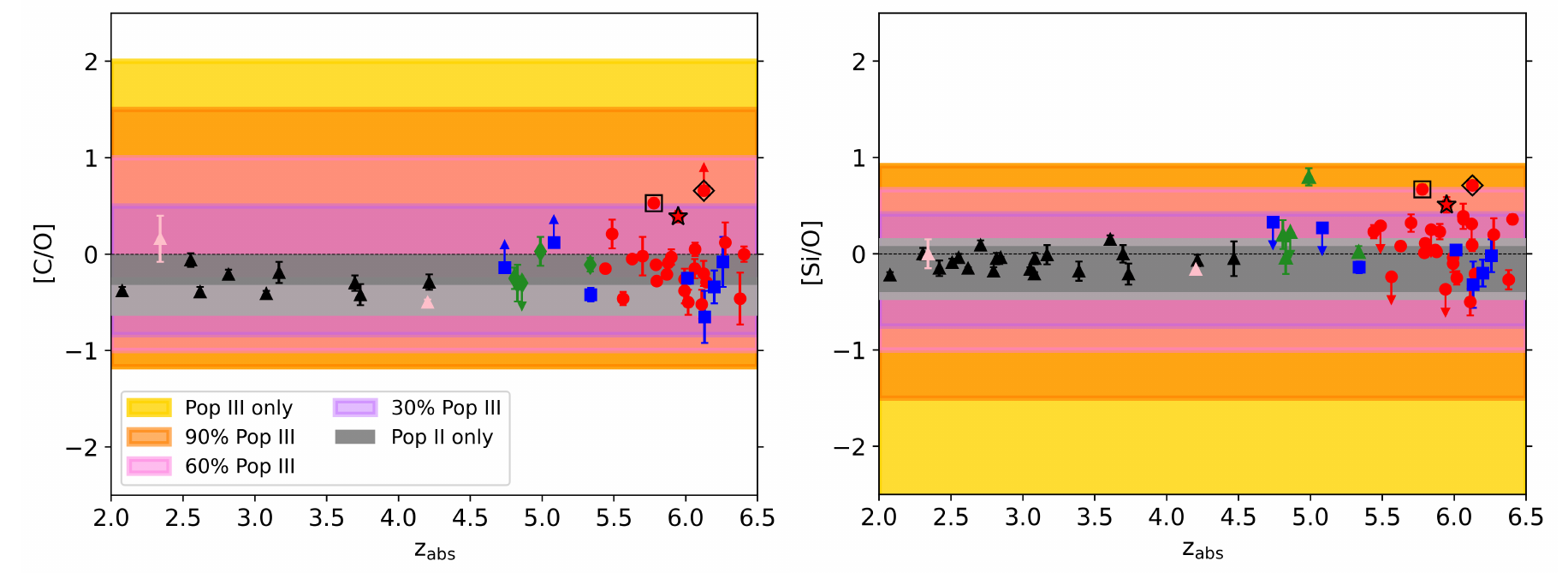}
    \caption{Comparison between the maximum extent of [C/O] (left panel) and [Si/O] (right panel) ISM ratios predicted in the model by \citet[][shaded area]{Vanni2024} and the measured abundance ratios in absorbers at different redshifts (points with errorbars as in Fig.~ \ref{fig:redshift}). In each panel the colored areas show the contribution of Pop~III stars to the chemical enrichment: $100\%$ (yellow), $90\%$ (orange), $60\%$ (pink), $30\%$ (purple) and $\leq 0.01\%$ (dark grey: \citealt{Limongi2018}; light grey: \citealt{woosley1995}). Only the environments with [O/H]$>-4$ are considered.
    The three remarkable systems with [C/O]$>0.35$ and [Si/O]$>0.50$ are highlighted using different symbols and black contours: PSO J025-11 at z=5.7763 (square), SDSS J0100+2802 at z=5.9450 (star) and PSO J065-26 at z=6.1263 (diamond).}
    \label{fig:scatter_points}
\end{figure*}

\begin{figure*}
    \centering
    \includegraphics[width=\linewidth]{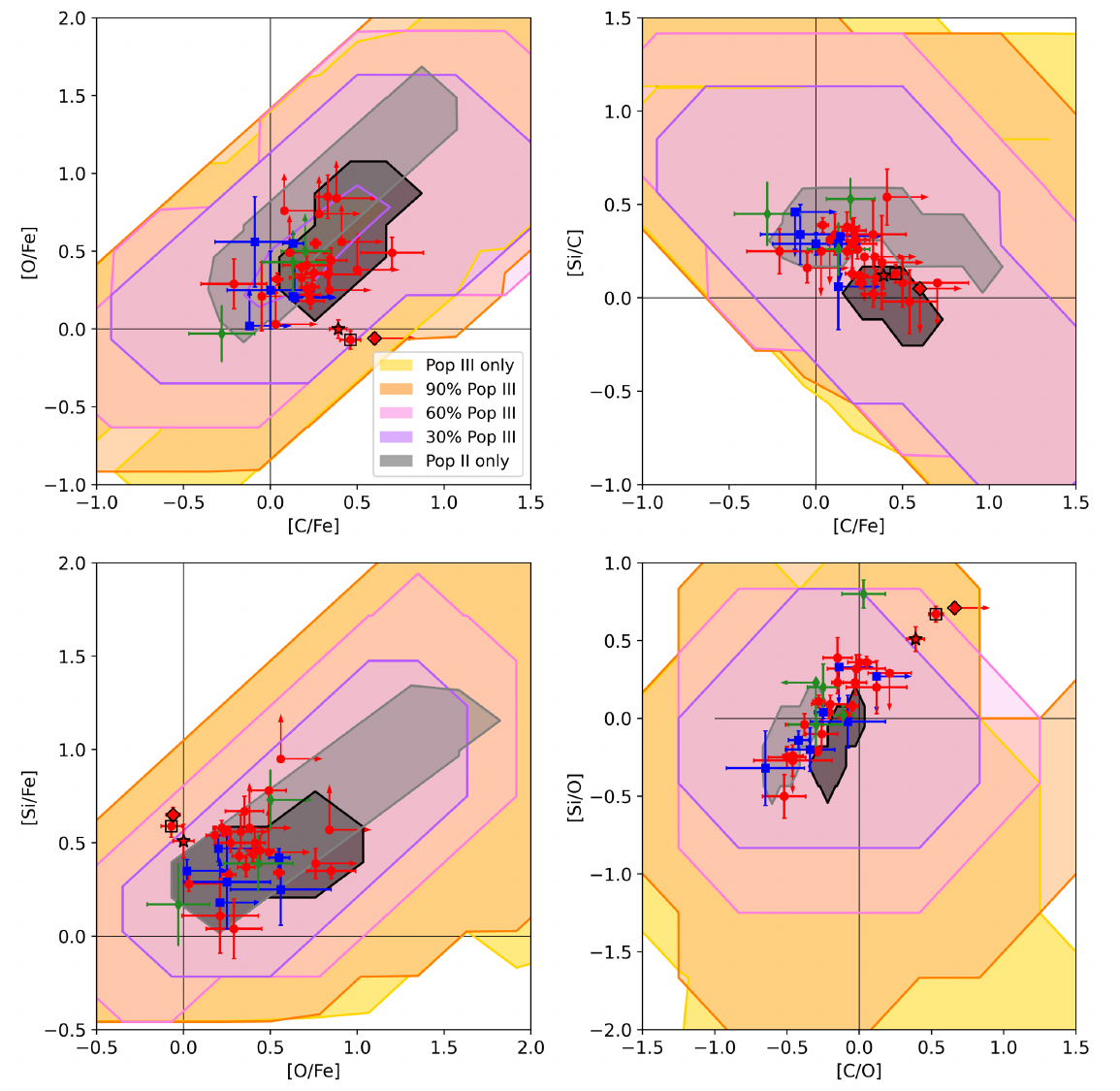}
    \caption{Comparison between different chemical abundance ratios measured in the considered sample of $z> 4.5$ DLA-analogs (points with errorbars as in Fig.~\ref{fig:literatureconf}) and predicted in the model by \citet[][shaded area]{Vanni2024}. Shaded areas are colored depending upon the contribution level of Pop III stars to the chemical enrichment: $100\%$ (yellow), $90\%$ (orange), $60\%$ (pink), $30\%$ (purple) and $\leq 0.01\%$ (dark grey: \citealt{Limongi2018}; light grey: \citealt{woosley1995}).  Only the environments with [O/H]$>-4$ are considered. The three remarkable systems with [C/O]$>0.35$ and [Si/O]$>0.50$ are highlighted using different symbols and  black contours: PSO J025-11 at z=5.7763 (square), SDSS J0100+2802 at z=5.9450 (star) and PSO J065-26 at z=6.1263 (diamond).}
    \label{fig:maps}
\end{figure*}

In Figs.~\ref{fig:scatter_points} and \ref{fig:maps}, we compare the observed results with the simple and general parametric study first developed by \cite{Salvadori2019} and recently refined by \citet{Vanni2023,Vanni2023b,Vanni2024}. 

In a few words, the model investigates the chemical abundance pattern of an ISM polluted by a Pop~III SN by exploring the full parameter space for explosion energies, ${\rm E=(0.3-100)\times 10^{51}}$~erg, and progenitor masses, $m_*=10-1000\,\rm M_{\odot}$ using the yields by \citet{Heger2002} and \citet{Heger2010}. The choice of a single Pop~III SN explosion is motivated by the results of hydro-dynamical cosmological simulations \citep[e.g.][]{Hirano2014,hirano2015} and further corroborated by the identification of present-day stars mono-enriched by Pop~III SNe \citep[e.g.][]{Skuladottir21,Rossi2024}.  

The model is very simple, it incorporates the main unknowns of early cosmic star formation into three free parameters: the star-formation efficiency, the metal dilution factor, and the mass fraction of metals provided by Pop~III stars with respect to the total one in the ISM, ${\rm f_{PopIII}}$. 
The chemical enrichment of the ISM is evaluated after the injection of heavy elements by Pop~III SNe only (${\rm f_{PopIII}=100\%}$) and after the subsequent contribution 
of Pop~II stars (${\rm f_{PopIII} < 100\%}$), which can promptly form after the Pop~III SN explosions when $Z_{\rm crit}> 10^{-4.5} Z_{\odot}$ \citep{Vanni2023}. Pop~II stars are assumed to form according to a standard Larson IMF, i.e., with masses $m_*=\rm 0.1-100\,M_{\odot}$, a peak at $\rm 0.35 \,M_{\odot}$, and a Salpeter-like slope at higher masses.

Two different sets of yields for Pop~II SNe are adopted, \citet{woosley1995} and \citet[set R, non-rotating]{Limongi2018}, to alleviate the model from uncertainties arising in yields' calculation \citep[see][Fig.~A1]{Vanni2023b}. In order to get consistent results for the two sets of yields, we assume also for \citet{woosley1995} that all Pop~II stars with $m_*\rm \ge 25 M_{\odot}$ evolve into black holes, as assumed by \citet{Limongi2018}. To take into account the enrichment due to SNe exploding at different timescales, we compute the IMF-integrated contribution of Pop II stars above $23, 21, 18, 15, 13 \rm M_{\odot}$ \citep[see][for details]{Salvadori2019,Vanni2024}. In Figs.~\ref{fig:scatter_points} and \ref{fig:maps}, we distinguish between the abundances determined from the set of yields by \citet{Limongi2018} (darker grey) and those based on the yields by \citet{woosley1995} (lighter grey) only in the case with $f_{\rm PopIII}=0.01$ (Pop~II only), while for all the other $f_{\rm PopIII}$ values they are plotted together. 

\begin{table*}
    \centering
    \caption{Properties of the three systems in our sample presenting the largest [C/O] and [Si/O] abundance ratios. }
    \label{tab:threesys}
    \begin{tabular}{cccccccc}
        \hline
        QSO & $z_{\rm abs}$ & $v_{\rm abs}$ (\kms) & $\log N(\mathrm{\HI})$ & [C/O] & [Si/O] & $\log(N_{\rm CII}/N_{\rm CIV})$ & $\log(N_{\rm SiII}/N_{\rm SiIV})$\\ 
        \hline
        PSO J025-11$^*$ & 5.7763 & 2866 & $20.35\pm0.1$ & $0.53\pm0.05$ & $0.67\pm0.05$ & $0.09\pm0.05$ & $-0.10\pm0.08$ \\
        SDSSJ0100+2802 & 5.9450 & 16029 & $>18.45$ & $0.39\pm0.04$ & $0.51\pm0.08$ & $0.68\pm0.09$ & $0.18\pm0.11$ \\
        PSO J065-26$^*$ & 6.1263 & 2543 & $>19.45$ & $>0.66$ & $>0.71$ & $>1.41$ & $0.71\pm0.04$ \\
        \hline 
    \end{tabular}
\end{table*}

One of the key predictions of the model by \citet[][see their Fig.~12]{Vanni2023} is that the scatter in the abundance ratios of different chemical species, such as [C/Fe] and [Mg/Fe], is maximum if we consider the pollution from Pop~III SNe only, and then it progressively decreases as the contribution of Pop~II SNe increases. In other words, gaseous absorbers predominantly polluted by first stars (${\rm f_{PopIII} \geq 60\%}$), or long-lived first stars' descendants formed in these gas clouds, are expected to show a larger absorber-to-absorber scatter in their chemical abundance ratios with respect to those predominantly imprinted by normal Pop~II stars (${\rm f_{PopIII} \leq 30\%}$). This is in agreement with e.g., the small star-to-star scatter observed in C-normal halo stars by \citet{Cayrel2004} (see Sec.~4.4 in \citealt{Vanni2023}). A physical explanation for the large dispersion can be found in the prediction from simulations that a few Pop~III stars form in the first low-mass mini-halos \citep[e.g.][]{hirano2015}, while lower mass Pop~II stars are expected to form in groups more efficiently \citep[e.g.][]{Ritter2015}. The isolation of Pop~III stars, combined with the ample variety of progenitor masses and explosion energies characterizing Pop~III SNe, make the chemical enrichment of the first mini-halos stochastic and thus, more diverse. 

In Fig.~\ref{fig:scatter_points}, we compare the [C/O] and [Si/O] values measured in DLAs at various redshifts (corresponding to the sample shown in Fig.~\ref{fig:redshift}) with the maximum extent predicted for these abundance ratios by models accounting for different contribution of Pop~III SNe to the gas enrichment. In order to ensure consistent metallicities between predictions and observations, we select only models with [O/H]$\geq-4$. Indeed, both in our sample (see Tab.~\ref{tab:RelAbb}) and in the literature \citep[see][]{Welsh2023} the most metal-poor DLAs have [O/H]~$\sim-3$. 

We see that $z\lesssim4.5$ DLAs exhibit a small [C/O] and [Si/O] scatter among each other, which is consistent with an enrichment only due to Pop~II SNe 
(${\rm f_{PopIII} =0.01}$). Conversely, for $z\gtrsim 4.5$ we see that the scatter among different systems increases for both [C/O] and [Si/O]: several absorbers now require a ${\ge 30\%}$ imprint from Pop~III stars to explain the observed abundances.
In particular, we notice three systems with [C/O]$>0.35$, [Si/O]$>0.50$ and very small error bars, that can only be explained if Pop~III SNe provide $\geq 60\%$ of their metals. The characteristics of these systems are summarized in Tab.~\ref{tab:threesys}.

We can further investigate the nature of these systems and in general of all $z>4.5$ absorbers by comparing some of their relative abundances (Fig.~\ref{fig:literatureconf}) with theoretical model predictions. This comparison is reported in Fig.~\ref{fig:maps}, where the shaded areas illustrate the expectations for an ISM imprinted by: Pop~III stars only (larger yellow contours), and by Pop~III stars but with an increasing contribution from normal Pop~II SNe (gradually darker and smaller contours). The solid lines encloses the abundance ratios recovered in at least one model. 

It is clear from these maps that the $z>4.5$ DLA-analogs are not all consistent with an enrichment solely driven by Pop~II stars (gray contours) but they require a $\geq 30\%$ contribution from Pop~III stars. 
Furthermore, we see that the three systems with highest [C/O] values in Tab.~\ref{tab:threesys} are at the edge of the $60\%$ Pop~III enrichment areas, and consistent with higher Pop~III contribution, in the [O/Fe] vs [C/Fe], [Si/Fe] vs [O/Fe] and [Si/O] vs [C/O] planes. 

Ultimately, the large scatter in the chemical abundance ratios of $z>4.5$ DLA-analogs is a strong indication of an heterogeneous chemical enrichment driven by different stellar populations and most likely including Pop~III SNe, which might account for $\geq 60\%$ of the ISM metals. 

\section{Discussion}
\label{sec:discussion}

The comparison between our observational findings and model results for the early metal enrichment, shows that the larger scatter in the abundance ratios at $z>4.5$ is most likely due to an enrichment by Pop~III SNe, which are predicted to have a broader range of progenitor masses and explosion energies. However, from a theoretical perspective, we can ask ourselves if a larger abundance scatter can be also driven by other, more metal-rich stellar sources. From an observational point of view, furthermore, we might wonder if we are really sure about the nature of our absorbers and thus of the robustness of our measurements.

\subsection{Exploring alternative sources of enrichment}

To exclude the possibility of a larger [C/O] and [Si/O] scatter driven by “exotic” but more metal-rich Pop II SNe exploding as energetic hypernovae or PISNe (e.g., see Langer et al. 2007), we can use the following arguments. 
Firstly, because of their high masses, $140-260 \rm M_{\odot}$, PISNe are not expected to form in environments enriched above the critical metallicity value, $Z_{\rm crit}$. Indeed, the more efficient gas cooling due to metals and dust causes the formation of sub-solar proto-stellar gas clouds and the decrease of the gas accretion rate onto the proto-star \cite[e.g.][for a review]{Klessen2023}. 
Conversely, we cannot exclude that a fraction of normal Pop II stars with intermediate masses ($20-40 \rm M_{\odot}$) might explode as more energetic hypernovae \citep[e.g., see][]{Kobayashi2006}. However, if we account for 
the contribution of metal-rich Pop~II hypernovae using the yields of \citet{Kobayashi2006}, we find that the scatter of Pop~II-dominated environments increases only in the sub-solar regimes of [C/O] and [Si/O], while our observations show super-solar values. 
Note also that very specific progenitors of core-collapse SNe with $m_* = 15 \rm M_{\odot}$ and $Z=Z_{\odot}$ can provide high values of [Si/O]$\approx 0.9$ \citep{Tominaga2007,Nomoto2013}. However, given the high metallicity of these sources their effect should be predominantly observed at lower redshifts. Finally, we note that the scatter predicted for environments enriched by Pop~III core-collapse SNe and hypernovae is still larger than for the ones enriched by Pop~II SNe of the same kinds. This result holds even when excluding Pop~III stars exploding as SNe with different energies in the mass range $25-100 \rm M_{\odot}$, as assumed for Pop~II core-collapse SNe \citep[see][for a study of direct black hole formation with progenitors of different metallicities]{Ebinger2020}.

Higher values for the [C/O] abundances can be obtained by accounting for the contribution of AGB stars \citep[see e.g.][]{SF2012} or dust depletion \citep{Konstantopoulou2024}. On the other hand, the spread in relative abundance of [Si/O] could be due to the enrichment by SNe type Ia (SNIa). Thus, it could be argued that the increase in both [C/O] and [Si/O] abundance ratios is a consequence of the joint enrichment by AGB stars and SNIa. However, both these sources act on longer timescales (from 30~Myr up to 1 Gyr) than Pop~III/II SNe ($< 30$~Myr). Therefore, it is at the lower redshifts, i.e., at  z < 4.5, that %if that was the case, at redshift z < 4.5 
we should see the chemical imprint of both these sources in many absorption systems, which instead show an astonishing small abundance scatter.

Ultimately, the larger scatter of both [C/O] and [Si/O] abundance ratios towards high-$z$ is a strong indication that these outliers were most likely imprinted by pristine Pop III SNe, which were more common in the first billion year of cosmic evolution \citep[e.g., see][]{Pallottini2014,Jaacks2018}.

\subsection{Verifying the nature of the studied systems}
\label{sec:IonizState}

The direct conversion from the ionic abundance derived from the measured column density to the total abundance of an element, generally requires the application of ionization corrections and the correction for the depletion of the element onto dust grains. 
A part of the metals can in fact be enclosed in solid phase dust grains \citep[e.g.][]{Peroux2020}. 

The formation of dust grains tends to increase with metallicity, but is considered negligible in DLAs with $\rm [Fe/H]\lesssim -2$ \citep{Vladilo2004,Konstantopoulou2024}. The decreasing metallicity trend of DLAs with redshift \citep{Rafelski2014,Decia2018} suggests that the systems studied here may be in this metal-poor regime. However, the differential contribution of the various elements to dust formation could be a potential source of error in determining relative abundances.

As for ionization corrections, these are negligible in heavily self-shielded systems, such as DLAs, in which almost all hydrogen is in the neutral state, and all metals are predominantly in their neutral or prime excited states. 
For sub-DLAs, which have a lower \HI\ column density, the shielding offered by the neutral gas is significantly reduced and ionization corrections begin to become important \citep{Peroux2007}. However, when calculating relative abundances, it is the difference in ionization correction between the two that matters. Although different elements are ionized differently, the effects will almost always act in the same direction, which is to reduce the population of the lowest ionization state present in the gas (\OI, \CII, \SiII, etc.). Even for a partially ionized system, therefore, the ratio between two ions can be expected to be reasonably close to the ratio between the total elements \citep{Becker2012}.

\begin{table*}
    \centering
    \caption{Column density ratios of \CII\ to \CIV, and \SiII\ to \SiIV, in the detected \OI\ systems.}
    \label{tab:IonizState}
    \begin{tabular}{lcccc}
        \hline
        QSO & $z_{\rm abs}$ & $v_{\rm abs}$ (\kms) & $\log(N_{\rm CII}/N_{\rm CIV})$ & $\log(N_{\rm SiII}/N_{\rm SiIV})$\\
        \hline
        PSO J308-27 & $5.4400$ & $16245$ & $0.27\pm 0.08$ & $0.47\pm 0.09$\\
        PSO J308-27 & $5.6268$ & $7686$ & $0.87\pm 0.12$ & $0.55\pm 0.05$\\
        PSO J023-02 & $5.4869$ & $16308$ & $>0.76$ & $\cdots$\\
        PSO J025-11$^*$ & $5.7763$ & $2866$ & $0.09\pm 0.05$ & $-0.10\pm 0.08$\\
        PSO J025-11$^*$ & $5.8385$ & $127$ & $>1.31$ & $1.28\pm 0.09$\\
        PSO J108+08 & $5.5624$ & $17396$ & $>-0.19$ & $\cdots$\\
        SDSS J0818+1722 & $5.7912$ & $8945$ & $0.76\pm 0.07$ & $>0.75$\\
        SDSS J0818+1722 & $5.8767$ & $5198$ & $0.49\pm 0.06$ & $0.45\pm 0.07$\\
        PSO J007+04$^*$ & $5.9917$ & $418$ & $0.03\pm 0.16$ & $-0.09\pm 0.11$\\
        SDSS J2310+1855$^*$ & $5.9388$ & $2763$ & $>1.38$ & $>1.41$\\
        PSO J158-14 & $5.8986$ & $7290$ & $1.18\pm 0.09$ & $0.82\pm 0.07$\\
        PSO J239-07$^*$ & $5.9918$ & $5033$ & $\cdots$ & $>0.27$\\
        ULAS J1319+0950$^*$ & $6.0172$ & $4977$ & $>0.28$ & $>0.16$\\
        PSO J060+24 & $5.6993$ & $20324$ & $0.93\pm 0.2$ & $0.92\pm 0.09$\\
        PSO J065-26 & $5.8677$ & $13615$ & $>1.71$ & $>1.39$\\
        PSO J065-26$^*$ & $6.1208$ & $2775$ & $>2.25$ & $>1.81$\\
        PSO J065-26$^*$ & $6.1263$ & $2543$ & $>1.41$ & $0.71\pm 0.04$\\
        SDSS J0100+2802 & $5.7974$ & $22441$ & $1.09\pm 0.10$ & $0.91\pm 0.12$\\
        SDSS J0100+2802 & $5.9450$ & $16029$ & $0.68\pm 0.09$ & $0.18\pm 0.11$\\
        SDSS J0100+2802 & $6.1114$ & $8939$ & $>1.59$ & $>1.00$\\
        SDSS J0100+2802 & $6.1434$ & $7595$ & $>1.84$ & $>1.20$\\
        DELS J1535+1943 & $5.8990$ & $20211$ & $0.50\pm 0.16$ & $0.23\pm 0.08$\\
        PSO J183+05 & $6.0642$ & $15466$ & $-0.16\pm 0.07$ & $-0.18\pm 0.05$\\
        PSO J183+05$^*$ & $6.4041$ & $1394$ & $>1.46$ & $>1.13$\\
        WISEA J0439+1634 & $6.2743$ & $9906$ & $>0.53$ & $>0.03$\\
        VDES J0224-4711 & $6.1228$ & $16450$ & $>1.24$ & $>0.61$\\
        PSO J036+03 & $6.0611$ & $19664$ & $0.54\pm 0.17$ & $0.52\pm 0.16$\\
        DELS J0923+0402 & $6.3784$ & $10165$ & $\cdots$ & $\cdots$\\
        \hline
    \end{tabular}
    \tablefoot{Column $v_{\rm abs}$ reports the velocity separation of the system from the QSO systemic redshift. \\
    \tablefoottext{*}{These systems are PDLA.}}
\end{table*}

\begin{figure*}
\centering
	\includegraphics[width=8cm]{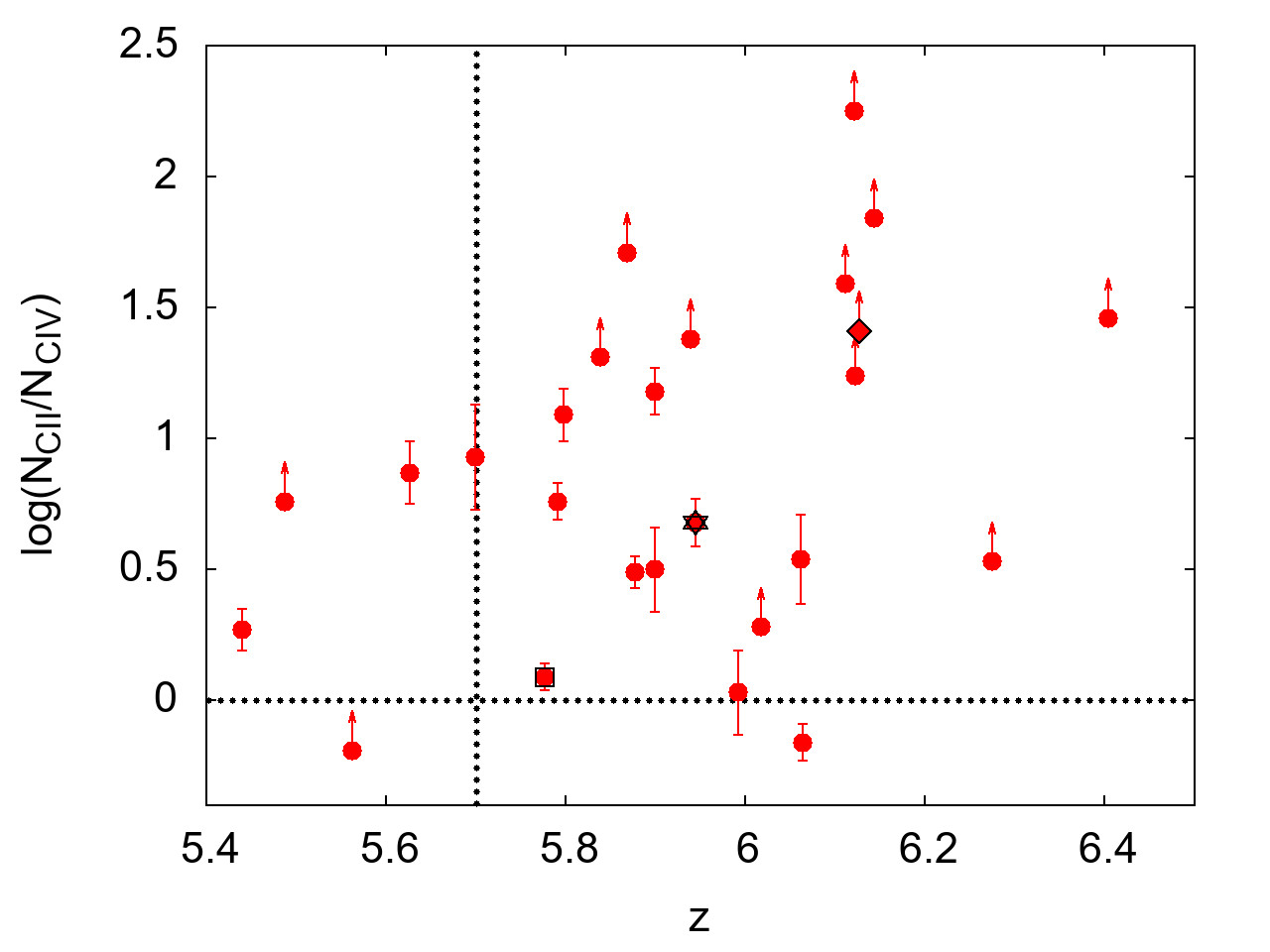}
	\includegraphics[width=8cm]{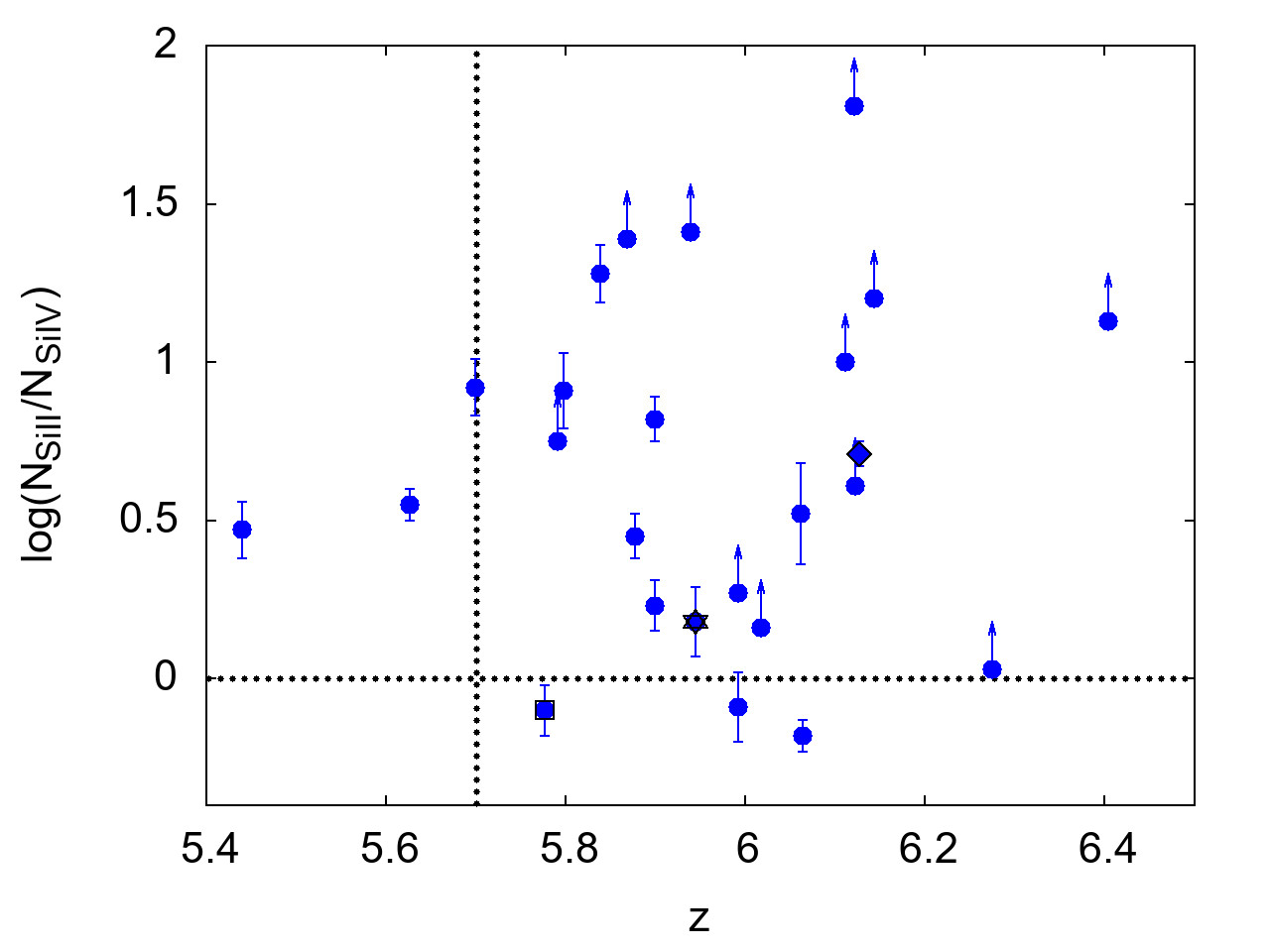}
	\includegraphics[width=8cm]{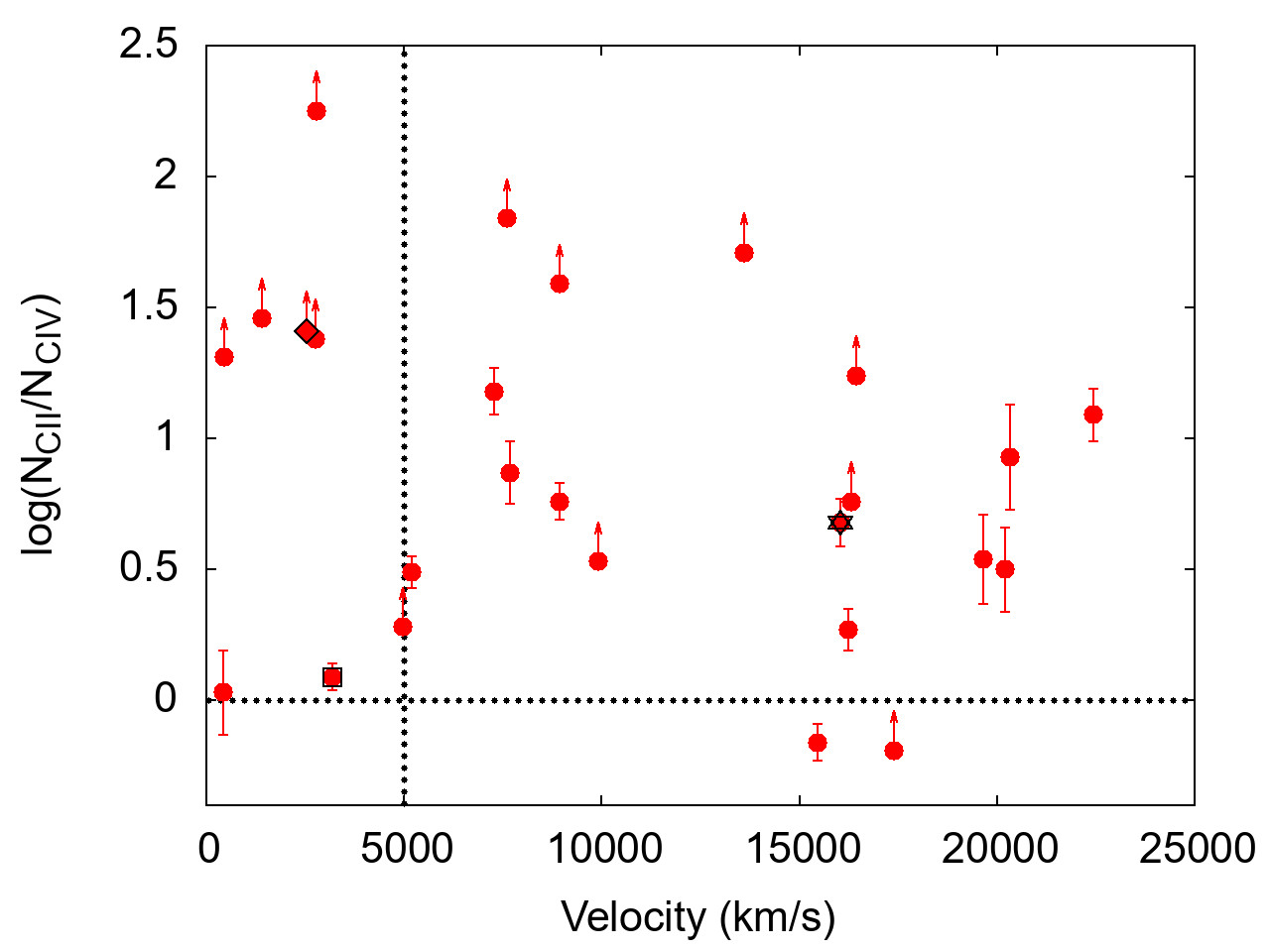}
	\includegraphics[width=8cm]{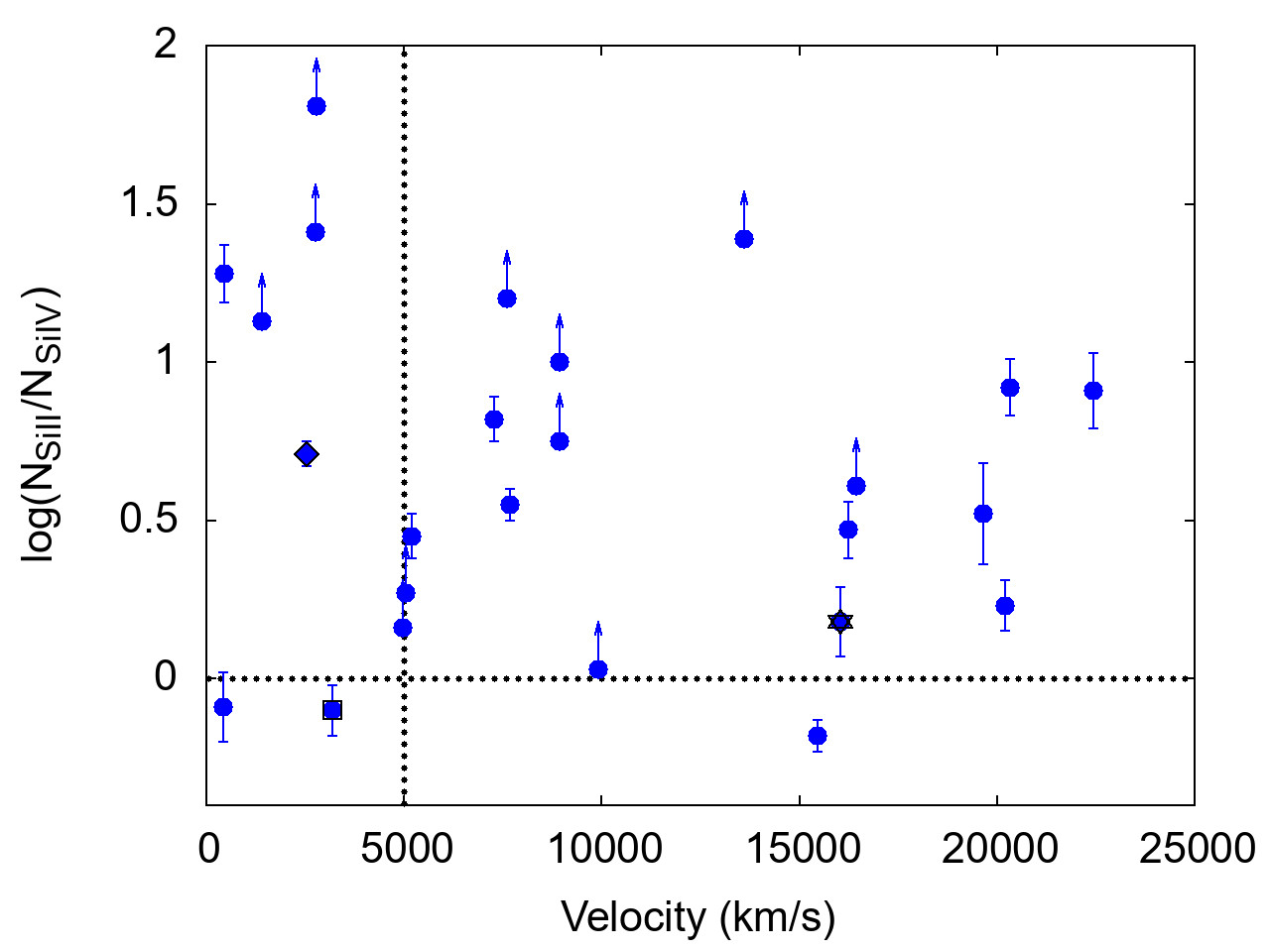}
    \caption{Ratio of \CII/\CIV\  (left panels) and \SiII/\SiIV\ (right panels) column densities as a function of redshift (upper panels) and velocity separation from the systemic redshift of the QSO (lower panels) for our sample of \OI\ systems. 
    The dotted vertical lines in the upper panels are drawn at $z=5.7$, the limit beyond which the low ionization phase is expected to dominate in the gas  \citep{Cooper2019}. 
    The vertical dotted lines in the lower panels mark a velocity separation of $5000$ \kms, which separates PDLAs from intervening DLAs. In each panel, the three systems with black contours (diamond, square and star) are the same highlighted in Fig.~\ref{fig:maps} and \ref{fig:scatter_points}.}
    \label{fig:ionizationstate}
\end{figure*}

The almost complete absence of transmitted flux in the Lyman forests of the E-XQR-30 QSOs does not allow us to directly and accurately estimate the \HI\ column density of the low ionization systems studied in this work and, as a consequence, their metallicity. 
As described in Sect.~\ref{sec:lya}, we could estimate lower limits to $\log N($\HI$)$ from the \OI\ column densities measured for our systems assuming they have a metallicity [Fe/H]~$\leq -2$ (see Fig.~\ref{fig:NHI_OI}) and for 7 systems we could measure $\log N($\HI$)$ from the spectrum. While the lower limits are in the vast majority of cases below $\log N($\HI$) \sim20$, in all cases in which we could perform a direct Voigt profile fitting we derived $\log N($\HI$) \geq 20$ confirming the predominantly neutral nature of the inspected gas. 

The ionization state of the considered systems has also been tested evaluating the ratio between the column densities of two ions of the same element in different ionization states. 
This comparison was only possible for carbon and silicon, which can be detected in singly (\CII\ and \SiII) and triply (\CIV\ and \SiIV) ionized state.  
All values are reported in Tab.~\ref{tab:IonizState}. 
In Fig.~\ref{fig:ionizationstate}, the values of $\log(N({\rm \CII})/N({\rm \CIV}))$ and $\log(N({\rm \SiII})/N({\rm SiIV}))$ are plotted as a function of the absorption redshift (upper panels), and of the velocity separation from the emission redshift of the QSO (lower panels).

At face value, we observe an increasing (decreasing) trend of the upper envelope of the data distribution with redshift (velocity separation). 
Previous works \citep[e.g.][]{Dodorico2013,Cooper2019} showed an increase of $\log(N({\rm \CII})/N({\rm \CIV}))$ for $z\gtrsim5.7$. This is suggested also by our data both for C and Si, although we have only a handful of systems at $z < 5.7$, and was also observed when considering the total sample of \CIV\ absorption systems detected in E-XQR-30  and the computation of the cosmic mass density of \CIV\ and \CII\ \citep{RDavies2023b}. A thorough analysis of the trends of $\log(N({\rm \CII})/N({\rm \CIV}))$ and $\log(N({\rm \SiII})/N({\rm \SiIV}))$ will be present in a future paper (Rowlands et al. in prep.).

Considering a conservative threshold of 0.5, the systems for which both $\log(N({\rm \CII})/N({\rm \CIV}))$ and $\log(N({\rm \SiII})/N({\rm \SiIV}))$ are below this threshold, that could be those with higher ionization state, are 6: 3 do have clearly detected high-ionization lines (PSO J308-27 $z=5.44$, PSO J025-11 $z=5.7763$ and PSO J183+05 $z=6.0642$) and the other 3 are instead very weak systems for which \CIV\ and \SiIV\ are either not detected or very weak (PSO J108+08 $z=5.5624$, PSO J007+04 $z=5.9917$ and ULAS J1319+0950 $z=6.0172$). 

In summary, we have reasons to believe that the majority of the \OI\ systems that form our sample are VMP-DLAs for which the corrections to chemical abundances due to ionization or dust depletion are negligible. 

\section{Conclusions}
\label{sec:Conclusion}

In this work, we looked for the traces of metal enrichment due to the first generation of stars that formed in the Universe (Pop III), which exploded as SNe and polluted the gas in their surroundings. 
To this aim, we analyzed the E-XQR-30 sample \citep{Dodorico2023}: a homogeneous sample of high SNR, intermediate resolution, optical and infrared spectra of 42 QSOs with emission redshift in the interval $5.8<z_{\rm em}<6.6$. 

The chemical signature of Pop III stars was looked for in neutral gas, likely tracing the ISM of primordial galaxies or satellite clumps in their vicinity. At lower redshift, this gas is identified through the strong \HI\ absorption of Damped \Lya\ systems (DLAs), while at $z>5$ the increasing saturation of the \Lya\ forest makes the measurement of the \HI\ column density and, consequently, the recognition of these systems more and more complicated. We overcome this problem through the detection of the neutral oxygen absorption line at $1302$ \AA, which is a very good tracer of neutral gas and of DLAs. 

In our sample of QSO spectra, we detected 29 \OI\ systems (or DLA-analogs) at $z \ge 5.4$, along the line of sight to 19 QSOs; 10 of these were classified as proximate DLA (PDLA, i.e., within $\sim5000$ \kms\ from the QSO emission redshift), while the other 19 as intervening DLAs. In each of these \OI\ systems, we looked for associated low ionization lines due to e.g., \CII, \SiII, \AlII, \FeII\ and \MgII, and to high ionization ones mainly due to \CIV\ and \SiIV, and performed fits of the velocity profiles with Voigt functions to derive ionic column densities. 
A lower limit on the \HI\ column density was derived from an empirical relation with \OI\ column density, in the hypothesis that the studied absorbers are very metal poor systems, i.e., with [Fe/H]~$\leq -2$. In the case of PDLAs, we performed also a Voigt profile fitting of the red damping wing of the \HI\ absorptions and derived direct estimates of the \HI\ column densities. 

Relative chemical abundances were obtained for 28 systems\footnote{One problematic PDLA was excluded from the analysis.} from the ratios of measured column densities, assuming that ionization and dust depletion had negligible effects. Our sample increases by a factor of 4 the number of studied \OI\ systems at $z \ge 5.4$. All abundances with respect to iron and oxygen are reported in Tabs.~\ref{tab:RelAbbFe} and \ref{tab:RelAbbO}, respectively.

Finally, to understand whether the observed high-redshift absorption systems were enriched by the first generation of stars, we compared the measured chemical abundance ratios with those predicted by two theoretical models: the semi-analytical model coupled with N-body simulations of \citet{kulkarni2013} and the general parametric study by \citet{Vanni2024}.

The main results of this work are the following.

\begin{itemize}
\item[-] We found no systematic differences in the average chemical abundances of PDLAs and intervening DLAs in our sample (see Fig.~\ref{fig:DLAconf}). 
Five PDLA systems have $\log(N({\rm \CII})/N({\rm \CIV}))$ and $\log(N({\rm \SiII})/N({\rm \SiIV}))$ ratios larger than 0.5 which should imply a low ionization state, while the other 4 have moderately low values but all $\gtrsim 0.0$ (see Fig.~\ref{fig:ionizationstate}). In summary, the presence of the QSO does not seem to have a strong impact on this class of systems, possibly due to their very large \HI\ column density.  

\item[-] The chemical abundances measured in our sample are consistent with those of the literature systems at  $z>4.5$ selected with similar criteria \citep[][see Fig.~\ref{fig:literatureconf}]{Becker2012,Poudel2018,Poudel2020}. Although, the high quality of our spectra allows to extend the sample toward lower column density values. This suggests that the gas was similarly enriched in every DLA at that time.

\item[-] We compared the relative abundances of the sample of \OI\ systems at $4.5 < z < 6.5$ (E-XQR-30 $+$ literature), with those of the Very Metal-Poor DLAs at $2<z<4.5$ identified by \citet{Cooke2011b} and \citet{Dutta2014} to highlight any possible evolution with redshift. The average abundances computed in the two redshift bins are in general agreement within uncertainties except for  [C/O], [Si/Fe] and [Si/O] which are significantly larger for the high-$z$ sample (see Fig.~\ref{fig:redshift}). 
However, the dispersion of the measurements in the higher redshift bin is always larger, and the Epps-Singleton test carried out for [Si/Fe], [Si/O] and [O/Fe] indicates that the low and high-redshift measurements are not drawn from the same distribution. 

\item[-] The comparison of the observational results with theoretical predictions suggests that the increase in the scatter of the relative chemical abundances at $z>4.5$ could be indicative of gas that partially retains the signature of Pop~III enrichment. In particular, the comparison with the model by \citet{Vanni2024} singles out three of our systems that could trace gas enriched for more than 60 \% by Pop~III stars (see Figs.~\ref{fig:scatter_points} and \ref{fig:maps}). On the other hand, absorbers at $z<4.5$ are all compatible with an enrichment predominantly by Pop~II stars. 
\end{itemize}

This work presents the largest sample of low ionization absorption systems at $z \ge 5.4$, tracing highly neutral gas in the ISM of primordial galaxies or in their vicinity, and possibly probing gas which still shows the imprint of Pop~III metal enrichment. 

The main, insurmountable limitation of this kind of studies is the impossibility of measuring the \HI\ column density of the analyzed systems. However, all the tests we have carried out seem to indicate that they are indeed DLA or subDLA systems with metallicities of [Fe/H]~$\lesssim -2$.  

There are other weaknesses affecting this work which however could be solved in the relatively near future. The resolving power of the E-XQR-30 survey is still not sufficient to optimally resolve QSO metal absorption lines; this can introduce systematic uncertainties in the column density measurement due to the ``hidden'' saturation of absorption lines. We have faced this problem as much as we could, but we will have to wait for the ANDES high-resolution spectrograph for the ELT \citep{Marconi2022} to carry out extremely accurate measurements of column densities and detect fundamental elements (like Zn) that are beyond reach in the present spectra.      

Another weak point is the paucity of systems studied in the redshift range $4.0 < z \le 5.5$ and the heterogeneous and generally poor quality of the available data at those redshifts (due to low resolution, low SNR and limited wavelength range).  This is the highest redshift range at which the \HI\ lines  can still be detected:  a large sample of DLA systems with high-quality, homogeneous spectroscopic observations would be key to clarify the evolution of the average chemical abundances and of their dispersion. 

\begin{acknowledgements}
The authors thank the anonymous referee for suggestions and comments that improved the clarity of this manuscript. S.S. and I.V. acknowledge support from the European Research Council (ERC) under the European Union’s Horizon 2020 research and innovation programme (grant agreement No 804240). V.D., A.F., S.S., and I.V. acknowledge support from the PRIN-MIUR17, The quest for the first stars, prot. n. 2017T4ARJ5.
%% Acknowledgements for Emma, Rebecca and Alma
This research was supported by the Australian Research Council Centre of Excellence for All Sky Astrophysics in 3 Dimensions (ASTRO 3D), through project number CE170100013.
%% Acknowl. Ema
E.P.F. is supported by the international Gemini Observatory, a program of NSF’s NOIRLab, which is managed by the Association of Universities for Research in Astronomy (AURA) under a cooperative agreement with the National Science Foundation, on behalf of the Gemini partnership of Argentina, Brazil, Canada, Chile, the Republic of Korea, and the United States of America.
\end{acknowledgements}


\bibliographystyle{aa}
\bibliography{Reference_sodini} % if your bibtex file is called example.bib

%\begin{thebibliography}{}

%  \bibitem[1966]{baker} Baker, N. 1966,
%      in Stellar Evolution,
%      ed.\ R. F. Stein,\& A. G. W. Cameron
%      (Plenum, New York) 333

%  \bibitem[1988]{balluch} Balluch, M. 1988,
%      A\&A, 200, 58

%\end{thebibliography}

\begin{appendix}
\section{Properties of the QSO and DLA-analog samples}
In Tab.~\ref{tab:QSO},  we report the properties of the E-XQR-30 QSOs and of their X-SHOOTER spectra. In particular, the complete name of the QSO is in the first column; the emission redshift, the line used for its determination and the reference paper are in the second and third columns. The fourth and the fifth columns report the resolving power in the VIS and NIR part of the X-SHOOTER spectrum measured from the observed frames \citep[see][for the details of how they were measured]{Dodorico2023}. Column 6 lists the signal-to-noise ratio (SNR) per 10 \kms\ bin, measured at $\lambda=1285$ \AA\ in rest frame. The last column gives the reference for the X-SHOOTER spectrum.     

Table~\ref{tab:LIS} summarizes the properties of all identified DLA-analogs by listing the total column densities measured for each detected ion. The second column lists the column-density weighted average redshifts of each system. Column 3 gives the lower limit on the \HI\ logarithmic column density determined with Eq.~\ref{eq:relOH} assuming a metallicity [Fe/H]$\leq-2$ (see also Fig.~\ref{fig:NHI_OI}). While column 4 reports the \HI\  logarithmic column density measured on the spectrum for PDLAs showing a well defined red damping wing of the \Lya\ velocity profile.    

\begin{table*}
    \centering
    \caption{List of the E-XQR-30 QSOs. }
    \label{tab:QSO}
    \begin{tabular}{lcccccc}
        \hline
        QSO & $z_{\rm em}$ & Ref. & $\rm R_{VIS}$ & $\rm R_{NIR}$ & SNR & Ref.\\
        \hline
        SDSS J092721.82+200123.7 & 5.7722 & CO,1 & 12900 & 9400 & 36.6 & 2\\
        PSO J308.4829-27.6485 & 5.799 & \MgII,2,3 & 11800 & 10600 & 31.2 & 1\\
        SDSS J083643.85+005453.3 & 5.810 & \Lya,4 & 13100 & 10200 & 37.5 & 3\\
        PSO J065.9589+01.7235 & 5.8348 & [\CII],5 & 10700 & 9700 & 27.4 & 1\\
        PSO J025.2376-11.6831 & 5.8414 & [\CII],5 & 10500 & 9700 & 27.5 & 1\\
        PSO J242.4397-12.9816 & 5.8468 & [\CII],5 & 10700 & 9700 & 15.0 & 1\\
        PSO J023.0071-02.2675 & 5.850 & \Lya,6 & 10900 & 9700 & 18.6 & 1\\
        PSO J183.2991-12.7676 & 5.893 & \MgII,2,3 & 11900 & 10000 & 34.1 & 1\\
        PSO J108.4429+08.9257 & 5.955 & \MgII,2,3 & 12200 & 9800 & 36.8 & 1\\
        PSO J089.9394-15.5833 & 5.972 & \MgII,2,3 & 12000 & 9700 & 28.9 & 1\\
        ULAS J014837.63+060020.0 & 5.977 & \MgII,3 & 13300 & 10000 & 60.7 & 4\\
        PSO J029.5172-29.0886 & 5.981 & \Lya,4 & 10800 & 9900 & 26.5 & 1\\
        VDES J2250-5015 & 5.985 & \MgII,2,3 & 10800 & 9600 & 20.9 & 1\\
        SDSS J081827.40+172251.8 & 5.997 & \Lya,4 & 11000 & 7600 & 52.8 & 3\\
        PSO J007.0273+04.9571 & 6.0015 & [\CII],7 & 11500 & 9700 & 28.4 & 1\\
        SDSS J231038.88+185519.7 & 6.0031 & [\CII],8 & 12700 & 9800 & 40.7 & 1\\
        PSO J009.7355-10.4316 & 6.0040 & [\CII],7 & 11500 & 10500 & 24.1 & 1\\
        VST-ATLAS J029.9915-36.5658 & 6.013 & \MgII,2,3 & 10100 & 9200 & 25.4 & 1\\
        SDSS J130608.26+035626.3 & 6.0330 & [\CII],7 & 12000 & 9600 & 34.0 & 3\\
        VDES J0408-5632 & 6.033 & \MgII,2,3 & 11300 & 9700 & 38.9 & 1\\
        PSO J158.6937-14.4210 & 6.0685 & [\CII],9 & 10800 & 9100 & 29.6 & 1\\
        SDSS J084229.43+121850.4 & 6.0754 & [\CII],7 & 11400 & 10000 & 37.7 & 1\\
        PSO J239.7124-07.4026 & 6.1102 & [\CII],9 & 11300 & 11000 & 34.7 & 1\\
        CFHQS J150941-174926 & 6.1225 & [\CII],10 & 11800 & 8000 & 27.4 & 3\\
        ULAS J131911.29+095051.4 & 6.1347 & [\CII],7 & 13700 & 9800 & 41.1 & 3\\
        PSO J217.0891-16.0453 & 6.1498 & [\CII],10 & 11500 & 10200 & 36.2 & 1\\
        PSO J217.9185-07.4120 & 6.166 & \MgII,2,3 & 11200 & 9900 & 24.1 & 1\\
        PSO J060.5529+24.8567 & 6.170 & \MgII,2,3 & 11500 & 10300 & 29.0 & 1\\
        PSO J359.1352-06.3831 & 6.1722 & [\CII],9 & 11300 & 10000 & 35.8 & 1\\
        PSO J065.4085-26.9543 & 6.1871 & [\CII],7 & 11700 & 10500 & 41.2 & 1\\
        SDSS J103027.09+052455.0 & 6.304 & \MgII,3 & 12300 & 8400 & 16.8 & 3\\
        SDSS J010013.02+280225.8 & 6.3268 & [\CII],7 & 11400 & 10300 & 103.2 & 7\\
        VST-ATLAS J025.6821-33.4627 & 6.3373 & [\CII],7 & 11200 & 9300 & 29.6 & 5,6\\
        VDES J2211-3206 & 6.3394 & [\CII],10 & 10600 & 9100 & 12.5 & 1\\
        DELS J153532.87+194320.1 & 6.381 & \Lya,4 & 11600 & 9700 & 15.9 & 1\\
        PSO J183.1124+05.0926 & 6.4386 & [\CII],7 & 11900 & 10200 & 21.8 & 1\\
        WISEA J043947.09+163415.8 & 6.5188 & [\CII],11 & 9500 & 8200 & 114.4 & 8\\
        VDES J0224-4711 & 6.525 & \MgII,2,3 & 11200 & 9400 & 15.1 & 1\\
        PSO J036.5078+03.0498 & 6.5405 & [\CII],7 & 10700 & 9200 & 18.4 & 5,6\\
        PSO J231.6576-20.8335 & 6.5869 & [\CII],7 & 10800 & 9800 & 18.0 & 1\\
        PSO J323.1382+12.2986 & 6.5872 & [\CII],7 & 10900 & 9800 & 15.9 & 1\\
        DELS J092347.12+040254.4 & 6.6330 & [\CII],11 & 11600 & 10200 & 10.8 & 1\\
        \hline
    \end{tabular}
    \tablefoot{The various columns show: the QSO name, the emission redshift, the emission line used to measure it and its reference, the measured resolution of the optical and infrared regions of the spectrum, the  SNR calculated at $1285$ \AA\ in the rest frame, and the reference for the X-SHOOTER spectrum.}
    \tablebib{Redshift: 1 - \citet{Carilli2007}, 2 - \citet{Bischetti2022}, 3 - \citet{Dodorico2023}, 4 - \citet{Zhu2021}, 5 - Bosman et al. (in prep.), 6 - This work, 7 - \citet{Venemans2020}, 8 - \citet{Wang2013}, 9 - \citet{Eilers2021}, 10 - \citet{Decarli2018}, 11 - \citet{Yang2021}.}
    \tablebib{X-SHOOTER spectrum: 1 - \citet{Dodorico2023}, 2 - \citet{Codorenau2017}, 3 - \citet{Dodorico2013}, 4 - \citet{Becker2015}, 5 - \citet{Becker2019}, 6 - \citet{Dodorico2022}, 7 - \citet{Bosman2018}, 8 - \citet{Fan2019}.}
\end{table*}

\begin{sidewaystable*}
    \centering
    \caption{Logarithm of the total column densities (in cm$^{-2}$) of ions detected in the observed \OI\ systems. }
    \label{tab:LIS}
    \resizebox{\textwidth}{!}{%
    \begin{tabular}{lccccccccccc}
        \hline
        QSO & $z_{\rm abs}$ & ${\log N_{\rm HI}}^a$ & ${\log N_{\rm HI}}^b$ & $\log N_{\rm OI}$ & $\log N_{\rm CII}$ & $\log N_{\rm SiII}$ & $\log N_{\rm AlII}$ & $\log N_{\rm FeII}$ & $\log N_{\rm MgII}$ & $\log N_{\rm SiIV}$ & $\log N_{\rm CIV}$\\
        \hline
        PSO J308-27 & $5.4400$ & $>19.10$ & $\cdots$ & $14.09\pm 0.02$ & $13.68\pm 0.03$ & $13.14\pm 0.07$ & $<11.86$ & $12.57\pm 0.06$ & $13.06\pm 0.06$ & $12.67\pm 0.05$ & $13.41\pm 0.07$\\
        PSO J308-27 & $5.6268$ & $>19.79$ & $\cdots$ & $14.78\pm 0.03$ & $14.47\pm 0.03$ & $13.67\pm 0.02$ & $12.45\pm 0.13$ & $13.33\pm 0.03$ & $13.74\pm 0.05$ & $13.12\pm 0.05$ & $13.60\pm 0.12$\\
        PSO J023-02 & $5.4869$ & $>19.33$ & $\cdots$ & $14.32\pm 0.03$ & $14.27\pm 0.15$ & $<13.43$ & $<12.33$ & $12.64\pm 0.10$ & $13.18\pm 0.11$ & $<12.65$ & $<13.51$\\
        PSO J023-02$^{**}$ & $5.8441$ & $\cdots$ & $\cdots$ & $14.35\pm 0.04$ & $14.32\pm 0.04$ & $13.14\pm 0.04$ & $12.84\pm 0.10$ & $12.95\pm 0.07$ & $\cdots$ & $>14.90$ & $>16.39$\\
        PSO J025-11$^*$ & $5.7763$ & $>19.23$ & $20.35\pm 0.10$ & $14.22\pm 0.04$ & $14.49\pm 0.03$ & $13.71\pm 0.04$ & $12.58\pm 0.16$ & $13.10\pm 0.05$ & $13.70\pm 0.11$ & $13.80\pm 0.07$ & $14.40\pm 0.05$\\
        PSO J025-11$^*$ & $5.8385$ & $>20.73$ & $21.25\pm 0.10$ & $>15.72$ & $>15.46$ & $14.79\pm 0.09$ & $13.43\pm 0.20$ & $14.50\pm 0.04$ & $>14.58$ & $13.51\pm 0.03$ & $14.15\pm 0.05$\\
        PSO J108+08 & $5.5624$ & $>18.85$ & $\cdots$ & $13.84\pm 0.03$ & $13.12\pm 0.06$ & $<12.42$ & $<11.56$ & $<11.91$ & $\cdots$ & $<12.39$ & $<13.31$\\
        SDSS J0818+1722 & $5.7912$ & $>19.51$ & $\cdots$ & $14.50\pm 0.02$ & $14.14\pm 0.01$ & $13.33\pm 0.01$ & $\cdots$ & $12.96\pm 0.05$ & $\cdots$ & $<12.58$ & $13.38\pm 0.07$\\
        SDSS J0818+1722 & $5.8767$ & $>19.13$ & $20.20\pm 0.10$ & $14.12\pm 0.02$ & $13.77\pm 0.03$ & $12.96\pm 0.02$ & $<11.90$ & $12.49\pm 0.08$ & $12.96\pm 0.09$ & $12.51\pm 0.07$ & $13.28\pm 0.05$\\
        PSO J007+04$^*$ & $5.9917$ & $>18.98$ & $20.40\pm 0.10$ & $13.97\pm 0.08$ & $13.45\pm 0.07$ & $12.69\pm 0.04$ & $<12.10$ & $12.57\pm 0.20$ & $<12.84$ & $12.78\pm 0.10$ & $13.42\pm 0.14$\\
        SDSS J2310+1855$^*$ & $5.9388$ & $>20.28$ & $21.00\pm 0.10$ & $>15.27$ & $>14.33$ & $13.72\pm 0.07$ & $12.17\pm 0.38$ & $13.32\pm 0.07$ & $>13.49$ & $<12.31$ & $<12.95$\\
        PSO J158-14 & $5.8986$ & $>19.61$ & $\cdots$ & $14.60\pm 0.07$ & $14.31\pm 0.03$ & $13.65\pm 0.04$ & $12.26\pm 0.21$ & $13.14\pm 0.03$ & $\cdots$ & $12.83\pm 0.06$ & $13.13\pm 0.09$\\
        PSO J239-07$^*$ & $5.9918$ & $>18.83$ & $\cdots$ & $13.82\pm 0.06$ & $13.18\pm 0.11$ & $12.60\pm 0.03$ & $<11.75$ & $<12.14$ & $12,47\pm 0.11$ & $<12.33$ & $\cdots$\\
        ULAS J1319+0950$^*$ & $6.0172$ & $>18.94$ & $\cdots$ & $13.93\pm 0.06$ & $13.17\pm 0.11$ & $12.50\pm 0.04$ & $11.84\pm 0.23$ & $12.45\pm 0.15$ & $12.61\pm 0.16$ & $<12.34$ & $<12.89$\\
        PSO J060+24 & $5.6993$ & $>19.51$ & $\cdots$ & $14.50\pm 0.09$ & $14.22\pm 0.18$ & $13.64\pm 0.03$ & $12.30\pm 0.23$ & $12.96\pm 0.07$ & $\cdots$ & $12.72\pm 0.08$ & $13.29\pm 0.09$\\
        PSO J065-26 & $5.8677$ & $>20.02$ & $\cdots$ & $15.01\pm 0.02$ & $14.53\pm 0.02$ & $13.87\pm 0.03$ & $12.47\pm 0.15$ & $13.42\pm 0.03$ & $13.64\pm 0.05$ & $<12.48$ & $<12.83$\\
        PSO J065-26$^*$ & $6.1208$ & $>20.42$ & $\cdots$ & $>15.41$ & $>15.24$ & $14.54\pm 0.02$ & $13.21\pm 0.04$ & $13.97\pm 0.03$ & $>14.43$ & $<12.73$ & $<12.99$\\
        PSO J065-26$^*$ & $6.1263$ & $>19.52$ & $\cdots$ & $14.51\pm 0.03$ & $>14.91$ & $14.04\pm 0.03$ & $13.11\pm 0.05$ & $13.38\pm 0.03$ & $>14.35$ & $13.33\pm 0.03$ & $13.50\pm 0.03$\\
        SDSS J0100+2802 & $5.7974$ & $>19.64$ & $\cdots$ & $14.63\pm 0.03$ & $14.09\pm 0.02$ & $13.56\pm 0.03$ & $12.06\pm 0.08$ & $13.12\pm 0.02$ & $\cdots$ & $12.65\pm 0.12$ & $13.00\pm 0.10$\\
        SDSS J0100+2802 & $5.9450$ & $>18.45$ & $\cdots$ & $13.44\pm 0.04$ & $13.57\pm 0.02$ & $12.77\pm 0.07$ & $11.45\pm 0.14$ & $12.25\pm 0.05$ & $12.64\pm 0.03$ & $12.59\pm 0.08$ & $12.89\pm 0.09$\\
        SDSS J0100+2802 & $6.1114$ & $>19.70$ & $\cdots$ & $14.69\pm 0.14$ & $13.91\pm 0.07$ & $13.01\pm 0.03$ & $11.56\pm 0.10$ & $12.65\pm 0.03$ & $13.10\pm 0.04$ & $<12.01$ & $<12.32$\\
        SDSS J0100+2802 & $6.1434$ & $>19.67$ & $20.20\pm 0.10$ & $14.66\pm 0.02$ & $14.11\pm 0.02$ & $13.27\pm 0.02$ & $11.74\pm 0.11$ & $12.92\pm 0.02$ & $13.26\pm 0.03$ & $<12.07$ & $<12.27$\\
        DELS J1535+1943 & $5.8990$ & $\cdots$ & $\cdots$ & $\cdots$ & $14.64\pm 0.16$ & $13.69\pm 0.07$ & $\cdots$ & $13.17\pm 0.04$ & $13.66\pm 0.19$ & $13.47\pm 0.04$ & $14.14\pm 0.05$\\
        PSO J183+05 & $6.0642$ & $>19.53$ & $\cdots$ & $14.52\pm 0.03$ & $14.31\pm 0.06$ & $13.70\pm 0.04$ & $12.55\pm 0.07$ & $13.15\pm 0.04$ & $13.83\pm 0.19$ & $13.88\pm 0.03$ & $14.47\pm 0.03$\\
        PSO J183+05$^*$ & $6.4041$ & $>19.79$ & $20.60\pm 0.10$ & $14.78\pm 0.04$ & $14.52\pm 0.07$ & $13.96\pm 0.03$ & $12.39\pm 0.07$ & $13.36\pm 0.03$ & $>14.07$ & $<12.83$ & $<13.06$\\
        WISEA J0439+1634 & $6.2743$ & $>18.18$ & $\cdots$ & $13.17\pm 0.17$ & $13.03\pm 0.12$ & $12.19\pm 0.04$ & $<11.05$ & $<11.60$ & $<11.78$ & $<12.16$ & $<12.50$\\
        VDES J0224-4711 & $6.1228$ & $>19.47$ & $\cdots$ & $14.46\pm 0.02$ & $14.00\pm 0.13$ & $13.37\pm 0.06$ & $<11.94$ & $12.86\pm 0.06$ & $13.14\pm 0.04$ & $<12.76$ & $<12.76$\\
        PSO J036+03 & $6.0611$ & $>19.89$ & $\cdots$ & $13.88\pm 0.04$ & $13.47\pm 0.09$ & $13.09\pm 0.12$ & $11.81\pm 0.23$ & $<12.13$ & $12.89\pm 0.13$ & $12.57\pm 0.11$ & $12.93\pm 0.15$\\
        DELS J0923+0402 & $6.3784$ & $>19.57$ & $\cdots$ & $14.56\pm 0.09$ & $13.84\pm 0.25$ & $13.11\pm 0.05$ & $<12.23$ & $<12.53$ & $12.86\pm 0.14$ & $\cdots$ & $\cdots$\\
        \hline
    \end{tabular}}
    \tablefoot{Errors on measurements and lower limits are $1\sigma$ while upper limits are $3\sigma$. \\
    \tablefoottext{a}{Estimate of the \HI\ column density obtained from eq.~\ref{eq:relOH} assuming a metallicity [Fe/H]$\leq-2$. } \\
    \tablefoottext{b}{Estimate of the \HI\ column density derived from the fit of the \Lya\ line.} \\
    \tablefoottext{*}{These systems are PDLA.} \\
    \tablefoottext{**}{This system was excluded from the analysis, see its description in Appendix~\ref{app:LIS}.}
}
\end{sidewaystable*}

\FloatBarrier

\section{Chemical abundances of the studied systems}
\label{app:abundances}

Table~\ref{tab:RelAbb} reports the absolute abundances for the 7 DLA systems for which it was possible to measure the \HI\ column density directly from the fit of the red damping wing of the \Lya\ transition.  On the other hand, in Tab.~\ref{tab:RelAbbO} we list the abundances relative to O for all the systems in our sample. 

\begin{table*}
    \centering
    \caption{Chemical abundance of elements detected in low ionization systems where the column density of \HI\ could be estimated.}
    \label{tab:RelAbb}
    \begin{tabular}{lccccccc}
        \hline
        QSO & $z_{\rm abs}$ & $\rm [C/H]$ & $\rm [O/H]$ & $\rm [Mg/H]$ & $\rm [Al/H]$ & $\rm [Si/H]$ & $\rm [Fe/H]$\\
        \hline
        PSO J025-11$^*$ & $5.7763$ & $-2.29\pm 0.10$ & $-2.82\pm 0.11$ & $-2.25\pm 0.15$ & $-2.22\pm 0.19$ & $-2.15\pm 0.11$ & $-2.75\pm 0.11$\\
        PSO J025-11$^*$ & $5.8385$ & $>-2.22$ & $>-2.22$ & $>-2.27$ & $-2.27\pm 0.22$ & $-1.97\pm 0.13$ & $-2.25\pm 0.11$\\
        SDSS J0818+1722 & $5.8767$ & $-2.86\pm 0.10$ & $-2.77\pm 0.10$ & $-2.84\pm 0.13$ & $-2.92\pm 0.37$ & $-2.75\pm 0.10$ & $-3.21\pm 0.13$\\
        PSO J007+04$^*$ & $5.9917$ & $-3.38\pm 0.12$ & $-3.12\pm 0.13$ & $<-3.16$ & $<-2.75$ & $-3.22\pm 0.11$ & $-3.33\pm 0.22$\\
        SDSS J2310+1855$^*$ & $5.9388$ & $>-3.10$ & $>-2.42$ & $>-3.11$ & $-3.28\pm 0.39$ & $-2.79\pm 0.12$ & $-3.18\pm 0.12$\\
        SDSS J0100+2802 & $6.1434$ & $-2.52\pm 0.10$ & $-2.23\pm 0.10$ & $-2.54\pm 0.10$ & $-2.91\pm 0.15$ & $-2.44\pm 0.10$ & $-2.78\pm 0.10$\\
        PSO J183+05$^*$ & $6.4041$ & $-2.51\pm 0.12$ & $-2.51\pm 0.11$ & $>-2.13$ & $-2.66\pm 0.12$ & $-2.15\pm 0.10$ & $-2.74\pm 0.10$\\
        \hline
    \end{tabular}
    \tablefoot{$^*$ These systems are PDLA.}
\end{table*}

\begin{table*}
    \centering
    \caption{Chemical abundances relative to oxygen of the elements detected in the observed low ionization systems.}
    \label{tab:RelAbbO}
    \begin{tabular}{lcccccc}
        \hline
        QSO & $z_{\rm abs}$ & $\rm [C/O]$ & $\rm [Mg/O]$ & $\rm [Al/O]$ & $\rm [Si/O]$ & $\rm [Fe/O]$\\
        \hline
        PSO J308-27 & $5.4400$ & $-0.15\pm 0.04$ & $0.06\pm 0.06$ & $<0.01$ & $0.23\pm 0.07$ & $-0.33\pm 0.06$\\
        PSO J308-27 & $5.6268$ & $-0.05\pm 0.04$ & $0.05\pm 0.06$ & $-0.08\pm 0.13$ & $0.08\pm 0.03$ & $-0.26\pm 0.04$\\
        PSO J023-02 & $5.4869$ & $0.21\pm 0.15$ & $-0.05\pm 0.11$ & $<0.25$ & $<0.29$ & $-0.49\pm 0.10$\\
        PSO J025-11$^*$ & $5.7763$ & $0.53\pm 0.05$ & $0.57\pm 0.12$ & $0.60\pm 0.17$ & $0.67\pm 0.05$ & $0.07\pm 0.06$\\
        PSO J025-11$^*$ & $5.8385$ & $\cdots$ & $\cdots$ & $<-0.05$ & $<0.25$ & $<-0.03$\\
        PSO J108+08 & $5.5624$ & $-0.46\pm 0.07$ & $\cdots$ & $<-0.04$ & $<-0.24$ & $<-0.74$\\
        SDSS J0818+1722 & $5.7912$ & $-0.11\pm 0.03$ & $\cdots$ & $\cdots$ & $0.01\pm 0.02$ & $-0.36\pm 0.05$\\
        SDSS J0818+1722 & $5.8767$ & $-0.09\pm 0.04$ & $-0.07\pm 0.09$ & $<0.02$ & $0.02\pm 0.03$ & $-0.44\pm 0.08$\\
        PSO J007+04$^*$ & $5.9917$ & $-0.26\pm 0.11$ & $<-0.04$ & $<0.37$ & $-0.10\pm 0.09$ & $-0.21\pm 0.22$\\
        SDSS J2310+1855$^*$ & $5.9388$ & $\cdots$ & $\cdots$ & $<-0.86$ & $<-0.37$ & $<-0.76$\\
        PSO J158-14 & $5.8986$ & $-0.03\pm 0.08$ & $\cdots$ & $-0.10\pm 0.22$ & $0.23\pm 0.08$ & $-0.27\pm 0.08$\\
        PSO J239-07$^*$ & $5.9918$ & $-0.38\pm 0.13$ & $-0.26\pm 0.13$ & $<0.17$ & $-0.04\pm 0.07$ & $<-0.49$\\
        ULAS J1319+0950$^*$ & $6.0172$ & $-0.50\pm 0.13$ & $-0.23\pm 0.17$ & $0.15\pm 0.24$ & $-0.25\pm 0.07$ & $-0.29\pm 0.16$\\
        PSO J060+24 & $5.6993$ & $-0.02\pm 0.20$ & $\cdots$ & $0.04\pm 0.25$ & $0.32\pm 0.09$ & $-0.35\pm 0.11$\\
        PSO J065-26 & $5.8677$ & $-0.21\pm 0.03$ & $-0.28\pm 0.05$ & $-0.30\pm 0.16$ & $0.04\pm 0.04$ & $-0.40\pm 0.03$\\
        PSO J065-26$^*$ & $6.1208$ & $\cdots$ & $\cdots$ & $<0.04$ & $<0.31$ & $<-0.25$\\
        PSO J065-26$^*$ & $6.1263$ & $>0.66$ & $>0.93$ & $0.83\pm 0.06$ & $0.71\pm 0.04$ & $0.06\pm 0.04$\\
        SDSS J0100+2802 & $5.7974$ & $-0.28\pm 0.04$ & $\cdots$ & $-0.33\pm 0.09$ & $0.11\pm 0.04$ & $-0.32\pm 0.04$\\
        SDSS J0100+2802 & $5.9450$ & $0.39\pm 0.04$ & $0.29\pm 0.05$ & $0.25\pm 0.15$ & $0.51\pm 0.08$ & $0.00\pm 0.06$\\
        SDSS J0100+2802 & $6.1114$ & $-0.52\pm 0.15$ & $-0.50\pm 0.15$ & $-0.89\pm 0.17$ & $-0.50\pm 0.14$ & $-0.85\pm 0.14$\\
        SDSS J0100+2802 & $6.1434$ & $-0.29\pm 0.03$ & $-0.31\pm 0.04$ & $-0.68\pm 0.11$ & $-0.21\pm 0.03$ & $-0.55\pm 0.03$\\
        DELS J1535+1943 & $5.8990$ & $\cdots$ & $\cdots$ & $\cdots$ & $\cdots$ & $\cdots$\\
        PSO J183+05 & $6.0642$ & $0.05\pm 0.07$ & $0.40\pm 0.20$ & $0.27\pm 0.07$ & $0.36\pm 0.04$ & $-0.18\pm 0.05$\\
        PSO J183+05$^*$ & $6.4041$ & $0.00\pm 0.08$ & $>0.38$ & $-0.15\pm 0.08$ & $0.36\pm 0.05$ & $-0.22\pm 0.05$\\
        WISEA J0439+1634 & $6.2743$ & $0.12\pm 0.21$ & $<-0.30$ & $<0.12$ & $0.20\pm 0.17$ & $<-0.38$\\
        VDES J0224-4711 & $6.1228$ & $-0.20\pm 0.13$ & $-0.23\pm 0.04$ & $<-0.28$ & $0.09\pm 0.06$ & $-0.41\pm 0.06$\\
        PSO J036+03 & $6.0611$ & $-0.15\pm 0.10$ & $0.10\pm 0.14$ & $0.17\pm 0.23$ & $0.39\pm 0.13$ & $<-0.56$\\
        DELS J0923+0402 & $6.3784$ & $-0.46\pm 0.27$ & $-0.61\pm 0.17$ & $<-0.09$ & $-0.27\pm 0.10$ & $<-0.84$\\
        \hline
    \end{tabular}
    \tablefoot{$^*$ These systems are PDLA.}
\end{table*}

\FloatBarrier

\section{Description of the studied systems}
\label{app:LIS}
In this section, we describe in details every \OI\ system that we have analyzed. For brevity, in the following we refer to them as DLA systems, although only for 9 of them we have the confirmation that they are actually DLA. All reported wavelengths are in \AA. 

\subsection*{C.1. PSO J308-27 $z_{\rm abs}=5.4400$}
\begin{figure}
\centering
	\includegraphics[width=\columnwidth]{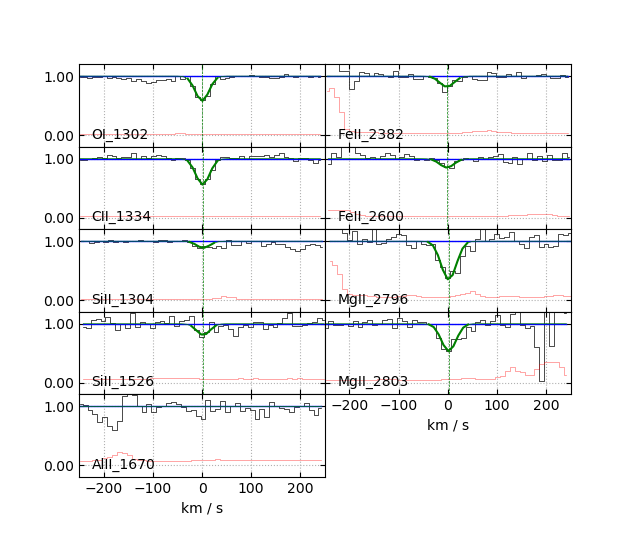}
	\includegraphics[width=\columnwidth]{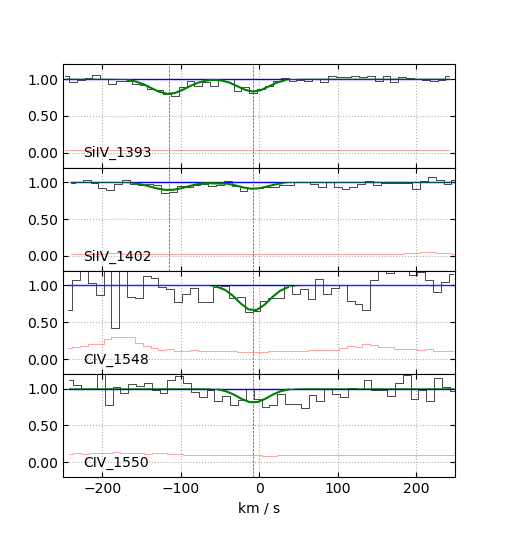}
    \caption{System at $z_{\rm abs}=5.4400$ in the spectrum of PSO J308-27. Upper panel: fit of the detected low-ionization metal lines, and region where \AlII\ $1670$ absorption would fall. Lower panel: fit of the detected high-ionization lines.}
    \label{fig:5.4400}
\end{figure}

This system is an intervening DLA, with a separation from the QSO emission redshift of $\sim 16250$ \kms. It is characterized by a simple velocity structure consisting of a single component. We detected the following ions: \OI\  $1302$, \CII\  $1334$, \SiII\  $1304$ and $1526$, \FeII\ $2382$ and $2600$, and \MgII\ $2796$,$2803$, as shown in Fig.~\ref{fig:5.4400}. The value of the parameters obtained from the fit are shown in Tab.~\ref{tab:5.4400}. We performed the fit by linking the $b$ parameter of all ions, and we also derived an upper limit on \AlII\ at $1670$.

In the system, we have also detected high ionization lines. At the redshift of the low ionization lines both \SiIV\ and \CIV\ are detected and have been fitted by linking the redshift and the Doppler parameter of the two ions (see Fig. \ref{fig:5.4400} and Tab.~\ref{tab:5.4400}). The profile of \SiIV\ shows a further component at $\Delta v \simeq -107.5$ \kms,  which is not considered in the total column density shown in Tab.~\ref{tab:LIS}. 
%have a velocity structure consisting of two \SiIV\ components and a single \CIV\ component . For the second component of \SiIV\ and for \CIV, we performed the fit by linking the redshift and the Doppler parameter of these ions.%, obtaining a value of $b=(22\pm 3)$ \kms.

Finally, we estimated a lower limit of $\log N_{\rm HI}>19.10$ for the \HI\ column density based on eq.~\ref{eq:relOH}.

\begin{table}
    \centering
    \caption{Voigt parameters obtained from the fit of the metal lines in the system at $z_{\rm abs}=5.4400$ in the spectrum of PSO J308-27.}
    \label{tab:5.4400}
    \begin{tabular}{lccc}
        \hline
        Ion & $z_{\rm abs}$ & $\log (N_{\rm X}/{\rm cm^{-2}})$ & $b$ (\kms)\\
        \hline
        $\OI$ & $5.44005\pm 0.00001$ & $14.09\pm 0.02$ & $11\pm 1$\\
        $\CII$ & $5.44007\pm 0.00002$ & $13.68\pm 0.03$ & $11\pm 1$\\
        $\SiII$ & $5.44010\pm 0.00006$ & $13.14\pm 0.07$ & $11\pm 1$\\
        $\AlII$ & $5.4400$ & $<11.86$ & $11$\\
        $\FeII$ & $5.44001\pm 0.00006$ & $12.57\pm 0.06$ & $11\pm 1$\\
        $\MgII$ & $5.44009\pm 0.00003$ & $13.06\pm 0.06$ & $11\pm 1$\\
        \hline
        \hline
        $\SiIV$ & $5.43757\pm 0.00005$ & $12.81\pm 0.04$ & $27\pm 4$\\
        \hline
        $\SiIV$ & $5.43988\pm 0.00005$ & $12.67\pm 0.05$ & $22\pm 3$\\
        $\CIV$ & $5.43988\pm 0.00005$ & $13.41\pm 0.07$ & $22\pm 3$\\
        \hline
    \end{tabular}
\end{table}

\subsection*{C.2. PSO J308-27 $z_{\rm abs}=5.6268$}
\begin{figure}
\centering
	\includegraphics[width=\columnwidth]{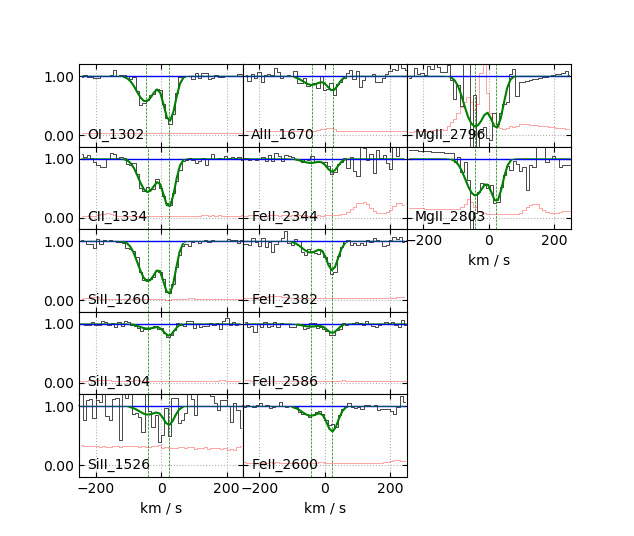}
	\includegraphics[width=\columnwidth]{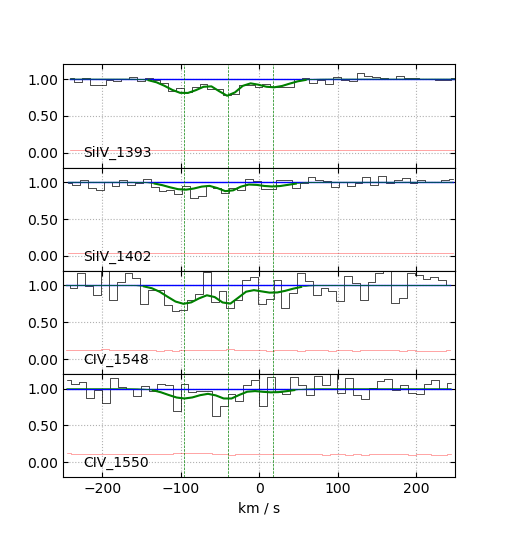}
    \caption{System at $z_{\rm abs}=5.6268$ in the spectrum of PSO J308-27. Upper panel: fit of the detected low-ionization metal lines. Lower panel: fit of the detected high-ionization lines.}
    \label{fig:5.6268}
\end{figure}

This system is an intervening DLA, with a separation from the QSO emission redshift of  $\sim 7690$ \kms, which presents a velocity structure characterized by two components. We detected the following ions: \OI\ $1302$, \CII\ $1334$, \SiII\ $1260$, $1304$ and $1526$, \AlII\ $1670$, \FeII\ $2344$, $2382$, $2586$ and $2600$, and \MgII\ $2796$, $2803$, as shown in Fig.~\ref{fig:5.6268}. The value of the parameters obtained from the fits of the low ionization lines are shown in Tab.~ \ref{tab:5.6268}. We have linked the $b$ parameters of all the ions in each of the two components, with the exception of that of \MgII\ which falls in the telluric band and is therefore strongly affected by the absorption lines of the sky (even if they have been partially corrected). %For the first component we obtained $b=(31\pm 1)$ \kms, while for the second $b=(14.5\pm 0.7)$ \kms. We also fit \SiII\ to $1526$, linking the redshift and column density to the other two transitions of the same ion. 
Since the \AlII\ at $1670$ line is very weak, we performed its fit by linking the redshifts of the two components with those of \SiII. As regards \MgII, on the other hand, for each of the two components we have fixed the redshift to a value equal to that of \SiII\ and the parameter $b$ to a value equal to that of the other ions; we then performed the fit with the column density as the only free parameter.

In this system, we have also detected high ionization lines, which have a velocity structure characterized by three components of both \SiIV\ and \CIV\ covering the same velocity range of the low-ionization line profiles (see Fig. \ref{fig:5.6268} and Tab.~\ref{tab:5.6268}). We performed the fit of all the components by linking the redshift and the $b$ parameter of the two ions.%; for the Doppler parameter we have obtained values of $b=(24\pm 7)$ \kms, $b=(9\pm 7)$ \kms, and $b=(22\pm 12)$ \kms.

Finally, we estimated a lower limit of $\log N_{\rm HI}>19.79$ for the \HI\ column density based on eq.~\ref{eq:relOH}.

\begin{table}
    \centering
    \caption{Voigt parameters obtained from the fit of the metal lines in the system at $z_{\rm abs}=5.6268$ in the spectrum of PSO J308-27.}
    \label{tab:5.6268}
    \begin{tabular}{lccc}
        \hline
        Ion & $z_{\rm abs}$ & $\log (N_{\rm X}/{\rm cm^{-2}})$ & $b$ (\kms)\\
        \hline
        $\OI$ & $5.62582\pm 0.00005$ & $14.32\pm 0.03$ & $31\pm 1$\\
        $\CII$ & $5.62597\pm 0.00004$ & $14.06\pm 0.03$ & $31\pm 1$\\
        $\SiII$ & $5.62597\pm 0.00002$ & $13.25\pm 0.01$ & $31\pm 1$\\
        $\AlII$ & $5.62597\pm 0.00002$ & $12.12\pm 0.16$ & $31\pm 1$\\
        $\FeII$ & $5.62591\pm 0.00009$ & $12.82\pm 0.05$ & $31\pm 1$\\
        $\MgII$ & $5.62597\pm 0.00000$ & $13.44\pm 0.08$ & $31$\\
        \hline
        $\OI$ & $5.62743\pm 0.00002$ & $14.59\pm 0.04$ & $15.5\pm 0.7$\\
        $\CII$ & $5.62739\pm 0.00002$ & $14.25\pm 0.04$ & $15.5\pm 0.7$\\
        $\SiII$ & $5.62739\pm 0.00001$ & $13.47\pm 0.03$ & $15.5\pm 0.7$\\
        $\AlII$ & $5.62739\pm 0.00001$ & $12.18\pm 0.20$ & $15.5\pm 0.7$\\
        $\FeII$ & $5.62737\pm 0.00003$ & $13.17\pm 0.03$ & $15.5\pm 0.7$\\
        $\MgII$ & $5.62739\pm 0.00000$ & $13.43\pm 0.06$ & $15.5$\\
        \hline
        \hline
        $\SiIV$ & $5.62475\pm 0.00008$ & $12.76\pm 0.07$ & $24\pm 7$\\
        $\CIV$ & $5.62475\pm 0.00008$ & $13.29\pm 0.15$ & $24\pm 7$\\
        \hline
        $\SiIV$ & $5.62599\pm 0.00006$ & $12.65\pm 0.07$ & $9\pm 7$\\
        $\CIV$ & $5.62599\pm 0.00006$ & $13.13\pm 0.21$ & $9\pm 7$\\
        \hline
        $\SiIV$ & $5.62724\pm 0.00014$ & $12.49\pm 0.13$ & $22\pm 12$\\
        $\CIV$ & $5.62724\pm 0.00014$ & $12.84\pm 0.35$ & $22\pm 12$\\
        \hline
    \end{tabular}
\end{table}

\subsection*{C.3. PSO J023-02 $z_{\rm abs}=5.4869$}
\begin{figure}
\centering
	\includegraphics[width=\columnwidth]{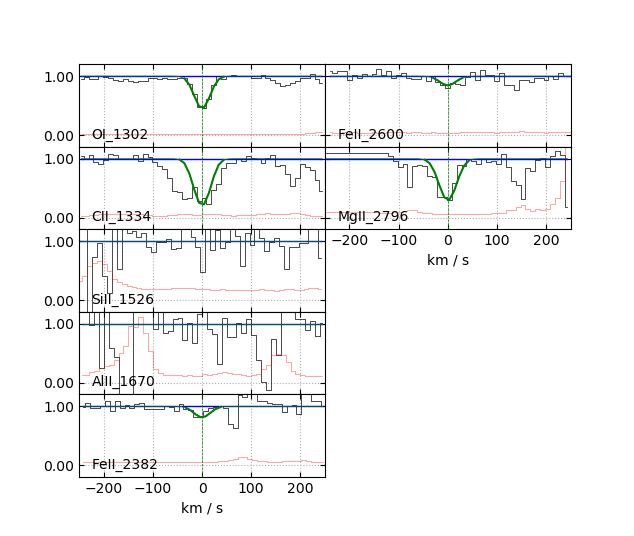}
	\includegraphics[width=\columnwidth]{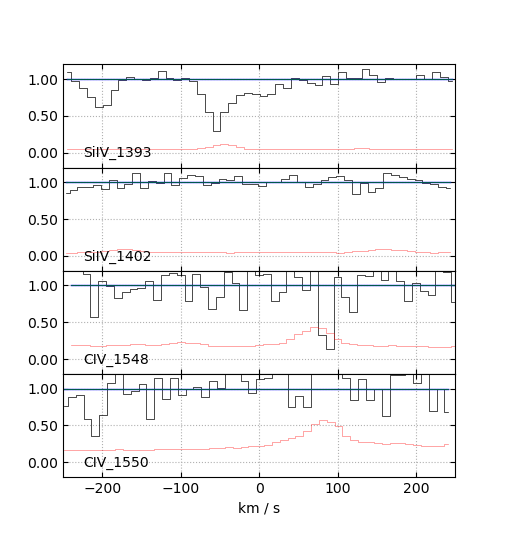}
    \caption{System at $z_{\rm abs}=5.4869$ in the spectrum of PSO J023-02. Upper panel: fit of the detected low-ionization metal lines, and the region where the absorptions of \SiII\ $1526$ and \AlII\ $1670$ would fall. Lower panel: region where the high ionization lines of \SiIV\ and \CIV\ would fall.}
    \label{fig:5.4869}
\end{figure}

This system is an intervening DLA composed of a single component, separated from the QSO emission redshift by $\sim 16310$ \kms. We detected the following ions: \OI\ $1302$, \CII\ $1334$, \FeII\ $2382$ and $2600$, and \MgII\ $2796$, as shown in Fig.~\ref{fig:5.4869}. The value of the parameters obtained from the fit of the low ionization lines are shown in Tab.~\ref{tab:5.4869}. The \CII\ line is blended with the fine structure line of \SiII* $1264$ at $z=5.8440$, while the lines of \MgII\ fall into the telluric band and  only the transition at $2796$ could be observed. We performed the fit by linking the redshift and the $b$ parameter of all ions.% and obtained for the latter the value of $b=(13\pm 2)$ \kms. 
We also derived an upper limit on \SiII\ at $1526$ (\SiII\ $1304$ is blended with a \CIV\ $1548$ at $z=4.4656$), and on \AlII\ at $1670$.

In this system, we did not detect high ionization lines (see Fig. \ref{fig:5.4869}) thus we derived upper limits on \SiIV\ and \CIV.
%using equation (\ref{eq:s_n}) and (\ref{eq:upperlimit}).% The results obtained from the calculation are shown in the Table.

Furthermore, a lower limit on the \HI\ column density of $\log N_{\rm HI}>19.33$ was estimated from eq.~\ref{eq:relOH}.

\begin{table}
    \centering
    \caption{Voigt parameters obtained from the fit of the metal lines in the system at $z_{\rm abs}=5.4869$ in the spectrum of PSO J023-02.}
    \label{tab:5.4869}
    \begin{tabular}{lccc}
        \hline
        Ion & $z_{\rm abs}$ & $\log (N_{\rm X}/{\rm cm^{-2}})$ & $b$ (\kms)\\
        \hline
        $\OI$ & $5.48698\pm 0.00002$ & $14.32\pm 0.03$ & $13\pm 2$\\
        $\CII$ & $5.48698\pm 0.00002$ & $14.27\pm 0.15$ & $13\pm 2$\\
        $\SiII$ & $5.4869$ & $<13.43$ & $13$\\
        $\AlII$ & $5.4869$ & $<12.33$ & $13$\\
        $\FeII$ & $5.48698\pm 0.00002$ & $12.64\pm 0.10$ & $13\pm 2$\\
        $\MgII$ & $5.48698\pm 0.00002$ & $13.18\pm 0.11$ & $13\pm 2$\\
        \hline
        \hline
        $\SiIV$ & $5.4869$ & $<12.65$ & $26$\\
        $\CIV$ & $5.4869$ & $<13.51$ & $26$\\
        \hline
    \end{tabular}
\end{table}

\subsection*{C.4. PSO J023-02 $z_{\rm abs}=5.8441$ PDLA/intrinsic}%associated
This system has a separation from the QSO emission redshift of $\sim 250$ \kms, and is characterized by a single-component velocity structure. Besides the common ions of \OI\ $1302$, \CII\ $1334$, \SiII\ $1260$, $1304$ and $1526$, \AlII\ $1670$ and \FeII\ $2344$, $2382$, $2586$ and $2600$, we could detect relatively rare (at this redshift) transitions of \SII\ $1250$, $1253$ and $1259$ and \MnII\ $2576$ and $2594$, as shown in Fig.~\ref{fig:5.8441}. The value of the parameters obtained from the fit of the low ionization lines are shown in the Tab.~\ref{tab:5.8441}. Furthermore, we observed absorption lines due to the fine-structure transitions \CII* $1335$ and  \SiII* $1264$ (Fig.~\ref{fig:5.8441fine}), which implies that the system could be very close to the QSO and possibly part of an outflow.
It was not possible to detect the \MgII\ lines as they fall in a region affected by strong telluric absorption that could not be corrected. The \CII\ $1334$ line is blended with an  \AlII\ at $z=4.4657$, while the fine structure line \SiII* is blended with the \CII\ line of the previous system at $z=5.4869$. We performed the fit by linking the $b$ parameters of all the detected ions.%, obtaining a value of $b=(26\pm 1)$ \kms.

In the system, we also detected high ionization lines due to \NV\, \SiIV\ and \CIV\ which have a broad velocity structure (see Fig. \ref{fig:5.8441}). We performed a fit with three components, linking the b parameters of these ions in each of them. %We have obtained: $b=(44\pm 2)$ \kms\ for the first component, $b=(28\pm 4)$ \kms\ for the second and $b=(66\pm 2)$ \kms\ for the third.
Note that this QSO show also a broad absorption line (BAL) system identified through the \CIV\ absorption in \citet{Bischetti2022}. 

\begin{figure}
\centering
    \includegraphics[width=\columnwidth]{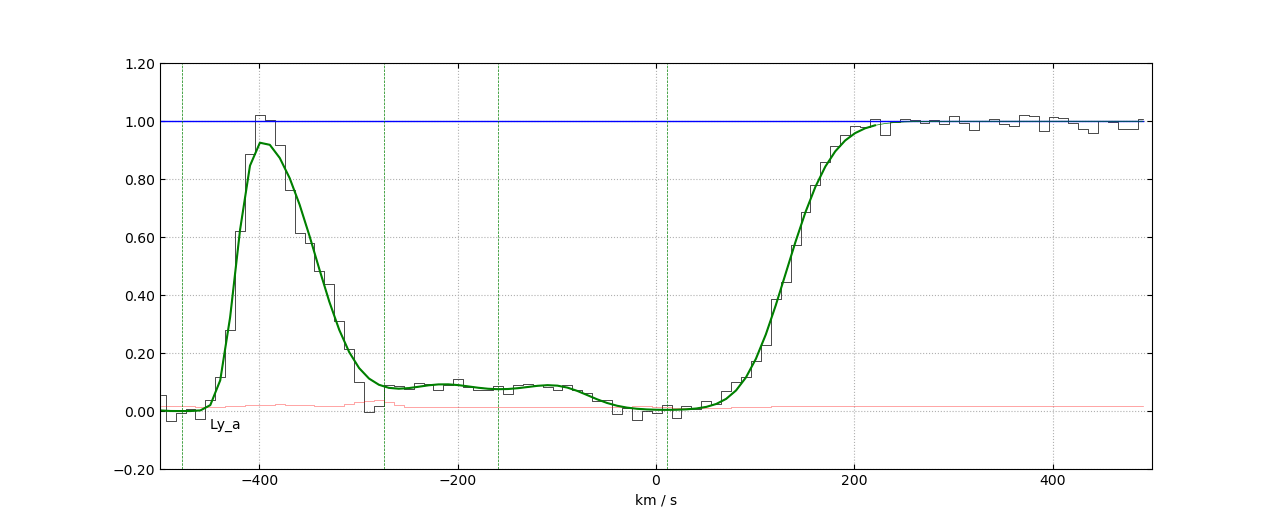}
\label{fig:5.8441Lya}
\caption{System at $z_{\rm abs}=5.8441$ in the spectrum of PSO J023-02. Fit of the \HI\ \Lya\ absorption line}
\end{figure}

\begin{figure}
\centering
    \includegraphics[width=\columnwidth]{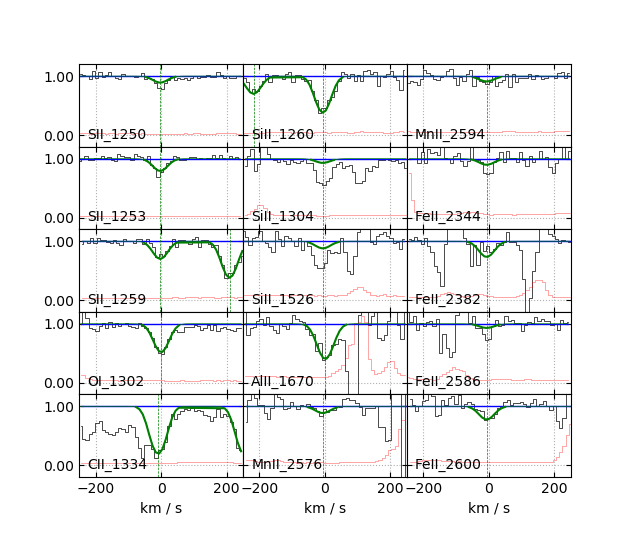}
    \includegraphics[width=\columnwidth]{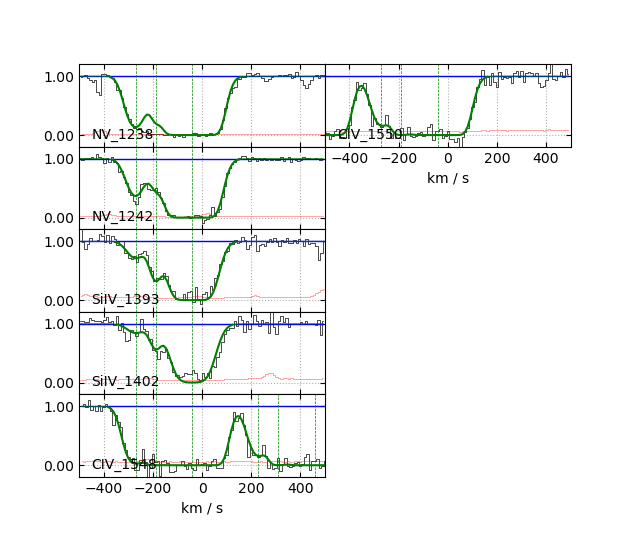}
    \caption{System at $z_{\rm abs}=5.8441$ in the spectrum of PSO J023-02. Upper panel: fit of the detected low-ionization metal lines. Lower panel: fit of the detected high-ionization lines.}
    \label{fig:5.8441}
\end{figure}

\begin{figure}
    \includegraphics[width=\columnwidth]{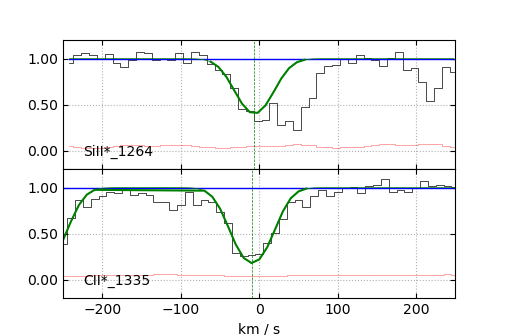}
    \caption{System at $z_{\rm abs}=5.8441$ in the spectrum of PSO J023-02. Fit of the detected fine-structure lines. }
    \label{fig:5.8441fine}
\end{figure}

Finally, we also observed the \Lya\ absorption line in the proximity region of the QSO (see Fig. \ref{fig:5.8441Lya}), which shows a possible signature of partial coverage, supporting the hypothesis that the system is arising in a clump of gas close to the ionizing source. Due to the peculiarity of this system, we decided to exclude it from the subsequent analysis and to study it in more details in a future work. 

%However, the presence of \NV\ suggests that this is an associated system with a different structure than the low ionization systems we are looking for. This system was therefore excluded from the subsequent analysis.

\begin{table}
    \centering
    \caption{Voigt parameters obtained from the fit of the metal lines in the system at $z_{\rm abs}=5.8441$ in the spectrum of PSO J023-02.}
    \label{tab:5.8441}
    \begin{tabular}{lccc}
        \hline
        Ion & $z_{\rm abs}$ & $\log (N_{\rm X}/{\rm cm^{-2}})$ & $b$ (\kms)\\
        \hline
        $\HI$ & $5.83793\pm 0.00022$ & $14.24\pm 0.16$ & $61\pm 5$\\
        \hline
        $\HI$ & $5.84055\pm 0.00023$ & $14.44\pm 0.14$ & $85\pm 26$\\
        \hline
        $\HI$ & $5.84444\pm 0.00015$ & $14.79\pm 0.05$ & $84\pm 3$\\
        \hline
        \hline
        $\SII$ & $5.84412\pm 0.00005$ & $14.55\pm 0.04$ & $26\pm 1$\\
        $\OI$ & $5.84419\pm 0.00005$ & $14.35\pm 0.04$ & $26\pm 1$\\
        $\CII$ & $5.84398\pm 0.00003$ & $14.32\pm 0.04$ & $26\pm 1$\\
        $\CII*$ & $5.84398\pm 0.00003$ & $14.38\pm 0.04$ & $26\pm 1$\\
        $\SiII$ & $5.84408\pm 0.00004$ & $13.14\pm 0.04$ & $26\pm 1$\\
        $\SiII*$ & $5.84408\pm 0.00004$ & $13.22\pm 0.05$ & $26\pm 1$\\
        $\AlII$ & $5.84421\pm 0.00013$ & $12.84\pm 0.10$ & $26\pm 1$\\
        $\FeII$ & $5.84404\pm 0.00010$ & $12.95\pm 0.07$ & $26\pm 1$\\
        $\MnII$ & $5.84403\pm 0.00021$ & $12.45\pm 0.13$ & $26\pm 1$\\
        \hline
        \hline
        $\NV$ & $5.83801\pm 0.00004$ & $14.52\pm 0.02$ & $44\pm 2$\\
        $\SiIV$ & $5.83801\pm 0.00004$ & $13.18\pm 0.08$ & $44\pm 2$\\
        $\CIV$ & $5.83801\pm 0.00004$ & $>14.60$ & $44\pm 2$\\
        \hline
        $\NV$ & $5.83987\pm 0.00006$ & $>14.11$ & $28\pm 4$\\
        $\SiIV$ & $5.83987\pm 0.00006$ & $13.51\pm 0.05$ & $28\pm 4$\\
        $\CIV$ & $5.83987\pm 0.00006$ & $>15.41$ & $28\pm 4$\\
        \hline
        $\NV$ & $5.84329\pm 0.00003$ & $>16.14$ & $66\pm 2$\\
        $\SiIV$ & $5.84329\pm 0.00003$ & $>14.87$ & $66\pm 2$\\
        $\CIV$ & $5.84329\pm 0.00003$ & $>16.33$ & $66\pm 2$\\
        \hline
    \end{tabular}
\end{table}

\subsection*{C.5. PSO J025-11 $z_{\rm abs}=5.7763$ PDLA}
\begin{figure}
\centering
    \includegraphics[width=\columnwidth]{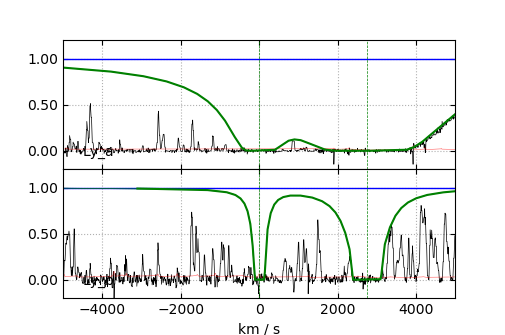}
	\includegraphics[width=\columnwidth]{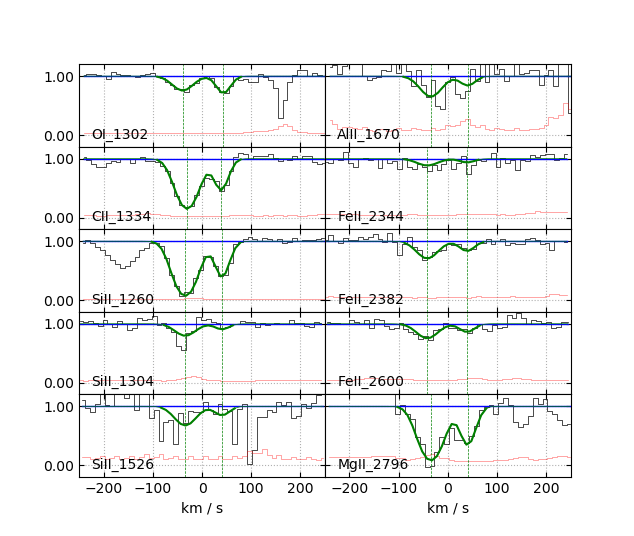}
	\includegraphics[width=\columnwidth]{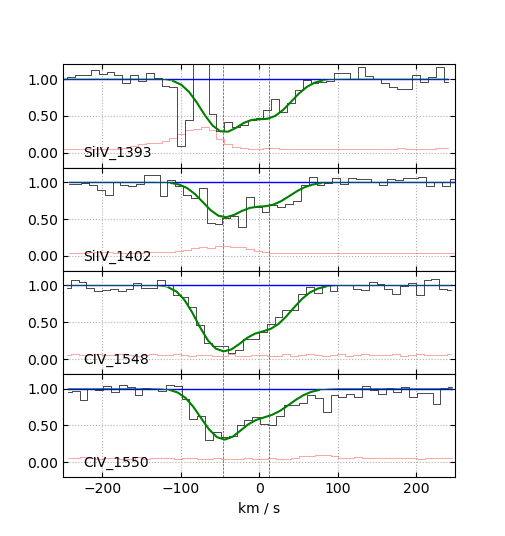}
    \caption{System at $z_{\rm abs}=5.7763$ in the spectrum of PSO J025-11. Upper panel: fit of the \HI\ \Lya\ and \Lyb\ absorption lines. Middle panel: fit of the detected low-ionization metal lines. Lower panel: fit of the detected high-ionization lines.}
    \label{fig:5.7763}
\end{figure}

This system is a PDLA with a two-component velocity structure and a separation from the QSO emission redshift of $\sim 2866$ \kms. We detected the following ions: \OI\ $1302$, \CII\ $1334$, \SiII\ $1260$, $1304$ and $1526$, \AlII\ at $1670$, \FeII\ at $2344$, $2382$ and $2600$, and \MgII\ at $2796$, as shown in Fig.~\ref{fig:5.7763}. The value of the parameters obtained from the fit of the low ionization lines are shown in Tab.~\ref{tab:5.7763}. We carried out the fit by linking the $b$ parameters of all the ions in each of the two components. 
%For the first component we got $b=(23\pm 1)$ \kms, while for the second $b=(10\pm 2)$ \kms. We also fit \SiII\ $1526$ linking the redshift, column density and Doppler parameter to the other two observed \SiII\ transitions. 
As for the \AlII, which falls in a region with low SNR, we have also linked the redshifts to those of the \SiII. The lines of \MgII, on the other hand, fall into the telluric band and are therefore strongly affected by the sky lines, even if they have been partially corrected; only the transition at $2796$ could be observed.

Since the system is a PDLA we have also performed the fit of the \Lya\ absorption line, as described in Sect.~\ref{sec:lya}. This line is blended with another \Lya\ absorption line present in the spectrum, associated with the absorption system at $z=5.8385$. From the fit we obtained a column density $\log N_{\rm HI}=20.35\pm 0.10$, in agreement with the lower limit determined from the rest equivalent width of \MgII\ $2796$ line.

In the system we have also detected high ionization lines, which present a velocity structure characterized by two components of \SiIV\ and \CIV\ (see Fig. \ref{fig:5.7763} and Tab.~\ref{tab:5.7763}). For both components we linked the redshifts and Doppler parameters of the two ions.
%, obtaining for the latter a value of $b=(28\pm 3)$ \kms\ for the first component and $b=(32\pm 6)$ \kms\ for the second.

\begin{table}
    \centering
    \caption{Voigt parameters obtained from the fit of the metal lines in the system at $z_{\rm abs}=5.7763$ in the spectrum of PSO J025-11.}
    \label{tab:5.7763}
    \begin{tabular}{lccc}
        \hline
        Ion & $z_{\rm abs}$ & $\log (N_{\rm X}/{\rm cm^{-2}})$ & $b$ (\kms)\\
        \hline
        $\HI$ & $5.7763$ & $20.35\pm 0.10$ & $40$\\
        \hline
        \hline
        $\OI$ & $5.77549\pm 0.00007$ & $13.94\pm 0.05$ & $23\pm 1$\\
        $\CII$ & $5.77569\pm 0.00002$ & $14.37\pm 0.03$ & $23\pm 1$\\
        $\SiII$ & $5.77560\pm 0.00002$ & $13.59\pm 0.04$ & $23\pm 1$\\
        $\AlII$ & $5.77560\pm 0.00002$ & $12.47\pm 0.11$ & $23\pm 1$\\
        $\FeII$ & $5.77544\pm 0.00005$ & $12.96\pm 0.05$ & $23\pm 1$\\
        $\MgII$ & $5.77560\pm 0.00006$ & $13.54\pm 0.14$ & $23\pm 1$\\
        \hline
        $\OI$ & $5.77736\pm 0.00005$ & $13.90\pm 0.06$ & $10\pm 2$\\
        $\CII$ & $5.77725\pm 0.00003$ & $13.89\pm 0.07$ & $10\pm 2$\\
        $\SiII$ & $5.77730\pm 0.00002$ & $13.08\pm 0.08$ & $10\pm 2$\\
        $\AlII$ & $5.77730\pm 0.00002$ & $11.95\pm 0.59$ & $10\pm 2$\\
        $\FeII$ & $5.77727\pm 0.00012$ & $12.55\pm 0.11$ & $10\pm 2$\\
        $\MgII$ & $5.77730\pm 0.00002$ & $13.19\pm 0.19$ & $10\pm 2$\\
        \hline
        \hline
        $\SiIV$ & $5.77534\pm 0.00007$ & $13.59\pm 0.10$ & $28\pm 3$\\
        $\CIV$ & $5.77534\pm 0.00007$ & $14.26\pm 0.05$ & $28\pm 3$\\
        \hline
        $\SiIV$ & $5.77665\pm 0.00014$ & $13.39\pm 0.10$ & $32\pm 6$\\
        $\CIV$ & $5.77665\pm 0.00014$ & $13.84\pm 0.10$ & $32\pm 6$\\
        \hline
    \end{tabular}
\end{table}

\subsection*{C.6. PSO J025-11 $z_{\rm abs}=5.8385$ PDLA}
This system is a PDLA with a separation from the QSO emission redshift of $\sim 127$ \kms. It is characterized by a complex velocity structure. We detected four components of \OI\ $1302$ and \CII\ $1334$, six components of \SiII\ at $1260$, $1304$ and $1526$, five components of \AlII\ $1670$, seven components of \FeII\ $2382$ and $2600$ (of which three also detected in \FeII\ $1608$, two also in \FeII\ $2374$ and five also in the transition \FeII\ $2586$), and three components of \MgII\ $2803$, as shown in Fig.~\ref{fig:5.8385}. The value of the parameters obtained from the fit of the low ionization lines are shown in Tab.~\ref{tab:5.8385}. We performed the fit by linking the Doppler parameter of the velocity profile components that matched in redshift. Oxygen, carbon and magnesium are saturated, therefore their column density was considered a lower limit.
The lines of \MgII\ fall in the telluric band and are therefore strongly affected by the lines of the sky, only three component of the transition at $2803$ could be observed, which correspond to the last three components of \OI\ and \CII. 
%We therefore freezed the $b$ parameter a $b=36$ \kms in the first component, which corresponds to the second of \OI\ and \CII, since it is located in a region with high noise.

We also performed the fit of the \HI\ \Lya\ absorption line, resulting in a column density $\log N_{\rm HI}=21.25\pm 0.10$. This line is blended with another \Lya\ absorption line present in the spectrum, due to the absorption system at $z=5.7763$. The \HI\ fit is in agreement with the lower limit for the \HI\ column density based on eq.~\ref{eq:relOH}.

In the system we have also detected high ionization lines, which have a velocity structure characterized by three components of \SiIV\ and \CIV%, as shown in the Figure. The value of the parameters obtained from the fit of these ions are shown in the Table
. We fit the first two components by linking the $b$ parameter of the two ions, while the third leaving this parameter free as well.
%obtaining $b=(28\pm 4)$ \kms\ for the first component and $b=(22\pm 4)$ \kms\ for the second.

\begin{table}
    \centering
    \caption{Voigt parameters obtained from the fit of the metal lines in the system at $z_{\rm abs}=5.8385$ in the spectrum of PSO J025-11.}
    \label{tab:5.8385}
    \begin{tabular}{lccc}
        \hline
        Ion & $z_{\rm abs}$ & $\log (N_{\rm X}/{\rm cm^{-2}})$ & $b$ (\kms)\\
        \hline
        $\HI$ & $5.8385$ & $21.25\pm 0.10$ & $100$\\
        \hline
        \hline
        $\OI$ & $5.83567\pm 0.00006$ & $14.42\pm 0.05$ & $24\pm 2$\\
        $\CII$ & $5.83578\pm 0.00007$ & $13.88\pm 0.06$ & $24\pm 2$\\
        $\SiII$ & $5.83548\pm 0.00004$ & $13.17\pm 0.04$ & $24\pm 2$\\
        $\FeII$ & $5.83545\pm 0.00009$ & $12.87\pm 0.14$ & $6$\\
        \hline
        $\AlII$ & $5.83643\pm 0.00011$ & $12.56\pm 0.12$ & $13\pm 6$\\
        $\FeII$ & $5.83644\pm 0.00007$ & $13.37\pm 0.08$ & $13\pm 6$\\
        \hline
        $\OI$ & $5.83737\pm 0.00004$ & $>15.45$ & $36\pm 2$\\
        $\CII$ & $5.83756\pm 0.00003$ & $>15.28$ & $36\pm 2$\\
        $\SiII$ & $5.83747\pm 0.00003$ & $14.45\pm 0.02$ & $36\pm 2$\\
        $\AlII$ & $5.83760\pm 0.00010$ & $12.83\pm 0.08$ & $19\pm 2$\\
        $\FeII$ & $5.83749\pm 0.00004$ & $14.05\pm 0.06$ & $19\pm 2$\\
        $\MgII$ & $5.83749\pm 0.00000$ & $>14.40$ & $36$\\
        \hline
        $\SiII$ & $5.83875\pm 0.00012$ & $13.11\pm 0.19$ & $18\pm 11$\\
        $\AlII$ & $5.83875\pm 0.00012$ & $12.29\pm 0.21$ & $18\pm 11$\\
        $\FeII$ & $5.83875\pm 0.00012$ & $12.87\pm 0.14$ & $18\pm 11$\\
        \hline
        $\OI$ & $5.84023\pm 0.00003$ & $>15.32$ & $35\pm 2$\\
        $\CII$ & $5.84023\pm 0.00002$ & $>14.92$ & $35\pm 2$\\
        $\SiII$ & $5.83967\pm 0.00002$ & $14.36\pm 0.24$ & $8\pm 1$\\
        $\AlII$ & $5.83972\pm 0.00009$ & $13.06\pm 0.46$ & $8\pm 1$\\
        $\FeII$ & $5.83967\pm 0.00002$ & $14.11\pm 0.08$ & $8\pm 1$\\
        $\MgII$ & $5.84002\pm 0.00007$ & $>14.10$ & $35\pm 2$\\
        \hline
        $\SiII$ & $5.84039\pm 0.00008$ & $13.88\pm 0.06$ & $32\pm 3$\\
        $\AlII$ & $5.84030\pm 0.00037$ & $12.53\pm 0.25$ & $32\pm 3$\\
        $\FeII$ & $5.84035\pm 0.00012$ & $13.54\pm 0.08$ & $32\pm 3$\\
        \hline
        $\OI$ & $5.84243\pm 0.00013$ & $14.11\pm 0.06$ & $39\pm 5$\\
        $\CII$ & $5.84263\pm 0.00009$ & $13.84\pm 0.05$ & $39\pm 5$\\
        $\SiII$ & $5.84251\pm 0.00006$ & $12.83\pm 0.04$ & $24\pm 4$\\
        $\FeII$ & $5.84251\pm 0.00006$ & $12.65\pm 0.10$ & $24\pm 4$\\
        $\MgII$ & $5.84267\pm 0.00054$ & $12.55\pm 0.23$ & $39\pm 5$\\
        \hline
        \hline
        $\SiIV$ & $5.83556\pm 0.00006$ & $13.15\pm 0.04$ & $28\pm 4$\\
        $\CIV$ & $5.83549\pm 0.00011$ & $13.73\pm 0.08$ & $28\pm 4$\\
        \hline
        $\SiIV$ & $5.83694\pm 0.00008$ & $12.94\pm 0.07$ & $22\pm 4$\\
        $\CIV$ & $5.83663\pm 0.00007$ & $13.88\pm 0.06$ & $22\pm 4$\\
        \hline
        $\SiIV$ & $5.83858\pm 0.00007$ & $12.98\pm 0.05$ & $27\pm 5$\\
        $\CIV$ & $5.83851\pm 0.00019$ & $13.11\pm 0.11$ & $19\pm 5$\\
        \hline
    \end{tabular}
\end{table}

\begin{figure}
\centering
    \includegraphics[width=\columnwidth]{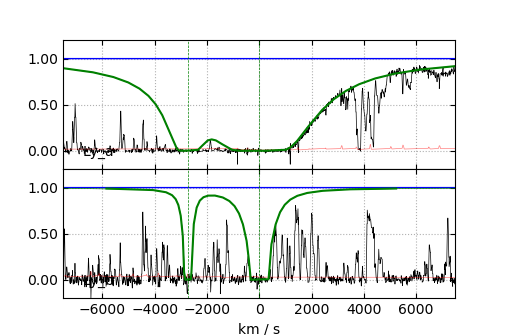}
    \caption{System at $z_{\rm abs}=5.8385$ in the spectrum of PSO J025-11. Fit of the \HI\ \Lya\ and \Lyb\ absorption lines.}
    \label{fig:5.8385Lya}
\end{figure}

\begin{figure}
	\includegraphics[width=\columnwidth]{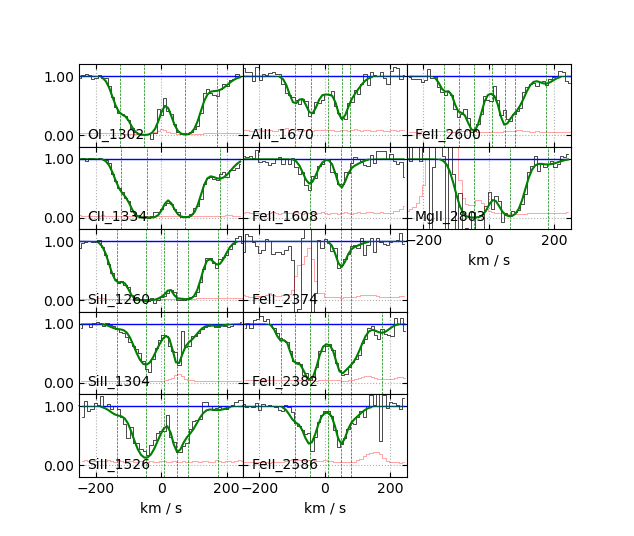}
	\includegraphics[width=\columnwidth]{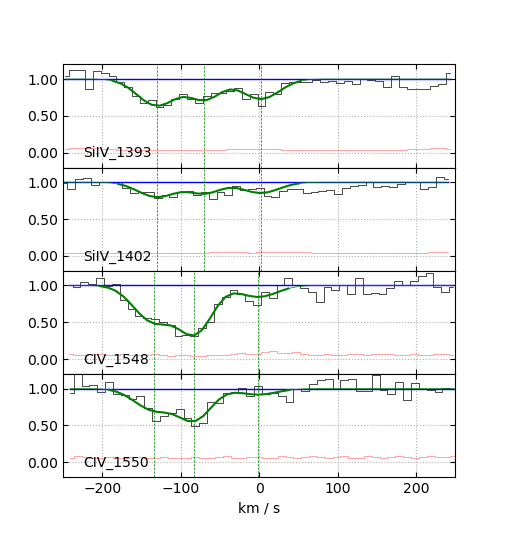}
    \caption{System at $z_{\rm abs}=5.8385$ in the spectrum of PSO J025-11. Upper panel: fit of the detected low-ionization metal lines. Lower panel: fit of the detected high-ionization lines.}
    \label{fig:5.8385}
\end{figure}

\subsection*{C.7. PSO J108+08 $z_{\rm abs}=5.5624$}
\begin{figure}
\centering
	\includegraphics[width=\columnwidth]{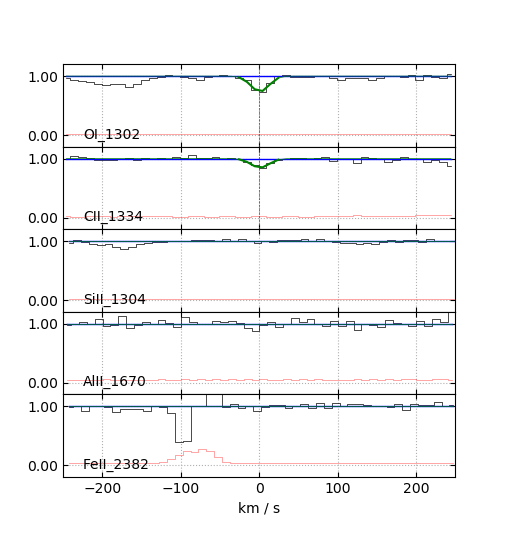}
	\includegraphics[width=\columnwidth]{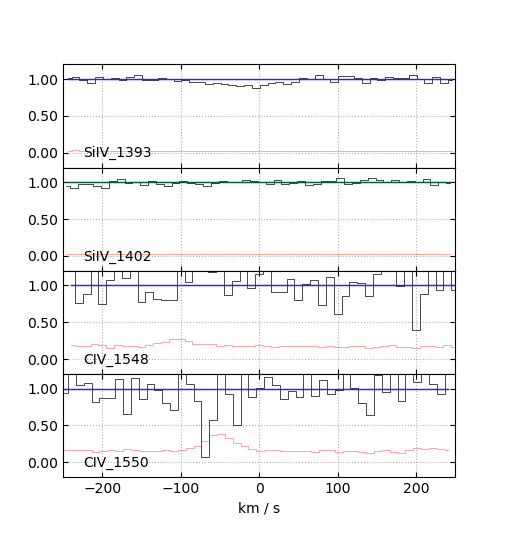}
    \caption{System at $z_{\rm abs}=5.5624$ in the spectrum of PSO J108+08. Upper panel: fit of the detected low-ionization metal lines, and the region where the absorptions of \SiII\ $1304$, \AlII\ $1670$ and \FeII\ $2382$ would fall. Lower panel: region where the high ionization lines of \SiIV\ and \CIV\ would fall.}
    \label{fig:5.5624}
\end{figure}

This system is an intervening DLA composed of a single component, separated from the QSO emission redshift by $\sim 17400$ \kms. We detected the following ions: \OI\ $1302$ and \CII\ $1334$, as shown in Fig.~\ref{fig:5.5624}. The value of the parameters obtained from the fit of the low ionization lines are shown in Tab.~\ref{tab:5.5624}. We performed the fit by linking the redshift of these ions and freezing the parameter $b$ to its minimum value of $5$ \kms for the visible region of the spectrum. We also derived an upper limit on \SiII\ $1304$, \AlII\ $1670$, and on \FeII\ $2382$. %using equation (\ref{eq:s_n}) and (\ref{eq:upperlimit}) and fixing the Doppler parameter to $b=13$ \kms
The \MgII\ lines fall in a telluric band, and it was not possible to correct the effect of the sky lines at the wavelength at which this absorption occurs.

In this system, we did not detect high ionization lines (see Fig. \ref{fig:5.5624}) thus we derived upper limits on \SiIV\ and \CIV.
The lower limit on the \HI\ column density derived from eq.~\ref{eq:relOH} is $\log N({\rm HI}) > 18.85$. 

\begin{table}
    \centering
    \caption{Voigt parameters obtained from the fit of the metal lines in the system at $z_{\rm abs}=5.5624$ in the spectrum of PSO J108+08.}
    \label{tab:5.5624}
    \begin{tabular}{lccc}
        \hline
        Ion & $z_{\rm abs}$ & $\log (N_{\rm X}/{\rm cm^{-2}})$ & $b$ (\kms)\\
        \hline
        $\OI$ & $5.56249\pm 0.00001$ & $13.84\pm 0.03$ & $5$\\
        $\CII$ & $5.56249\pm 0.00001$ & $13.12\pm 0.07$ & $5$\\
        $\SiII$ & $5.5624$ & $<12.42$ & $5$\\
        $\AlII$ & $5.5624$ & $<11.56$ & $6$\\
        $\FeII$ & $5.5624$ & $<11.91$ & $6$\\
        \hline
        \hline
        $\SiIV$ & $5.5624$ & $<12.39$ & $26$\\
        $\CIV$ & $5.5624$ & $<13.31$ & $26$\\
        \hline
    \end{tabular}
\end{table}

\subsection*{C.8. SDSS J0818+1722 $z_{\rm abs}=5.7912$}
\begin{figure}
\centering
	\includegraphics[width=\columnwidth]{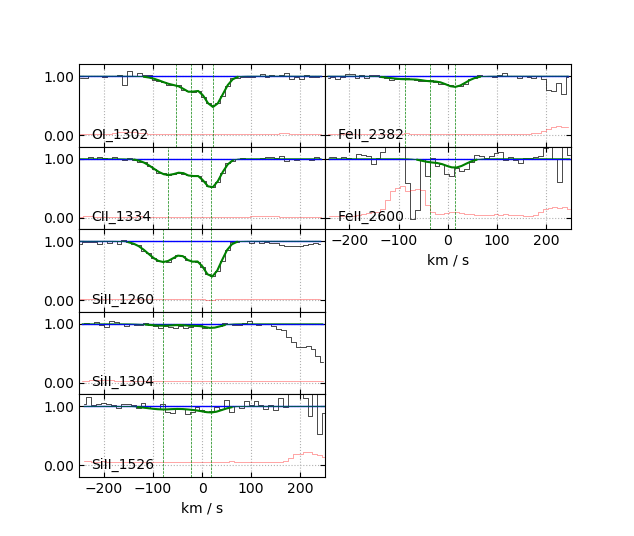}
	\includegraphics[width=\columnwidth]{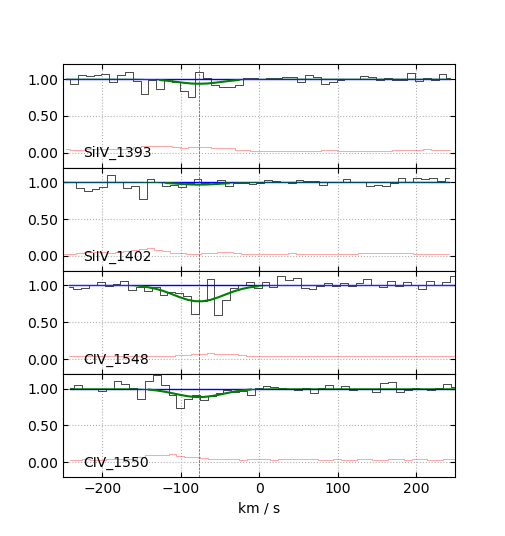}
    \caption{System at $z_{\rm abs}=5.7912$ in the spectrum of SDSS J0818+1722. Upper panel: fit of the detected low-ionization metal lines. Lower panel: fit of the detected high-ionization lines.}
    \label{fig:5.7912}
\end{figure}

This system is an intervening DLA with a velocity structure characterized by three components. The separation from the QSO emission redshift is of $\sim 8950$ \kms. We detected the following ions: \OI\ $1302$, \CII\ $1334$, \SiII\ $1260$, $1304$ and $1526$, \FeII\ $2382$ and $2600$, as shown in Fig.~\ref{fig:5.7912}. The value of the parameters obtained from the fit of the low ionization lines are shown in Tab.~\ref{tab:5.7912}. We linked the $b$ parameters of all the ions in each of the three components. %For the first component we got $b=(36\pm 2)$ \kms, for the second $b=(7\pm 6)$ \kms, while for the third $b=(19.1\pm 0.8)$ \kms. 
%Finally, it was not possible to detect 
The \MgII\ lines fall in a telluric band and its column density cannot be recovered. 

As regards the high ionization lines, in this system we detected the \CIV\ doublet, while we obtained an upper limit for the column density of \SiIV, using the transition at $1402$ and setting the parameter $b$ at $36$ \kms\ like that of \CIV\ (see Fig. \ref{fig:5.7912} and Tab.~ \ref{tab:5.7912}).

The lower limit on the \HI\ column density derived from eq.~\ref{eq:relOH} is $\log N({\rm HI}) > 19.51$. 

This system was already identified by \citet{Becker2011} in a high-resolution spectrum obtained with HIRES at the Keck telescope. They detected \OI, \CII\ and \SiII, while they did not cover the wavelength region for \FeII\ and \MgII. The SNR in the regions where \CIV\ and \SiIV\ fall was too low to detect the lines so they determined upper limits. The total column densities for the ions in common between their work and this one are in very good agreement.

\begin{table}
 \centering
    \caption{Voigt parameters obtained from the fit of the metal lines in the system at $z_{\rm abs}=5.7912$ in the spectrum of SDSS J0818+1722.}
    \label{tab:5.7912}
    \begin{tabular}{lccc}
        \hline
        Ion & $z_{\rm abs}$ & $\log (N_{\rm X}/{\rm cm^{-2}})$ & $b$ (\kms)\\
        \hline
        $\OI$ & $5.79004\pm 0.00013$ & $13.84\pm 0.06$ & $36\pm 2$\\
        $\CII$ & $5.78968\pm 0.00005$ & $13.71\pm 0.02$ & $36\pm 2$\\
        $\SiII$ & $5.78945\pm 0.00003$ & $12.89\pm 0.01$ & $36\pm 2$\\
        $\FeII$ & $5.78928\pm 0.00033$ & $12.31\pm 0.13$ & $36\pm 2$\\
        \hline
        $\OI$ & $5.79071\pm 0.00005$ & $13.62\pm 0.10$ & $7\pm 6$\\
        $\CII$ & $5.79073\pm 0.00004$ & $13.24\pm 0.06$ & $7\pm 6$\\
        $\SiII$ & $5.79072\pm 0.00004$ & $12.46\pm 0.04$ & $7\pm 6$\\
        $\FeII$ & $5.79041\pm 0.00021$ & $12.17\pm 0.17$ & $7\pm 6$\\
        \hline
        $\OI$ & $5.79175\pm 0.00002$ & $14.32\pm 0.02$ & $19.1\pm 0.8$\\
        $\CII$ & $5.79166\pm 0.00002$ & $13.84\pm 0.02$ & $19.1\pm 0.8$\\
        $\SiII$ & $5.79166\pm 0.00001$ & $13.04\pm 0.01$ & $19.1\pm 0.8$\\
        $\FeII$ & $5.79156\pm 0.00006$ & $12.74\pm 0.04$ & $19.1\pm 0.8$\\
        \hline
        \hline
        $\SiIV$ & $5.7895$ & $<12.58$ & $36$\\
        $\CIV$ & $5.78950\pm 0.00013$ & $13.38\pm 0.07$ & $36\pm 10$\\
        \hline
    \end{tabular}
\end{table}

\subsection*{C.9. SDSS J0818+1722 $z_{\rm abs}=5.8767$}
This system is an intervening DLA, with a separation from the QSO emission redshift of $\sim 5200$ \kms, and a velocity structure consisting of a single component. We detected the ionic transitions due to: \OI\ $1302$, \CII\ $1334$, \SiII\ $1260$, $1304$ and $1526$, \FeII\ $2344$, $2382$, $2586$ and $2600$, and \MgII\ at $2796$, as shown in Fig.~\ref{fig:5.8767}. The value of the parameters obtained from the fit of the low ionization lines are shown in Tab.~\ref{tab:5.8767}. We performed the fit by linking the Doppler parameters of all the ions.%, obtaining a value of $b=(15.0\pm 0.9)$ \kms.
We also derived an upper limit on \AlII\ $1670$.
%using equation (\ref{eq:s_n}) and (\ref{eq:upperlimit}) and fixing the Doppler parameter to $b=15.0$ \kms. 
The lines of \MgII, on the other hand, fall into the telluric band and are therefore strongly affected by the lines of the sky; only the stronger transition of the doublet could be observed.

For this system it was possible to fit the \HI\ \Lya\ absorption line. The obtained column density, $\log N_{\rm HI}=20.20\pm 0.10$.

High ionization lines were also detected in the system, which have a velocity structure consisting of a single component of \SiIV\ and \CIV\ (see Fig. \ref{fig:5.8767} and Tab.~\ref{tab:5.8767}). We performed the fit by linking the redshift and the $b$ parameter of the two ions.%, obtaining for the latter a value of $b=(36\pm 6)$ \kms.

Also this system was identified and analyzed by \citet{Becker2011} in the high-resolution spectrum obtained with Keck-HIRES. They detected \OI, \CII\ and \SiII, and determined upper limits on \CIV\ and \SiIV\ lines. The total column densities they obtained are in general good agreement with our estimates. Only in the case of \SiII\ our determination of the column density is  $3.6\,\sigma$ larger than their estimate. However, we base our fit on three transitions of the doublet so we are confident that the determination is correct.

\begin{table}
    \centering
    \caption{Voigt parameters obtained from the fit of the metal lines in the system at $z_{\rm abs}=5.8767$ in the spectrum of SDSS J0818+1722.}
    \label{tab:5.8767}
    \begin{tabular}{lccc}
        \hline
        Ion & $z_{\rm abs}$ & $\log (N_{\rm X}/{\rm cm^{-2}})$ & $b$ (\kms)\\
        \hline
        $\HI$ & $5.8767$ & $20.20\pm 0.10$ & $55$\\
        \hline
        \hline
        $\OI$ & $5.87672\pm 0.00002$ & $14.12\pm 0.02$ & $15.0\pm 0.9$\\
        $\CII$ & $5.87676\pm 0.00002$ & $13.77\pm 0.03$ & $15.0\pm 0.9$\\
        $\SiII$ & $5.87673\pm 0.00002$ & $12.96\pm 0.02$ & $15.0\pm 0.9$\\
        $\AlII$ & $<11.90$ & $15.0$\\
        $\FeII$ & $5.87663\pm 0.00012$ & $12.49\pm 0.08$ & $15.0\pm 0.9$\\
        $\MgII$ & $12.96\pm 0.09$ & $15.0\pm 0.9$\\
        \hline
        \hline
        $\SiIV$ & $5.87705\pm 0.00009$ & $12.51\pm 0.07$ & $36\pm 6$\\
        $\CIV$ & $5.87705\pm 0.00009$ & $13.28\pm 0.05$ & $36\pm 6$\\
        \hline
    \end{tabular}
\end{table}

\begin{figure}
\centering
    \includegraphics[width=\columnwidth]{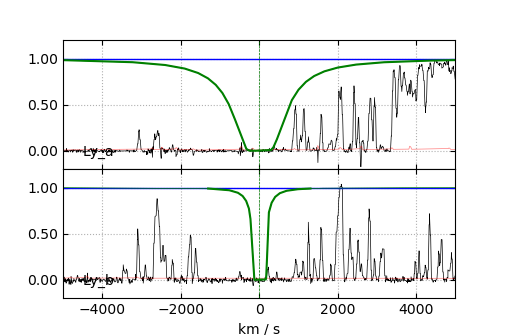}
    \caption{System at $z_{\rm abs}=5.8767$ in the spectrum of SDSS J0818+1722. Fit of the \HI\ \Lya\ and \Lyb\ absorption lines. }
    \label{fig:5.8767Lya}
\end{figure}

\begin{figure}
\centering
	\includegraphics[width=\columnwidth]{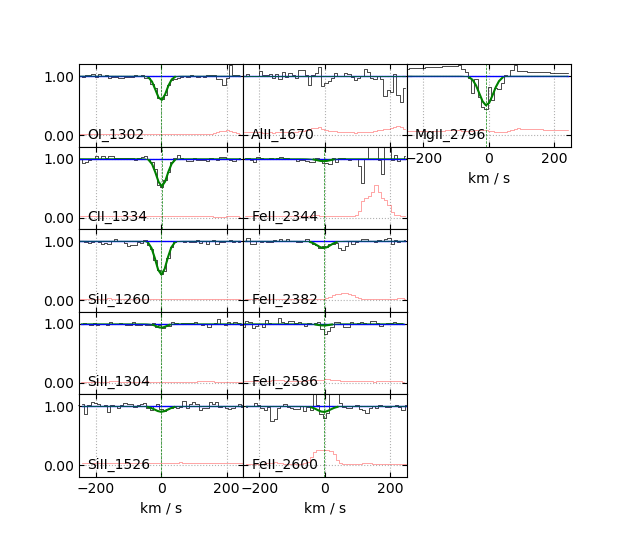}
	\includegraphics[width=\columnwidth]{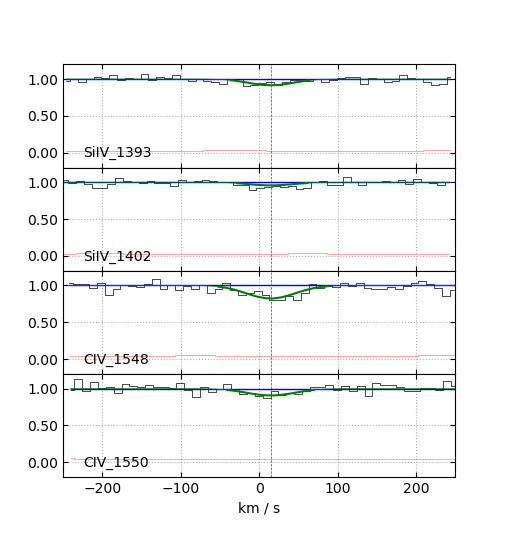}
    \caption{System at $z_{\rm abs}=5.8767$ in the spectrum of SDSS J0818+1722. Upper panel: fit of the detected low-ionization metal lines, and region where \AlII\ $1670$ absorption would fall. Lower panel: fit of the detected high-ionization lines.}
    \label{fig:5.8767}
\end{figure}

\subsection*{C.10. PSO J007+04 $z_{\rm abs}=5.9917$ PDLA}
This system is a single component PDLA, with a separation from the QSO emission redshift of $\sim 420$ \kms. We detected the ionic transitions due to: \OI\ $1302$, \CII\ $1334$, \SiII\ $1260$, $1304$ and $1526$, \FeII\ $2382$, as shown in Fig.~\ref{fig:5.9917}. The value of the parameters obtained from the fit of the low ionization lines are shown in the Tab.~\ref{tab:5.9917}. We performed the fit by linking the $b$ parameters of all the detected ions. 
%detected, obtaining a value of $b=(13\pm 2)$ \kms. We also performed the fit for \SiII\ at $1526$, also tying the redshift and column density to those of the other \SiII\ transitions. 
We also derived an upper limit on \AlII\ $1670$ and on \MgII\ $2796$.
%using equation (\ref{eq:s_n}) and (\ref{eq:upperlimit}) and fixing the Doppler parameter to $b=13$ \kms. 
Finally, since \MgII\ is located in a noisy region of the spectrum, the fit of this ion was performed by also linking the redshift with that of \SiII.

\begin{figure}
\centering
    \includegraphics[width=\columnwidth]{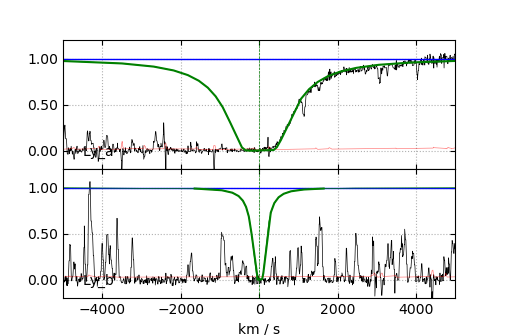}
    \caption{System at $z_{\rm abs}=5.9917$ in the spectrum of PSO J007+04. Fit of the \HI\ \Lya\ and \Lyb\ absorption lines. }
    \label{fig:5.9917Lya}
\end{figure}

\begin{figure}
\centering
	\includegraphics[width=\columnwidth]{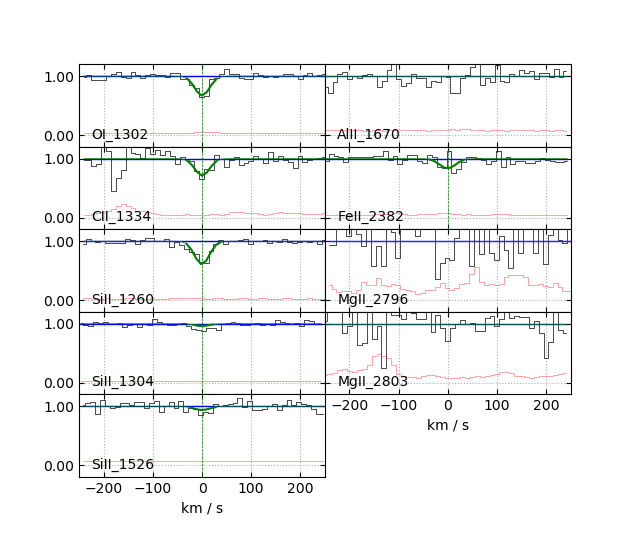}
	\includegraphics[width=\columnwidth]{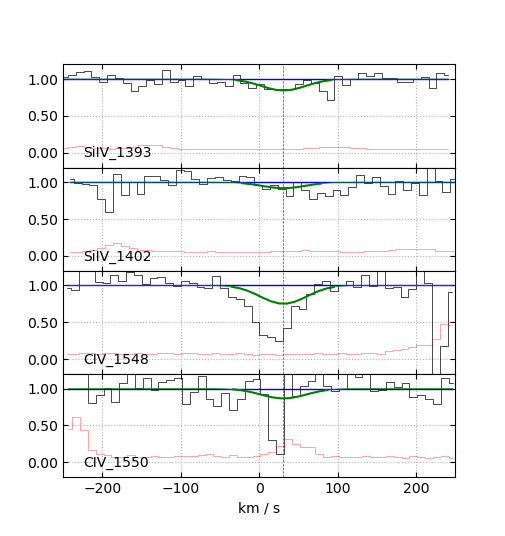}
    \caption{System at $z_{\rm abs}=5.9917$ in the spectrum of PSO J007+04. Upper panel: fit of the detected low-ionization metal lines, and region where the \AlII\ $1670$ and \MgII\ $2796$, $2803$ absorptions would fall. Lower panel: fit of the detected high-ionization lines. The strong line at the wavelength of \CIV\ 1548 is a \MgII\ 2803 at $z=2.861$. }
    \label{fig:5.9917}
\end{figure}

Since the system is a PDLA we have also performed the fit of the \HI\ \Lya\ absorption line (see Fig.~\ref{fig:5.9917Lya}, as described in Sect.~\ref{sec:lya}, obtaining a value of $\log N_{\rm HI}=20.40\pm 0.10$.

High ionization lines were also detected in the system, with a velocity structure characterized by a single component of both \SiIV\ and \CIV\ (see Fig. \ref{fig:5.9917} and Tab.~\ref{tab:5.9917}). As for the \CIV, the transition at $1548$ is blended with a line of MgII $2803$ at $z=2.8613$, while the transition at $1550$ is blended with a sky line. We then performed the fit by linking the redshift and the Doppler parameter of \SiIV\ and \CIV.
%; for the parameter $b$ we obtain a value $b=(35\pm 9)$ \kms.

\begin{table}
    \centering
    \caption{Voigt parameters obtained from the fit of the metal lines in the system at $z_{\rm abs}=5.9917$ in the spectrum of PSO J007+04.}
    \label{tab:5.9917}
    \begin{tabular}{lccc}
        \hline
        Ion & $z_{\rm abs}$ & $\log (N_{\rm X}/{\rm cm^{-2}})$ & $b$ (\kms)\\
        \hline
        $\HI$ & $5.9917$ & $20.40\pm 0.10$ & $20$\\
        \hline
        \hline
        $\OI$ & $5.99174\pm 0.00007$ & $13.97\pm 0.08$ & $13\pm 2$\\
        $\CII$ & $5.99174\pm 0.00008$ & $13.45\pm 0.07$ & $13\pm 2$\\
        $\SiII$ & $5.99173\pm 0.00004$ & $12.69\pm 0.04$ & $13\pm 2$\\
        $\AlII$ & $5.99173\pm 0.00004$ & $<12.10$ & $13$\\
        $\FeII$ & $5.99173\pm 0.00004$ & $12.57\pm 0.20$ & $13\pm 2$\\
        $\MgII$ & $5.99173\pm 0.00004$ & $<12.84$ & $13\pm 2$\\
        \hline
        \hline
        $\SiIV$ & $5.99246\pm 0.00015$ & $12.78\pm 0.10$ & $35\pm 9$\\
        $\CIV$ & $5.99246\pm 0.00015$ & $13.42\pm 0.14$ & $35\pm 9$\\
        \hline
    \end{tabular}
\end{table}

\subsection*{C.11. SDSS J2310+1855 $z_{\rm abs}=5.9388$ PDLA}
\begin{figure}
\centering
    \includegraphics[width=\columnwidth]{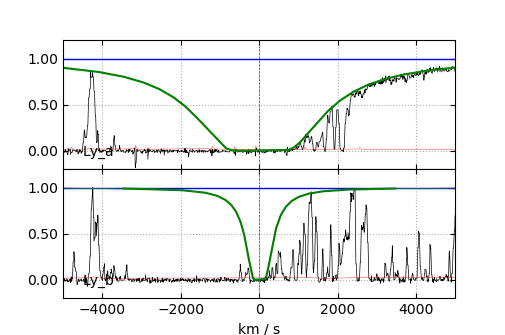}
	\includegraphics[width=\columnwidth]{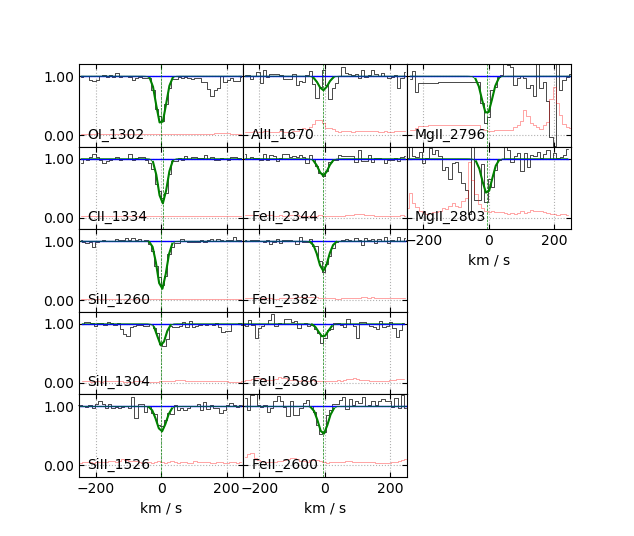}
	\includegraphics[width=\columnwidth]{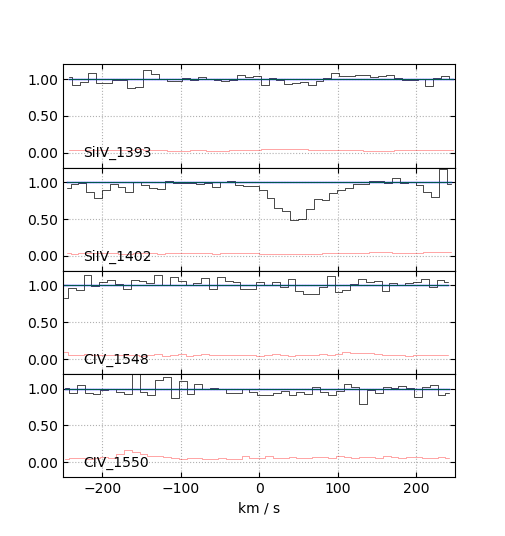}
    \caption{System at $z_{\rm abs}=5.9388$ in the spectrum of SDSS J2310+1855. Upper panel: fit of the \HI\ \Lya\ and \Lyb\ absorption lines. Middle panel: fit of the detected low-ionization metal lines. Lower panel: region where the high ionization lines of \SiIV\ and \CIV\ would fall.}
    \label{fig:5.9388}
\end{figure}

This system is a PDLA, with a separation from the QSO emission redshift of $\sim2760$ \kms, characterized by a single-component velocity structure. We detected the following ions: \OI\ $1302$, \CII\ $1334$, \SiII\ $1260$, $1304$ and $1526$, \AlII\ $1670$, \FeII\ $2344$, $2382$, $2586$, $2600$, and \MgII\ $2796$, $2803$, as shown in Fig.~\ref{fig:5.9388}. The value of the parameters obtained from the fit of the low ionization lines are shown in Tab.~\ref{tab:5.9388}. We performed the fit by linking the Doppler parameter of all ions, with the exception of \AlII\ and \MgII.%, obtaining a value of $b=(7.9\pm 0.5)$ \kms.
The \AlII\ line is superimposed on a sky line while that of \MgII\ is in a high-noise region, so we fit these ions by fixing the redshift to the same value obtained for \FeII\ and the $b$ parameter to $7.9$ \kms as for the other ions. The lines of \OI, \CII\ and \MgII\ are saturated, so we considered their column densities as lower limits.

We have also performed the fit of the \HI\ \Lya\ and \Lyb\ absorption lines of this PDLA (see Fig.~\ref{fig:5.9388}). The obtained column density, $\log N_{\rm HI}=21.00\pm 0.10$, is in agreement with the lower limit derived from eq,~\ref{eq:relOH}.

In the system, on the other hand, we did not detect high ionization lines (see Fig. \ref{fig:5.9388}), so we derived upper limits on \SiIV\ and \CIV. %The results obtained from the calculation are shown in the Table.

This system was already analyzed in \citet{Dodorico2018} using a lower SNR spectrum, the derived column densities are generally in agreement within errors with the old determinations.

\begin{table}
    \centering
    \caption{Voigt parameters obtained from the fit of the metal lines in the system at $z_{\rm abs}=5.9388$ in the spectrum of SDSS J2310+1855.}
    \label{tab:5.9388}
    \begin{tabular}{lccc}
        \hline
        Ion & $z_{\rm abs}$ & $\log (N_{\rm X}/{\rm cm^{-2}})$ & $b$ (\kms)\\
        \hline
        $\HI$ & $5.9388$ & $21.00\pm 0.10$ & $20$\\
        \hline
        \hline
        $\OI$ & $5.93886\pm 0.00001$ & $>15.27$ & $7.9\pm 0.5$\\
        $\CII$ & $5.93896\pm 0.00002$ & $>14.33$ & $7.9\pm 0.5$\\
        $\SiII$ & $5.93888\pm 0.00001$ & $13.72\pm 0.07$ & $7.9\pm 0.5$\\
        $\AlII$ & $5.93874$ & $12.17\pm 0.38$ & $7.9$\\
        $\FeII$ & $5.93874\pm 0.00004$ & $13.32\pm 0.07$ & $7.9\pm 0.5$\\
        $\MgII$ & $5.93874$ & $>13.49$ & $7.9$\\
        \hline
        \hline
        $\SiIV$ & $5.9388$ & $<12.31$ & $26$\\
        $\CIV$ & $5.9388$ & $<12.95$ & $26$\\
        \hline
    \end{tabular}
\end{table}

\subsection*{C.12. PSO J158-14 $z_{\rm abs}=5.8986$}
\begin{figure}
\centering
	\includegraphics[width=\columnwidth]{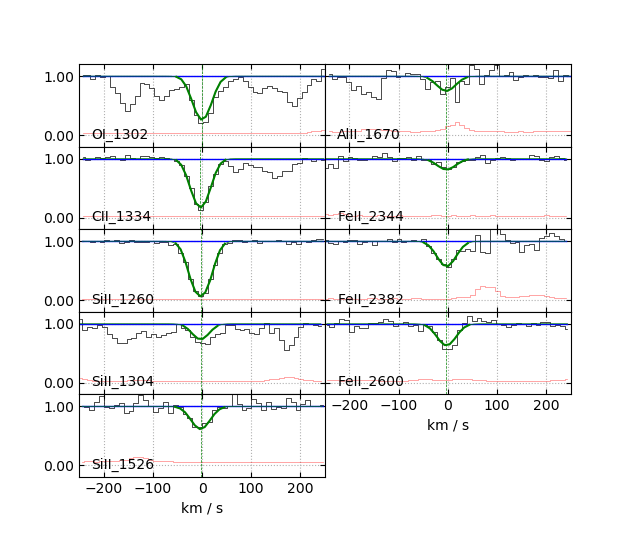}
	\includegraphics[width=\columnwidth]{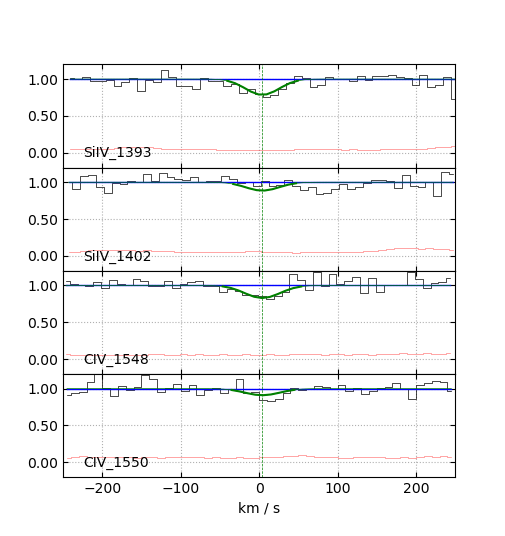}
    \caption{System at $z_{\rm abs}=5.8986$ in the spectrum of PSO J158-14. Upper panel: fit of the detected low-ionization metal lines. Lower panel: fit of the detected high-ionization lines.}
    \label{fig:5.8985}
\end{figure}

This system is an intervening DLA with a single-component velocity structure, separated from the QSO emission redshift by $\sim 7290$ \kms. The following ions were detected: \OI\ $1302$, \CII\ $1334$, \SiII\ $1260$, $1304$, $1526$, \AlII\ $1670$, \FeII\ $2344$, $2382$, and $2600$, as shown in Fig.~\ref{fig:5.8985}. The values of the parameters obtained from the fit of the low ionization lines are shown in Tab.~ \ref{tab:5.8985}. The \OI\ line is partially blended with a \CIV\ $1548$ line at $z=4.80$. We performed the fit by linking the Doppler parameters of all the ions.%, obtaining a value of $b=(17.3\pm 0.6)$ \kms.
The \AlII\ line is located in a noisy region of the spectrum, so we fitted this ion by also linking the redshift with that of \SiII. Finally, it was not possible to detect the \MgII\ 2796, 2803 lines as they fall in a strong telluric band which could not be corrected. 
%and it was not possible to correct the effect of the sky lines at the wavelength at which this absorption occurs.

A lower limit on \HI\ column density of $\log N_{\rm HI}>19.61$ is obtained from eq.~\ref{eq:relOH}. 

High ionization lines were also detected in the system, with a velocity structure consisting of a single component of \SiIV\ and a very weak \CIV\ (see Fig. \ref{fig:5.8985} and Tab.~ \ref{tab:5.8985}); we performed the fit by linking the redshift and the parameter $b$ of the two ions.%, obtaining for the latter a value of $b=(24\pm 5)$ \kms.

\begin{table}
    \centering
    \caption{Voigt parameters obtained from the fit of the metal lines in the system at $z_{\rm abs}=5.8986$ in the spectrum of PSO J158-14.}
    \label{tab:5.8985}
    \begin{tabular}{llccc}
        \hline
        Ion & $z_{\rm abs}$ & $\log (N_{\rm X}/{\rm cm^{-2}})$ & $b$ (\kms)\\
        \hline
        $\OI$ & $5.89865\pm 0.00004$ & $14.60\pm 0.07$ & $17.3\pm 0.6$\\
        $\CII$ & $5.89857\pm 0.00002$ & $14.31\pm 0.03$ & $17.3\pm 0.6$\\
        $\SiII$ & $5.89858\pm 0.00001$ & $13.65\pm 0.04$ & $17.3\pm 0.6$\\
        $\AlII$ & $5.89858\pm 0.00001$ & $12.26\pm 0.21$ & $17.3\pm 0.6$\\
        $\FeII$ & $5.89859\pm 0.00004$ & $13.14\pm 0.03$ & $17.3\pm 0.6$\\
        \hline
        \hline
        $\SiIV$ & $5.89874\pm 0.00007$ & $12.83\pm 0.06$ & $24\pm 5$\\
        $\CIV$ & $5.89874\pm 0.00007$ & $13.13\pm 0.09$ & $24\pm 5$\\
        \hline
    \end{tabular}
\end{table}

\subsection*{C.13. PSO J239-07 $z_{\rm abs}=5.9918$ PDLA}
\begin{figure}
\centering
        \includegraphics[width=\columnwidth]{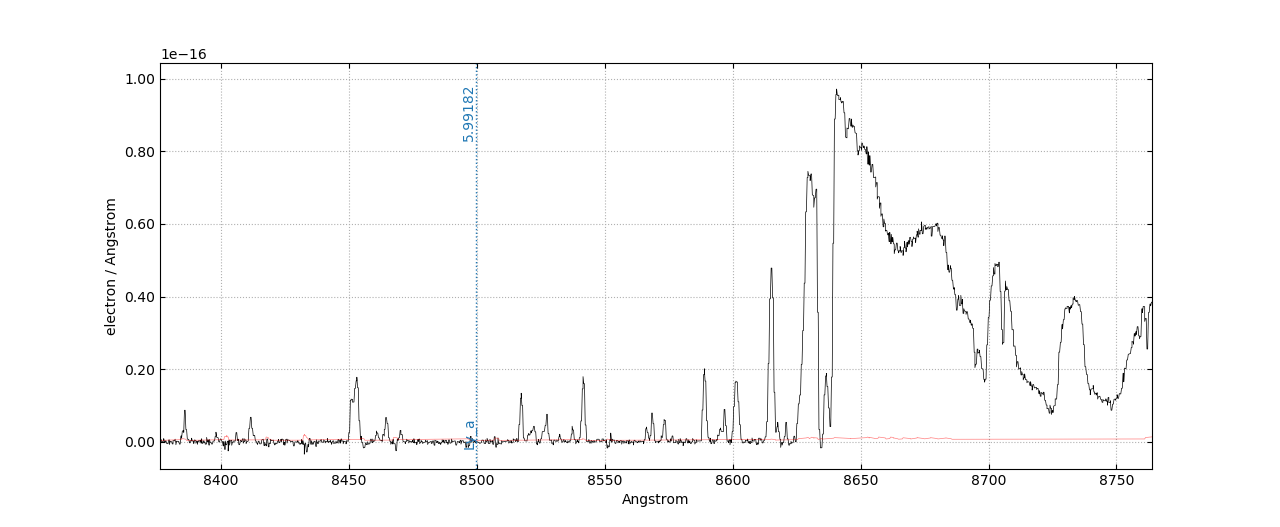}
	\includegraphics[width=\columnwidth]{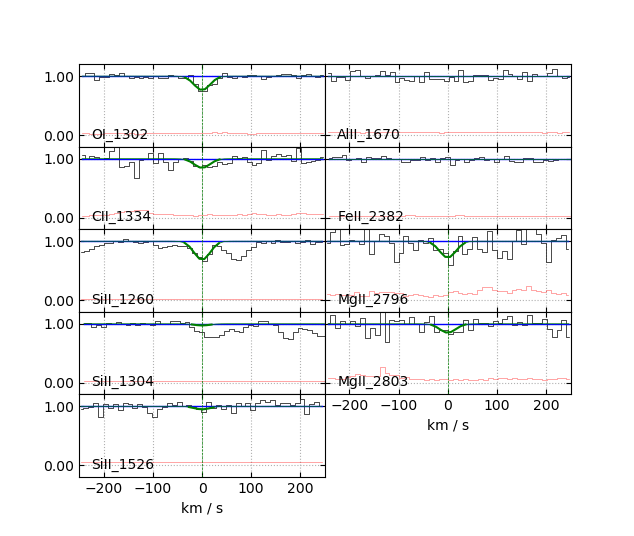}
	\includegraphics[width=\columnwidth]{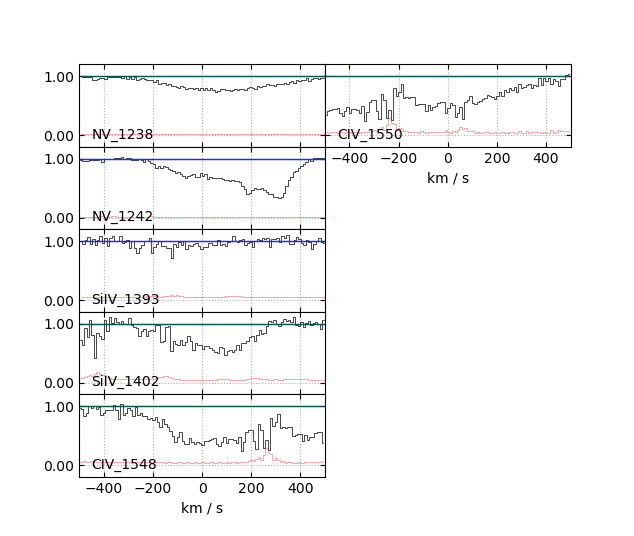}
    \caption{System at $z_{\rm abs}=5.9918$ in the spectrum of PSO J239-07. Upper panel: region of the spectrum where the HI \Lya\ would fall. Middle panel: fit of the detected low-ionization metal lines. Lower panel: region where the high-ionization lines would fall showing the broad absorptions due to \NV\ and \CIV.}
    \label{fig:5.9918}
\end{figure}

This system has a velocity separation of $\sim 5030$ \kms, at the assumed threshold between intervening and proximate DLAs. It has a single-component velocity structure. We detected the following ions: \OI\ $1302$, \CII\ $1334$, \SiII\ $1260$, $1304$, and \MgII\ $2796$, $2803$, as shown in Fig.~\ref{fig:5.9918}. The value of the parameters obtained from the fit of the low ionization lines are shown in Tab.~ \ref{tab:5.9918}. The \SiII\ line $1304$ is blended with \AlIII\ $1862$ at $z=3.8962$. We performed the fit by linking the redshift and the Doppler parameter of all detected ions.%, obtaining for the latter a value of $b=(16\pm 2)$ \kms.
We also derived an upper limit on \AlII\ $1670$ and \FeII\ $2382$. 
%, using equation (\ref{eq:s_n}) and (\ref{eq:upperlimit}) and fixing the Doppler parameter to $b=16$ \kms.

At the redshift of the low ionization lines, we also detected broad absorption lines due to \NV\ and \CIV\ (see Fig. \ref{fig:5.9918}) and we derived an upper limit on \SiIV. This QSO was classified as BAL by \citet{Bischetti2022}, it shows a complex structure of broad and narrow high-ionization associated systems spanning the velocity range between the QSO systemic redshift and $\sim5350$ \kms. 
%using equation (\ref{eq:s_n}) and (\ref{eq:upperlimit}). 
The present system could be part of the outflow traced by the high-ionization broad lines, or it could be a neutral intervening system that by chance falls at the same redshift of the outflow. From these observations it is not possible to distinguish between these two scenarios.  We decided to choose the latter hypothesis and keep this system in the sample; as a consequence, the detected broad \CIV\ absorption was not associated with the low ionization system and its column density is not considered in the subsequent analysis (e.g., for the computation of the \CII/\CIV\ ratio).  For the same reason, we did not carry out the fit of the \HI\ \Lya\ absorption, which, furthermore, would have very limited constraints from the spectrum (see Fig. \ref{fig:5.9918}). However, we estimated a lower limit for the \HI\ column density of $\log N_{\rm HI}>18.83$ from eq.~\ref{eq:relOH}).

\begin{table}
    \centering
    \caption{Voigt parameters obtained from the fit of the metal lines in the system at $z_{\rm abs}=5.9918$ in the spectrum of PSO J239-07.}
    \label{tab:5.9918}
    \begin{tabular}{lccc}
        \hline
        Ion & $z_{\rm abs}$ & $\log (N_{\rm X}/{\rm cm^{-2}})$ & $b$ (\kms)\\
        \hline
        $\OI$ & $5.99182\pm 0.00002$ & $13.82\pm 0.06$ & $16\pm 2$\\
        $\CII$ & $5.99182\pm 0.00002$ & $13.18\pm 0.11$ & $16\pm 2$\\
        $\SiII$ & $5.99182\pm 0.00002$ & $12.60\pm 0.03$ & $16\pm 2$\\
        $\AlII$ & $5.99182\pm 0.00002$ & $<11.75$ & $16$\\
        $\FeII$ & $5.99182\pm 0.00002$ & $<12.14$ & $16$\\
        $\MgII$ & $5.99182\pm 0.00002$ & $12.47\pm 0.11$ & $16\pm 2$\\
        \hline
        \hline
        $\SiIV$ & $5.9918$ & $<12.33$ & $26$\\
        \hline
    \end{tabular}
\end{table}

\subsection*{C.14. ULAS J1319+0950 $z_{\rm abs}=6.0172$ PDLA}
This system is a PDLA, with a separation from the QSO emission redshift of $\sim 4980$ \kms. 
We detected single-component velocity features due to the following ions: \OI\ $1302$, \CII\ $1334$, \SiII\ $1260$, $1304$, $1526$, \AlII\ $1670$, \FeII\ $2382$, and \MgII\ $2796$, $2803$, as shown in Fig.~\ref{fig:6.0172}. The value of the parameters obtained from the fit of the low ionization lines are shown in Tab.~\ref{tab:6.0172}. We performed the fit by linking the Doppler parameter of all detected ions, obtaining a value of $b=(7\pm 2)$ \kms.

In the system, on the other hand, we did not detect high ionization lines (see Fig. \ref{fig:6.0172}). We derived upper limits on \SiIV\ and \CIV.

Although the system is a PDLA, we could not perform the fit of the \HI\ \Lya\ absorption line, because the spectrum does not show a clear damping wing and the fits is not constrained (see Fig. \ref{fig:6.0172}). However, we estimated a lower limit for the \HI\ column density of $\log N_{\rm HI}>18.94$ from eq.~\ref{eq:relOH}. 

\citet{Simcoe2011} and \citet{Dodorico2013} used a FIRE and the same X-SHOOTER spectrum used here, respectively, to study the absorption systems along the line of sight to this QSO but they did not report the detection of this system.

\begin{table}
    \centering
    \caption{Voigt parameters obtained from the fit of the metal lines in the system at $z_{\rm abs}=6.0172$ in the spectrum of ULAS J1319+0950.}
    \label{tab:6.0172}
    \begin{tabular}{lccc}
        \hline
        Ion & $z_{\rm abs}$ & $\log (N_{\rm X}/{\rm cm^{-2}})$ & $b$ (\kms)\\
        \hline
        $\OI$ & $6.01721\pm 0.00002$ & $13.93\pm 0.06$ & $7\pm 2$\\
        $\CII$ & $6.01717\pm 0.00008$ & $13.17\pm 0.11$ & $7\pm 2$\\
        $\SiII$ & $6.01722\pm 0.00003$ & $12.50\pm 0.04$ & $7\pm 2$\\
        $\AlII$ & $6.01720\pm 0.00023$ & $11.84\pm 0.23$ & $7\pm 2$\\
        $\FeII$ & $6.01710\pm 0.00013$ & $12.45\pm 0.15$ & $7\pm 2$\\
        $\MgII$ & $6.01717\pm 0.00009$ & $12.61\pm 0.16$ & $7\pm 2$\\
        \hline
        \hline
        $\SiIV$ & $6.0172$ & $<12.34$ & $26$\\
        $\CIV$ & $6.0172$ & $<12.89$ & $26$\\
        \hline
    \end{tabular}
\end{table}

\begin{figure}
\centering
    \includegraphics[width=\columnwidth]{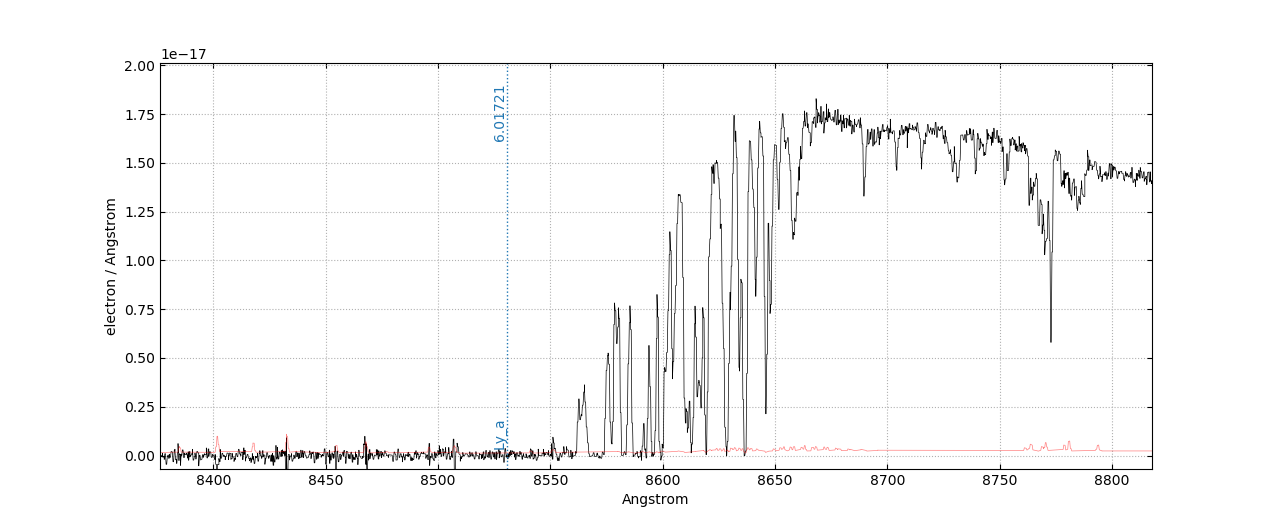}
    \caption{System at $z_{\rm abs}=6.0172$ in the spectrum of ULAS J1319+0950. Region of the spectrum where the HI \Lya\ falls. }
    \label{fig:6.0172Lya}
\end{figure}

\begin{figure}
\centering
	\includegraphics[width=\columnwidth]{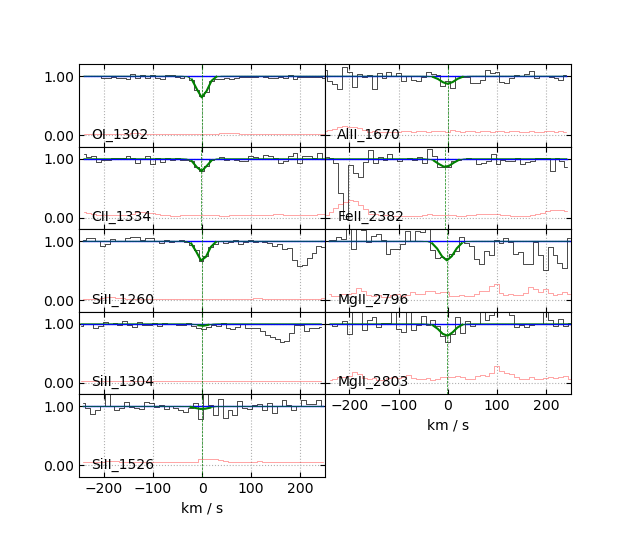}
	\includegraphics[width=\columnwidth]{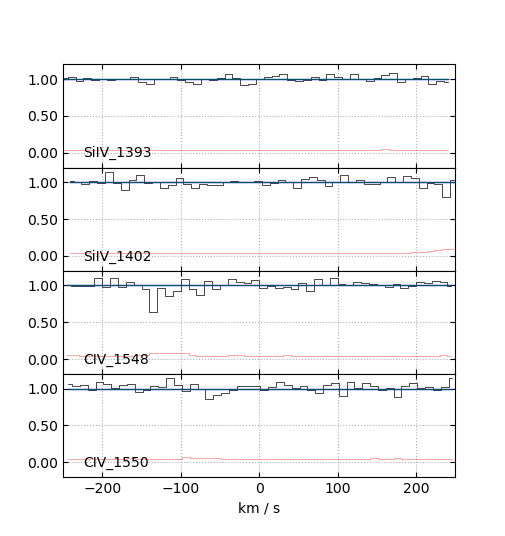}
    \caption{System at $z_{\rm abs}=6.0172$ in the spectrum of ULAS J1319+0950. Upper panel: fit of the detected low-ionization metal lines. Lower panel: region where the high ionization lines of \SiIV\ and \CIV\ would fall.}
    \label{fig:6.0172}
\end{figure}

\subsection*{C.15. PSO J060+24 $z_{\rm abs}=5.6993$}
\begin{figure}
\centering
	\includegraphics[width=\columnwidth]{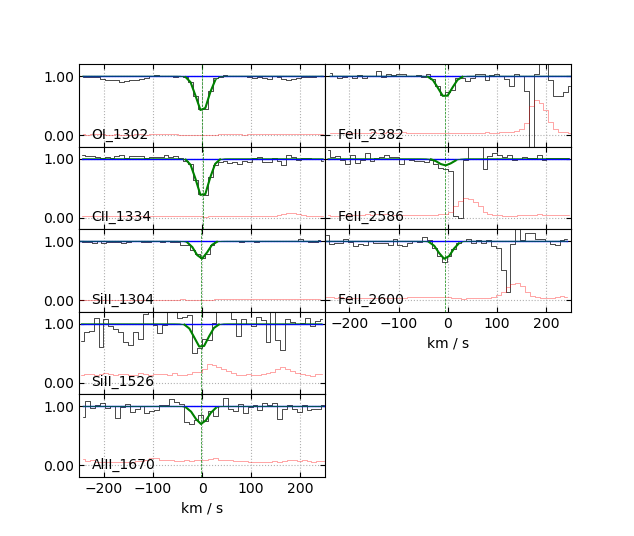}
	\includegraphics[width=\columnwidth]{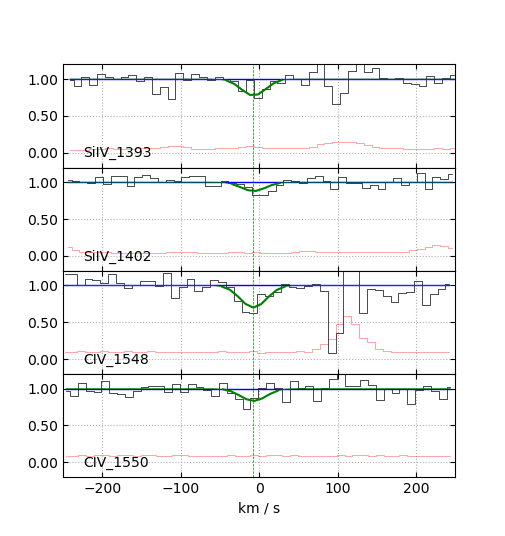}
    \caption{System at $z_{\rm abs}=5.6993$ in the spectrum of PSO J060+24. Upper panel: fit of the detected low-ionization metal lines. Lower panel: fit of the detected high-ionization lines.}
    \label{fig:5.6993}
\end{figure}

This system is an intervening DLA separated from the QSO emission redshift by $\sim 20320$ \kms, and characterized by a simple velocity structure consisting of a single component. We detected the following ions: \OI\ $1302$, \CII\ $1334$, \SiII\ $1304$, $1526$, \AlII\ $1670$, and \FeII\ $2382$, $2586$, $2600$, as shown in Fig.~\ref{fig:5.6993}. The value of the parameters obtained from the fit of the low ionization lines are shown in Tab.~ \ref{tab:5.6993}. We performed the fit by linking the Doppler parameters of all the ions.%, obtaining a value of $b=(7.2\pm 0.9)$ \kms.
The \AlII\ line falls in a noisy spectral region so we fitted it by also linking the redshift with that of \SiII. Finally, it was not possible to detect the \MgII\ doublet as it is lost  among the sky lines in a very strong telluric band. 

High ionization lines were also detected in the system, which have a velocity structure consisting of a single component of \SiIV\ and \CIV\ (see Fig. \ref{fig:5.6993} and Tab.~ \ref{tab:5.6993}); we performed the fit by linking the redshift and the parameter $b$ of the two ions.%, obtaining for the latter a value of $b=(15\pm 5)$ \kms.

The lower limit on \HI\ column density from eq.~\ref{eq:relOH} is $\log N_{\rm HI}>19.51$.  
\begin{table}
    \centering
    \caption{Voigt parameters obtained from the fit of the metal lines in the system at $z_{\rm abs}=5.6993$ in the spectrum of PSO J060+24.}
    \label{tab:5.6993}
    \begin{tabular}{lccc}
        \hline
        Ion & $z_{\rm abs}$ & $\log (N_{\rm X}/{\rm cm^{-2}})$ & $b$ (\kms)\\
        \hline
        $\OI$ & $5.69934\pm 0.00001$ & $14.50\pm 0.09$ & $7.2\pm 0.9$\\
        $\CII$ & $5.69936\pm 0.00002$ & $14.22\pm 0.18$ & $7.2\pm 0.9$\\
        $\SiII$ & $5.69931\pm 0.00002$ & $13.64\pm 0.03$ & $7.2\pm 0.9$\\
        $\AlII$ & $5.69931\pm 0.00002$ & $12.30\pm 0.23$ & $7.2\pm 0.9$\\
        $\FeII$ & $5.69922\pm 0.00005$ & $12.96\pm 0.07$ & $7.2\pm 0.9$\\
        \hline
        \hline
        $\SiIV$ & $5.69918\pm 0.00006$ & $12.72\pm 0.08$ & $15\pm 5$\\
        $\CIV$ & $5.69918\pm 0.00006$ & $13.29\pm 0.09$ & $15\pm 5$\\
        \hline
    \end{tabular}
\end{table}

\subsection*{C.16. PSO J065-26 $z_{\rm abs}=5.8677$}
\begin{figure}
\centering
	\includegraphics[width=\columnwidth]{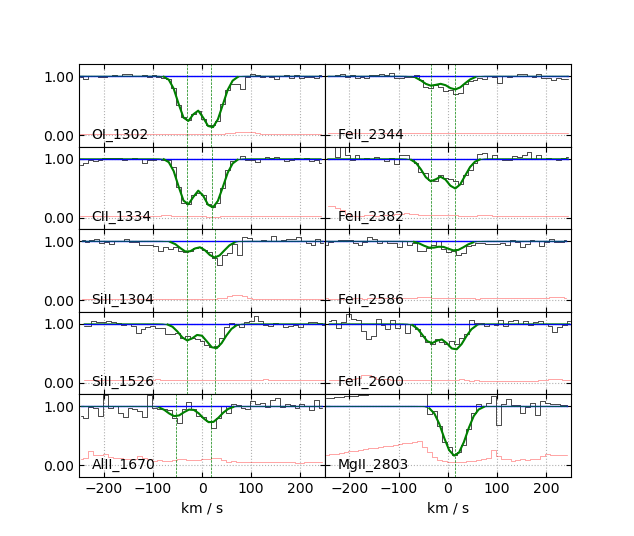}
	\includegraphics[width=\columnwidth]{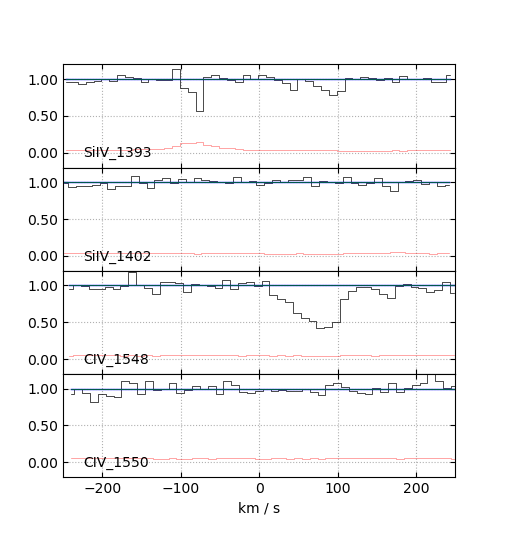}
    \caption{System at $z_{\rm abs}=5.8677$ in the spectrum of PSO J065-26. Upper panel: fit of the detected low-ionization metal lines. Lower panel: region where the high ionization lines of \SiIV\ and \CIV\ would fall.}
    \label{fig:5.8679}
\end{figure}

This intervening DLA has a velocity structure characterized by two components, with an average separation from the QSO emission redshift of $\sim 13620$ \kms. We detected the following ions: \OI\ $1302$, \CII\  $1334$, \SiII\ $1304$, $1526$, \AlII\ $1670$, \FeII\ $2344$, $2382$, $2586$, $2600$, and \MgII\ $2803$, as shown in Fig.~\ref{fig:5.8679}. The value of the parameters obtained from the fit of the low ionization lines are shown in Tab.~\ref{tab:5.8679}. We linked the $b$ parameters of all the ions in each of the two components, except the \MgII. 
%For the first component we got $b=(15\pm 1)$ \kms, while for the second $b=(19\pm 1)$ \kms. 
The lines of \MgII\ fall into the telluric band and are therefore strongly affected by the lines of the sky; for this reason we could detect only one component of the transition at $2803$.

\CIV\ and \SiIV\ doublets are not observed at the redshift of this system (see Fig. \ref{fig:5.8679}), therefore we derived upper limits for their column densities. 
%In the system, on the other hand, we did not detect high ionization lines. We derived an upper limit on \SiIV\ and \CIV\, using equation (\ref{eq:s_n}) and (\ref{eq:upperlimit}).% The results obtained from the calculation are shown in the Table.

Finally, we used eq.~\ref{eq:relOH} to estimate a lower limit on the \HI\ column density of $\log N_{\rm HI}>20.02$.

\begin{table}
    \centering
    \caption{Voigt parameters obtained from the fit of the metal lines in the system at $z_{\rm abs}=5.8677$ in the spectrum of PSO J065-26.}
    \label{tab:5.8679}
    \begin{tabular}{lccc}
        \hline
        Ion & $z_{\rm abs}$ & $\log (N_{\rm X}/{\rm cm^{-2}})$ & $b$ (\kms)\\
        \hline
        $\OI$ & $5.86707\pm 0.00002$ & $14.59\pm 0.03$ & $15\pm 1$\\
        $\CII$ & $5.86708\pm 0.00002$ & $14.18\pm 0.04$ & $15\pm 1$\\
        $\SiII$ & $5.86709\pm 0.00006$ & $13.41\pm 0.05$ & $15\pm 1$\\
        $\AlII$ & $5.86655\pm 0.00027$ & $11.99\pm 0.26$ & $15\pm 1$\\
        $\FeII$ & $5.86698\pm 0.00005$ & $12.99\pm 0.05$ & $15\pm 1$\\
        \hline
        $\OI$ & $5.86821\pm 0.00002$ & $14.80\pm 0.03$ & $19\pm 1$\\
        $\CII$ & $5.86824\pm 0.00002$ & $14.28\pm 0.03$ & $19\pm 1$\\
        $\SiII$ & $5.86838\pm 0.00004$ & $13.68\pm 0.04$ & $19\pm 1$\\
        $\AlII$ & $5.86818\pm 0.00020$ & $12.29\pm 0.19$ & $19\pm 1$\\
        $\FeII$ & $5.86812\pm 0.00004$ & $13.22\pm 0.03$ & $19\pm 1$\\
        $\MgII$ & $5.86811\pm 0.00002$ & $13.64\pm 0.05$ & $20\pm 2$\\
        \hline
        \hline
        $\SiIV$ & $5.8677$ & $<12.48$ & $26$\\
        $\CIV$ & $5.8677$ & $<12.83$ & $26$\\
        \hline
    \end{tabular}
\end{table}

\subsection*{C.17. PSO J065-26 $z_{\rm abs}=6.1208$ PDLA}
This system is a PDLA with a separation from the QSO emission redshift of about $\sim 2780$ \kms, and a velocity structure characterized by two components. 
We detected the following ions: \SII\ $1250$, $1253$, $1259$, \OI\ $1302$, \CII\ $1334$, \SiII\ $1260$, $1304$, $1526$, \AlII\ $1670$, \FeII\ $2344$, $2374$, $2382$, and \MgII\ $2796$, $2803$, as shown in Fig.~\ref{fig:6.1208}. The value of the parameters obtained from the fit of the low ionization lines are shown in Tab.~ \ref{tab:6.1208}. We linked the Doppler parameters of all the ions in each component. 
%, obtaining: $b=(25.6\pm 0.8)$ \kms\ for the first component, $b=(26\pm 2)$ \kms\ for the second, $b=(15\pm 1)$ \kms\ for the third, $b=(15.5\pm 0.6)$ \kms\ for the fourth and $b=(22\pm 2)$ \kms\ for the fifth. 
Oxygen, carbon and magnesium are saturated, therefore their column densities were considered as lower limits.

From eq.~\ref{eq:relOH}, we derived a lower limit on the \HI\ column density of $\log N_{\rm HI}>20.42$. Since this system is a PDLA, we performed also the fit of the \HI\ \Lya\ absorption line which however comprises both this system and the next one at $z=6.1263$, which are separated by less than $250$ \kms. Furthermore, it was not possible to obtain an \HI\ column density larger than $19.9$ independently of the adopted continuum. This suggests that these systems are affected by partial coverage, and therefore we have not considered the value obtained from this fit.

\CIV\ and \SiIV\ doublets are not observed at the redshift of this system (see Fig. \ref{fig:6.1208}), therefore we derived upper limits for their column densities.

\begin{table}
    \centering
    \caption{Voigt parameters obtained from the fit of the metal lines in the system at $z_{\rm abs}=6.1208$ in the spectrum of PSO J065-26.}
    \label{tab:6.1208}
    \begin{tabular}{lccc}
        \hline
        Ion & $z_{\rm abs}$ & $\log (N_{\rm X}/{\rm cm^{-2}})$ & $b$ (\kms)\\
        \hline
        $\SII$ & $6.12074\pm 0.00006$ & $14.35\pm 0.04$ & $25.6\pm 0.8$\\
        $\OI$ & $6.12054\pm 0.00003$ & $>15.28$ & $25.6\pm 0.8$\\
        $\CII$ & $6.12069\pm 0.00013$ & $>15.19$ & $25.6\pm 0.8$\\
        $\SiII$ & $6.12063\pm 0.00002$ & $14.47\pm 0.02$ & $25.6\pm 0.8$\\
        $\AlII$ & $6.12063\pm 0.00004$ & $13.13\pm 0.04$ & $25.6\pm 0.8$\\
        $\FeII$ & $6.12052\pm 0.00003$ & $13.87\pm 0.03$ & $25.6\pm 0.8$\\
        $\MgII$ & $6.12043\pm 0.00007$ & $>14.35$ & $25.6\pm 0.8$\\
        \hline
        $\SII$ & $6.12225\pm 0.00035$ & $13.60\pm 0.19$ & $26\pm 2$\\
        $\OI$ & $6.12183\pm 0.00005$ & $14.82\pm 0.04$ & $26\pm 2$\\
        $\CII$ & $6.12214\pm 0.00014$ & $14.26\pm 0.17$ & $26\pm 2$\\
        $\SiII$ & $6.12197\pm 0.00005$ & $13.73\pm 0.05$ & $26\pm 2$\\
        $\AlII$ & $6.12206\pm 0.00012$ & $12.42\pm 0.08$ & $26\pm 2$\\
        $\FeII$ & $6.12180\pm 0.00006$ & $13.29\pm 0.04$ & $26\pm 2$\\
        $\MgII$ & $6.12181\pm 0.00008$ & $13.63\pm 0.10$ & $26\pm 2$\\
        \hline
        \hline
        $\SiIV$ & $6.1208$ & $<12.73$ & $26$\\
        $\CIV$ & $6.1208$ & $<12.99$ & $26$\\
        \hline
    \end{tabular}
\end{table}

\begin{figure}
\centering
    \includegraphics[width=\columnwidth]{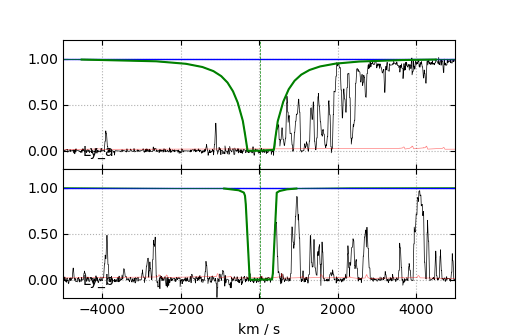}
    \caption{Fit of the \HI\ \Lya\ and \Lyb\ absorption lines corresponding to the systems at $z_{\rm abs}=6.1208$ and $6.1263$ in the spectrum of PSO J065-26.}
    \label{fig:6.1208Lya}
\end{figure}

\begin{figure}
\centering
	\includegraphics[width=\columnwidth]{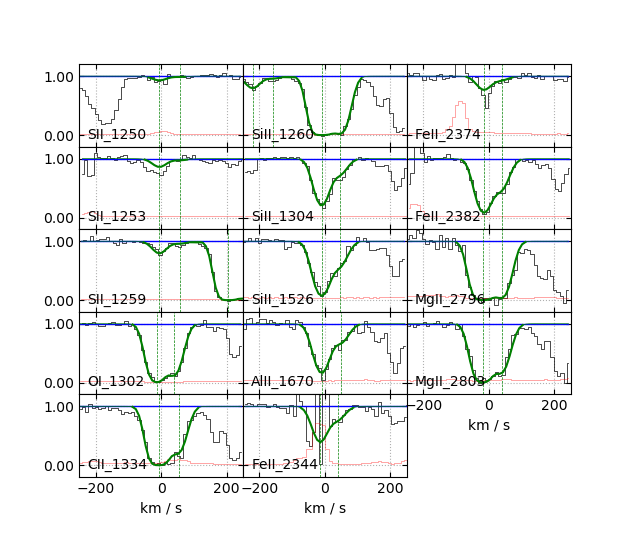}
	\includegraphics[width=\columnwidth]{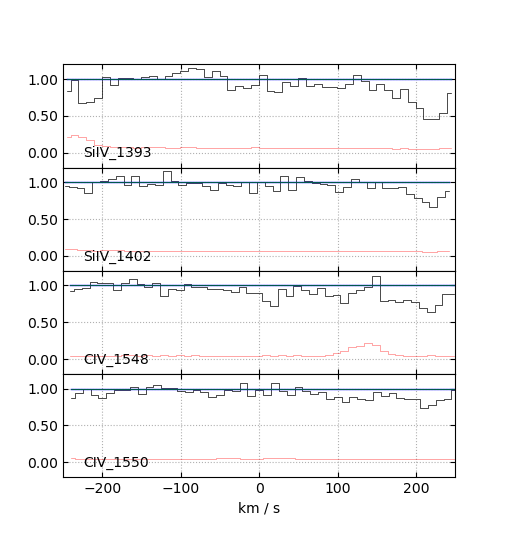}
    \caption{System at $z_{\rm abs}=6.1208$ in the spectrum of PSO J065-26. Upper panel: fit of the \HI\ \Lya\ and \Lyb\ absorption lines together with the $z=6.1263$ system. Middle panel: fit of the detected low-ionization metal lines. Lower panel: region where the high ionization lines of \SiIV\ and \CIV\ would fall.}
    \label{fig:6.1208}
\end{figure}

\subsection*{C.18. PSO J065-26 $z_{\rm abs}=6.1263$ PDLA}
\begin{figure}
\centering
	\includegraphics[width=\columnwidth]{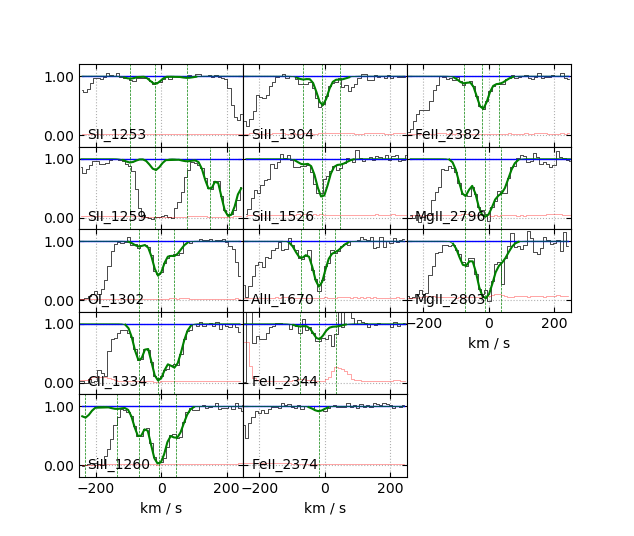}
	\includegraphics[width=\columnwidth]{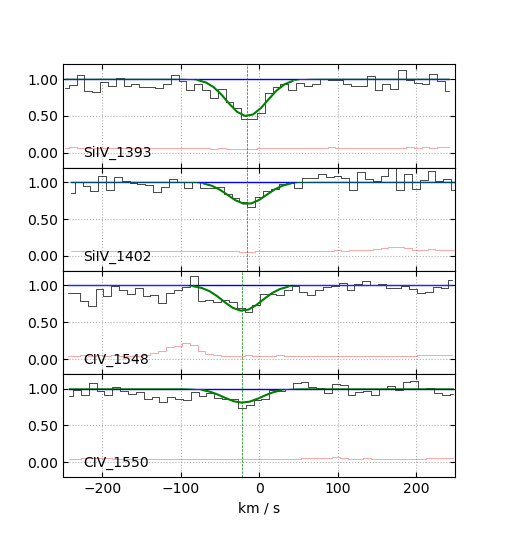}
    \caption{System at $z_{\rm abs}=6.1263$ in the spectrum of PSO J065-26. Upper panel: fit of the detected low-ionization metal lines. Lower panel: fit of the detected high-ionization lines.}
    \label{fig:6.1263}
\end{figure}

This system is PDLA with a separation from the QSO emission redshift of about $\sim 2540$ \kms, and a velocity structure characterized by three components. It is separated by only $\sim250$\kms\ from the system described in the previous section. Nonetheless, we decided to describe the two systems separately, in particular, because the one at larger velocities from the QSO did not show associated high-ionization absorption lines, while this one does. 
We detected the following ions: \SII\ $1253$, $1259$, \OI\ $1302$, \CII\ $1334$, \SiII\ $1260$, $1304$, $1526$, \AlII\ $1670$, \FeII\ $2344$, $2374$, $2382$, and \MgII\ $2796$, $2803$, as shown in Fig.~\ref{fig:6.1263}. The value of the parameters obtained from the fit of the low ionization lines are shown in Tab.~ \ref{tab:6.1263}. We linked the Doppler parameters of all the ions in each component. 
%, obtaining: $b=(25.6\pm 0.8)$ \kms\ for the first component, $b=(26\pm 2)$ \kms\ for the second, $b=(15\pm 1)$ \kms\ for the third, $b=(15.5\pm 0.6)$ \kms\ for the fourth and $b=(22\pm 2)$ \kms\ for the fifth. 
Carbon and magnesium are saturated, therefore their column densities were considered as lower limit.

Furthermore, we derived a lower limit on the \HI\ column density of $\log N_{\rm HI}>19.52$ from eq.~\ref{eq:relOH}, and we performed the fit of the \HI\ \Lya\ absorption line as described in the previous section. 

High ionization lines were also detected in the system, which have a velocity structure consisting of a single component of \SiIV\ and \CIV\ (see Fig. \ref{fig:6.1263} and Tab.~ \ref{tab:6.1263}); we performed the fit by linking the redshift and the parameter $b$ of the two ions.%, obtaining for the latter a value of $b=(15\pm 5)$ \kms.

\begin{table}
    \centering
    \caption{Voigt parameters obtained from the fit of the metal lines in the system at $z_{\rm abs}=6.1263$ in the spectrum of PSO J065-26.}
    \label{tab:6.1263}
    \begin{tabular}{lccc}
        \hline
        Ion & $z_{\rm abs}$ & $\log (N_{\rm X}/{\rm cm^{-2}})$ & $b$ (\kms)\\
        \hline
        $\SII$ & $6.12415\pm 0.00019$ & $13.67\pm 0.17$ & $15\pm 1$\\
        $\OI$ & $6.12478\pm 0.00015$ & $13.33\pm 0.13$ & $15\pm 1$\\
        $\CII$ & $6.12482\pm 0.00003$ & $13.96\pm 0.04$ & $15\pm 1$\\
        $\SiII$ & $6.12481\pm 0.00002$ & $12.88\pm 0.02$ & $15\pm 1$\\
        $\AlII$ & $6.12465\pm 0.00009$ & $12.26\pm 0.07$ & $15\pm 1$\\
        $\FeII$ & $6.12458\pm 0.00012$ & $12.42\pm 0.10$ & $15\pm 1$\\
        $\MgII$ & $6.12465\pm 0.00004$ & $13.02\pm 0.05$ & $15\pm 1$\\
        \hline
        $\SII$ & $6.12597\pm 0.00008$ & $14.18\pm 0.08$ & $15.5\pm 0.6$\\
        $\OI$ & $6.12620\pm 0.00003$ & $14.35\pm 0.03$ & $15.5\pm 0.6$\\
        $\CII$ & $6.12619\pm 0.00003$ & $>14.76$ & $15.5\pm 0.6$\\
        $\SiII$ & $6.12620\pm 0.00001$ & $13.97\pm 0.03$ & $15.5\pm 0.6$\\
        $\AlII$ & $6.12597\pm 0.00004$ & $12.99\pm 0.06$ & $15.5\pm 0.6$\\
        $\FeII$ & $6.12594\pm 0.00003$ & $13.26\pm 0.03$ & $15.5\pm 0.6$\\
        $\MgII$ & $6.12613\pm 0.00004$ & $>14.31$ & $15.5\pm 0.6$\\
        \hline
        $\SII$ & $6.12829\pm 0.00034$ & $13.59\pm 0.26$ & $22\pm 2$\\
        $\OI$ & $6.12733\pm 0.00008$ & $13.91\pm 0.06$ & $22\pm 2$\\
        $\CII$ & $6.12735\pm 0.00004$ & $14.16\pm 0.03$ & $22\pm 2$\\
        $\SiII$ & $6.12750\pm 0.00004$ & $12.97\pm 0.03$ & $22\pm 2$\\
        $\AlII$ & $6.12712\pm 0.00021$ & $12.08\pm 0.13$ & $22\pm 2$\\
        $\FeII$ & $6.12716\pm 0.00015$ & $12.51\pm 0.09$ & $22\pm 2$\\
        $\MgII$ & $6.12730\pm 0.00008$ & $12.96\pm 0.07$ & $22\pm 2$\\
        \hline
        \hline
        $\SiIV$ & $6.12602\pm 0.00004$ & $13.33\pm 0.03$ & $28\pm 2$\\
        $\CIV$ & $6.12588\pm 0.00005$ & $13.50\pm 0.03$ & $28\pm 2$\\
        \hline
    \end{tabular}
\end{table}

\subsection*{C.19 SDSS J0100+2802 $z_{\rm abs}=5.7974$}
\begin{figure}
\centering
	\includegraphics[width=\columnwidth]{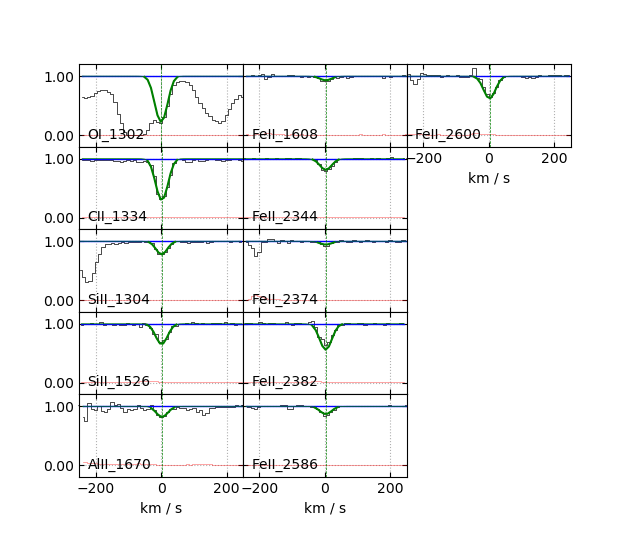}
	\includegraphics[width=\columnwidth]{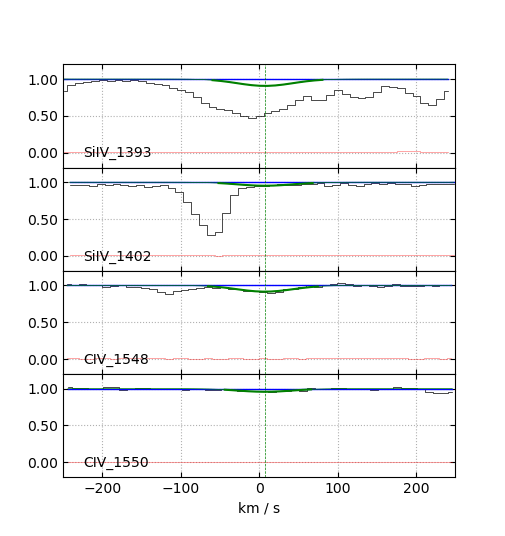}
    \caption{System at $z_{\rm abs}=5.7974$ in the spectrum of SDSS J0100+2802. Upper panel: fit of the detected low-ionization metal lines. Lower panel: fit of the detected high-ionization lines.}
    \label{fig:5.7974}
\end{figure}

This literature QSO has the highest SNR spectrum of the sample and is the only one which shows four \OI\ absorption systems. The system described here is an intervening DLA at $\sim 22440$ \kms from the QSO emission redshift, characterized by a simple velocity structure consisting of a single component. We detected the following ions: \OI\ $1302$, \CII\ $1334$, \SiII\ $1304$, $1526$, \AlII\ $1670$, and \FeII\ $1608$,$2344$,$2374$,$2382$, $2586$, $2600$, as shown in Fig.~\ref{fig:5.7974}. The values of the parameters obtained from the fit of the low ionization lines are shown in Tab.~ \ref{tab:5.7974}. The \OI\ line is blended with a \Lya\ line at $z=6.2792$, and we performed the fit by linking the Doppler parameters of all the ions.

High ionization lines were also detected in the system, which have a velocity structure consisting of a single component of \SiIV\ and \CIV\ (see Fig. \ref{fig:5.7974} and Tab.~ \ref{tab:5.7974}). We performed the fit by linking the redshift and the parameter $b$ of the two ions.

The lower limit on the \HI\ column density obtained from eq.~\ref{eq:relOH} is $\log N_{\rm HI}>19.64$. 

\begin{table}
    \centering
    \caption{Voigt parameters obtained from the fit of the metal lines in the system at $z_{\rm abs}=5.7974$ in the spectrum of SDSS J0100+2802.}
    \label{tab:5.7974}
    \begin{tabular}{lccc}
        \hline
        Ion & $z_{\rm abs}$ & $\log (N_{\rm X}/{\rm cm^{-2}})$ & $b$ (\kms)\\
        \hline
        $\OI$ & $5.79743\pm 0.00002$ & $14.63\pm 0.03$ & $17.5\pm 0.6$\\
        $\CII$ & $5.79748\pm 0.00001$ & $14.09\pm 0.02$ & $17.5\pm 0.6$\\
        $\SiII$ & $5.79746\pm 0.00003$ & $13.56\pm 0.03$ & $17.5\pm 0.6$\\
        $\AlII$ & $5.79749\pm 0.00009$ & $12.06\pm 0.08$ & $17.5\pm 0.6$\\
        $\FeII$ & $5.79749\pm 0.00002$ & $13.12\pm 0.02$ & $17.5\pm 0.6$\\
        \hline
        \hline
        $\SiIV$ & $5.79760\pm 0.00018$ & $12.65\pm 0.12$ & $47\pm 14$\\
        $\CIV$ & $5.79760\pm 0.00018$ & $13.00\pm 0.10$ & $47\pm 14$\\
        \hline
    \end{tabular}
\end{table}

\subsection*{C.20. SDSS J0100+2802 $z_{\rm abs}=5.9450$}
\begin{figure}
\centering
	\includegraphics[width=\columnwidth]{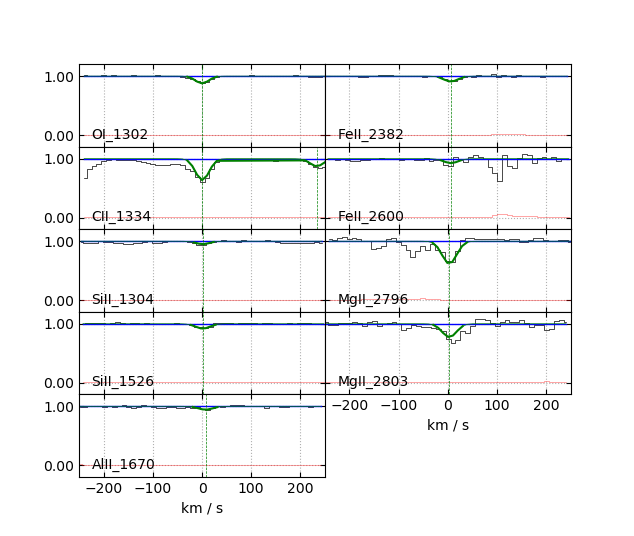}
	\includegraphics[width=\columnwidth]{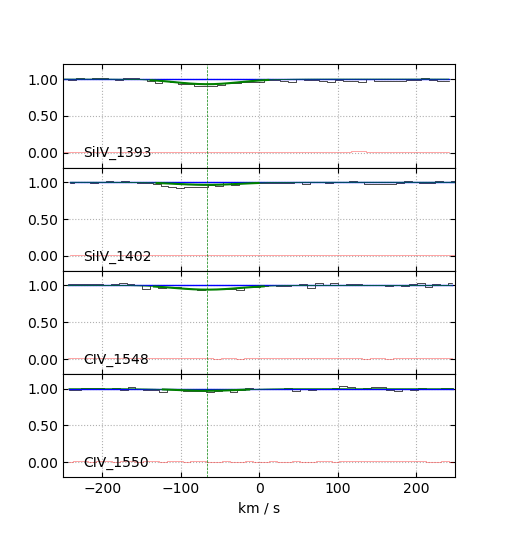}
    \caption{System at $z_{\rm abs}=5.9450$ in the spectrum of SDSS J0100+2802. Upper panel: fit of the detected low-ionization metal lines. Lower panel: fit of the detected high-ionization lines.}
    \label{fig:5.9450}
\end{figure}

The system described here is an intervening DLA at $\sim 16030$ \kms from the QSO emission redshift, characterized by a simple velocity structure consisting of a single component. We detected the following ions: \OI\ $1302$, \CII\ $1334$, \SiII\ $1304$, $1526$, \AlII\ $1670$, \FeII\ $2382$, $2600$, and \MgII\ $2796$, $2803$, as shown in Fig.~\ref{fig:5.9450}. The value of the parameters obtained from the fit of the low ionization lines are shown in Tab.~ \ref{tab:5.9450}. We performed the fit by linking the Doppler parameters of all the ions.%, obtaining a value of $b=(11\pm 1)$ \kms.

High ionization lines were also detected in the system, which have a velocity structure consisting of a single component of \SiIV\ and \CIV\ and a shift of $\sim66$ \kms\ with respect to the low-ionization lines (see Fig. \ref{fig:5.9450} and Tab.~ \ref{tab:5.9450}). We performed the fit by linking the redshift and the parameter $b$ of the two ions.%, obtaining for the latter a value of $b=(56\pm 12)$ \kms.

Finally, we estimated a lower limit for the \HI\ column density of $\log N_{\rm HI}>18.45$ from the eq.~\ref{eq:relOH}. 

\begin{table}
    \centering
    \caption{Voigt parameters obtained from the fit of the metal lines in the system at $z_{\rm abs}=5.9450$ in the spectrum of SDSS J0100+2802.}
    \label{tab:5.9450}
    \begin{tabular}{lccc}
        \hline
        Ion & $z_{\rm abs}$ & $\log (N_{\rm X}/{\rm cm^{-2}})$ & $b$ (\kms)\\
        \hline
        $\OI$ & $5.94500\pm 0.00004$ & $13.44\pm 0.04$ & $11\pm 1$\\
        $\CII$ & $5.94501\pm 0.00001$ & $13.57\pm 0.02$ & $11\pm 1$\\
        $\SiII$ & $5.94504\pm 0.00007$ & $12.77\pm 0.07$ & $11\pm 1$\\
        $\AlII$ & $5.94517\pm 0.00015$ & $11.45\pm 0.14$ & $11\pm 1$\\
        $\FeII$ & $5.94515\pm 0.00005$ & $12.25\pm 0.05$ & $11\pm 1$\\
        $\MgII$ & $5.94508\pm 0.00002$ & $12.64\pm 0.03$ & $11\pm 1$\\
        \hline
        \hline
        $\SiIV$ & $5.94347\pm 0.00016$ & $12.59\pm 0.08$ & $56\pm 12$\\
        $\CIV$ & $5.94347\pm 0.00016$ & $12.89\pm 0.09$ & $56\pm 12$\\
        \hline
    \end{tabular}
\end{table}

\subsection*{C.21. SDSS J0100+2802 $z_{\rm abs}=6.1114$}
\begin{figure}
\centering
	\includegraphics[width=\columnwidth]{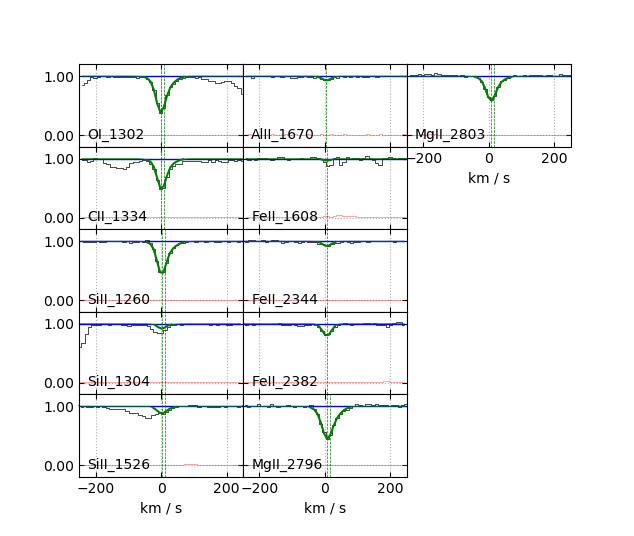}
	\includegraphics[width=\columnwidth]{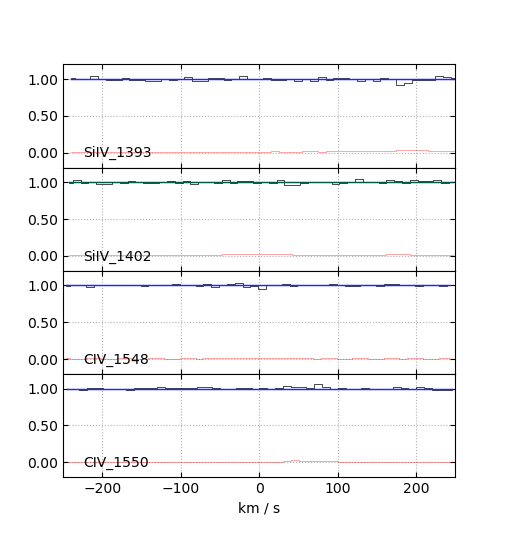}
    \caption{System at $z_{\rm abs}=6.1114$ in the spectrum of SDSS J0100+2802. Upper panel: fit of the detected low-ionization metal lines. Lower panel: region where the high ionization lines of \SiIV\ and \CIV\ would fall.}
    \label{fig:6.1114}
\end{figure}

This system is an intervening DLA, with a separation from the QSO emission redshift of $\sim 8940$ \kms. We detected two components of \OI\ $1302$, \CII\ $1334$, \SiII\ $1260$, $1304$, $1526$, and \MgII\ $2796$, $2803$, and a single component of \AlII\ $1670$, and \FeII\ $1608$, $2344$, $2382$, as shown in Fig.~\ref{fig:6.1114}. The value of the parameters obtained from the fit of the low ionization lines are shown in Tab.~ \ref{tab:6.1114}. The line of \SiII\ $1526$ is blended with the \CIV\ $1548$ line at $z=6.01$. We performed the fit by linking the $b$ parameters of all the ions in each of the two components. %For the first component we got $b=(6.1\pm 0.6)$ \kms, while for the second $b=(32\pm 2)$ \kms.

In this system, on the other hand, we did not detect high ionization lines, we have therefore derived an upper limit on \SiIV\ and \CIV. %, using equation (\ref{eq:s_n}) and (\ref{eq:upperlimit}). The results obtained from the calculation are shown in the Table.

Finally, from eq.~\ref{eq:relOH}, we derived a lower limit for the \HI\ column density of $\log N_{\rm HI}>19.70$.

\begin{table}
    \centering
    \caption{Voigt parameters obtained from the fit of the metal lines in the system at $z_{\rm abs}=6.1114$ in the spectrum of SDSS J0100+2802.}
    \label{tab:6.1114}
    \begin{tabular}{lccc}
        \hline
        Ion & $z_{\rm abs}$ & $\log (N_{\rm X}/{\rm cm^{-2}})$ & $b$ (\kms)\\
        \hline
        $\OI$ & $6.11145\pm 0.00001$ & $14.60\pm 0.17$ & $6.1\pm 0.6$\\
        $\CII$ & $6.11151\pm 0.00002$ & $13.77\pm 0.09$ & $6.1\pm 0.6$\\
        $\SiII$ & $6.11152\pm 0.00001$ & $12.86\pm 0.04$ & $6.1\pm 0.6$\\
        $\AlII$ & $6.11160\pm 0.00008$ & $11.56\pm 0.10$ & $6.1\pm 0.6$\\
        $\FeII$ & $6.11163\pm 0.00002$ & $12.65\pm 0.03$ & $6.1\pm 0.6$\\
        $\MgII$ & $6.11165\pm 0.00001$ & $12.94\pm 0.06$ & $6.1\pm 0.6$\\
        \hline
        $\OI$ & $6.11167\pm 0.00006$ & $13.96\pm 0.05$ & $32\pm 2$\\
        $\CII$ & $6.11170\pm 0.00010$ & $13.35\pm 0.07$ & $32\pm 2$\\
        $\SiII$ & $6.11179\pm 0.00006$ & $12.49\pm 0.04$ & $32\pm 2$\\
        $\MgII$ & $6.11184\pm 0.00005$ & $12.60\pm 0.05$ & $32\pm 2$\\
        \hline
        \hline
        $\SiIV$ & $6.1114$ & $<12.01$ & $26$\\
        $\CIV$ & $6.1114$ & $<12.32$ & $26$\\
        \hline
    \end{tabular}
\end{table}

\subsection*{C.22. SDSS J0100+2802 $z_{\rm abs}=6.1434$}
This system is separated from the QSO emission redshift by $\sim 7600$ \kms. We detected the following ions: \OI\ $1302$, \CII\ $1334$, \SiII\ $1260$, $1304$ and $1526$, \AlII\ $1670$, \FeII\ $1608$, $2344$, $2382$, and \MgII\ $2796$ and $2803$, as shown in Fig.~\ref{fig:6.1434}. The value of the parameters obtained from the fit of the low ionization lines are shown in Tab.~ \ref{tab:6.1434}. We performed the fit by linking the Doppler parameters of all the ions.%, obtaining a value of $b=(18.2\pm 0.5)$ \kms.

\begin{figure}
\centering
    \includegraphics[width=\columnwidth]{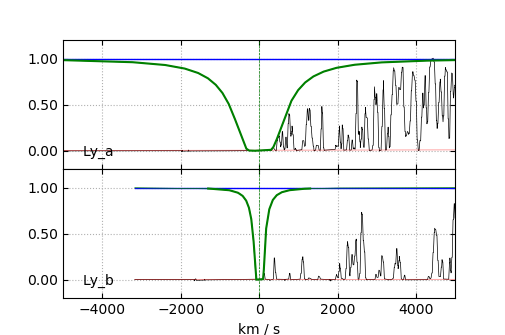}
    \caption{System at $z_{\rm abs}=6.1434$ in the spectrum of SDSS J0100+2802. Fit of the \HI\ \Lya\ and \Lyb\ absorption lines.}
    \label{fig:6.1434Lya}
\end{figure}

\begin{figure}
\centering
	\includegraphics[width=\columnwidth]{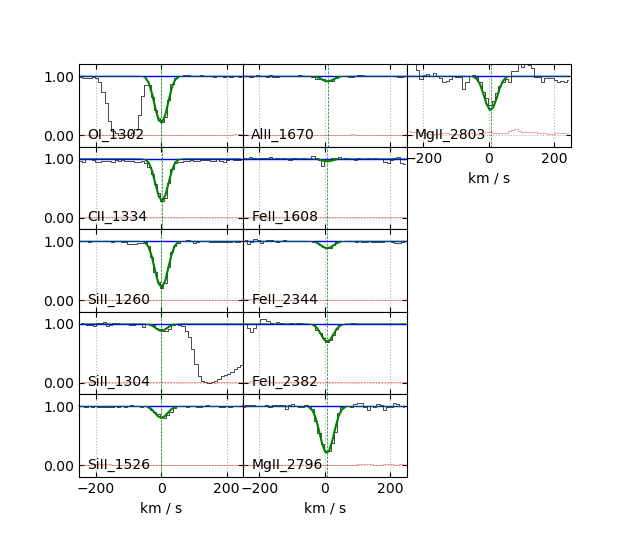}
	\includegraphics[width=\columnwidth]{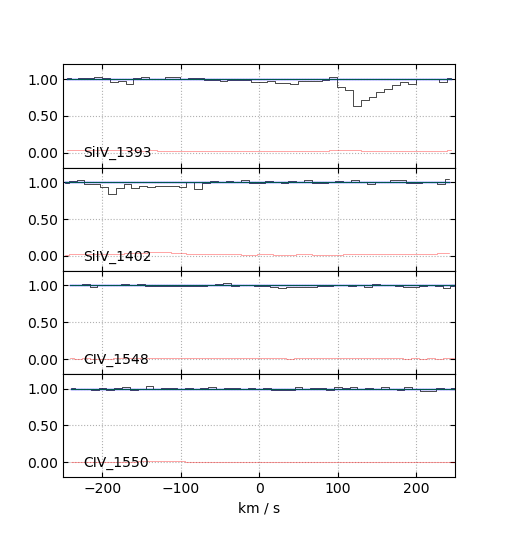}
    \caption{System at $z_{\rm abs}=6.1434$ in the spectrum of SDSS J0100+2802. Upper panel: fit of the detected low-ionization metal lines. Lower panel: region where the high ionization lines of \SiIV\ and \CIV\ would fall.}
    \label{fig:6.1434}
\end{figure}

For this system, we could carry out the fit of the \HI\ \Lya\ absorption line, which resulted in $\log N_{\rm HI}=20.20\pm 0.10$ (see Fig.~\ref{fig:6.1434Lya}. 

Also for this system, we did not detect high ionization lines, we have therefore derived an upper limit on \SiIV\ and \CIV. %, using equation (\ref{eq:s_n}) and (\ref{eq:upperlimit}). The results obtained from the calculation are shown in the Table.

\begin{table}
    \centering
    \caption{Voigt parameters obtained from the fit of the metal lines in the system at $z_{\rm abs}=6.1434$ in the spectrum of SDSS J0100+2802.}
    \label{tab:6.1434}
    \begin{tabular}{lccc}
        \hline
        Ion & $z_{\rm abs}$ & $\log (N_{\rm X}/{\rm cm^{-2}})$ & $b$ (\kms)\\
        \hline
        $\HI$ & $6.1434$ & $20.20\pm 0.10$ & $32$\\
        \hline
        \hline
        $\OI$ & $6.14347\pm 0.00001$ & $14.66\pm 0.02$ & $18.3\pm 0.5$\\
        $\CII$ & $6.14352\pm 0.00002$ & $14.11\pm 0.02$ & $18.3\pm 0.5$\\
        $\SiII$ & $6.14349\pm 0.00001$ & $13.27\pm 0.02$ & $18.3\pm 0.5$\\
        $\AlII$ & $6.14367\pm 0.00014$ & $11.74\pm 0.11$ & $18.3\pm 0.5$\\
        $\FeII$ & $6.14364\pm 0.00002$ & $12.92\pm 0.02$ & $18.3\pm 0.5$\\
        $\MgII$ & $6.14361\pm 0.00002$ & $13.26\pm 0.03$ & $18.3\pm 0.5$\\
        \hline
        \hline
        $\SiIV$ & $6.1434$ & $<12.07$ & $26$\\
        $\CIV$ & $6.1434$ & $<12.27$ & $26$\\
        \hline
    \end{tabular}
\end{table}

\subsection*{C.23. DELS J1535+1943 $z_{\rm abs}=5.8990$}
\begin{figure}
\centering
	\includegraphics[width=\columnwidth]{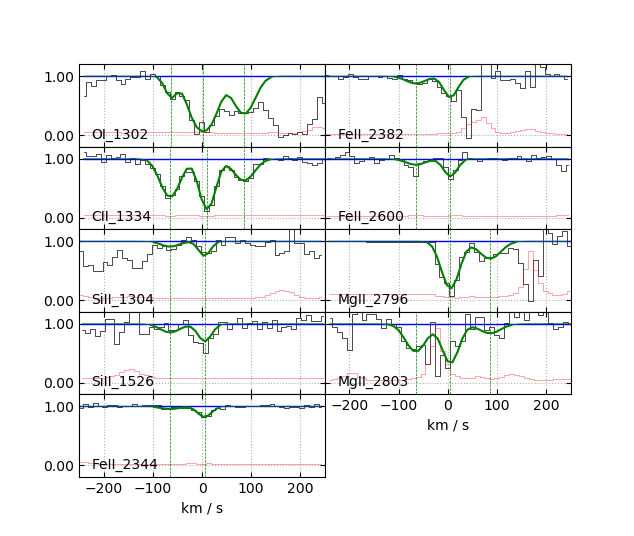}
	\includegraphics[width=\columnwidth]{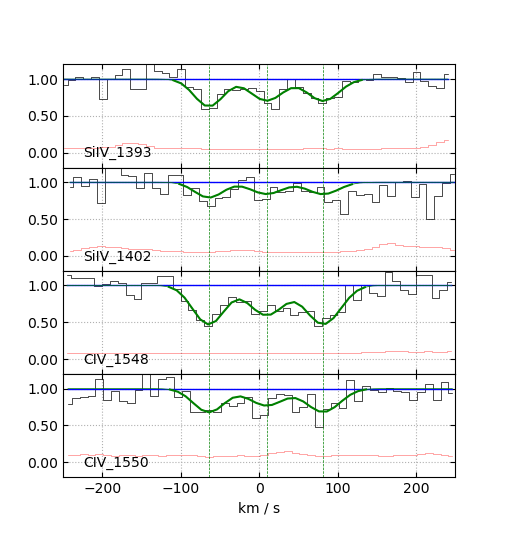}
    \caption{System at $z_{\rm abs}=5.8990$ in the spectrum of DELS J1535+1943. Upper panel: fit of the detected low-ionization metal lines. Lower panel: fit of the detected high-ionization lines.}
    \label{fig:5.8990}
\end{figure}

This system is an intervening DLA with a separation from the QSO emission redshift of $\sim 20210$ \kms. It has a complex velocity structure characterized by three components observed in \OI\ $1302$, \CII\ $1334$, and \MgII\ $2803$ (the profile of the \MgII\ $2796$ line is partially occulted by a telluric absorption), and two components in \SiII\ $1304$, $1526$, \FeII\ $2344$, $2382$ and $2600$, as shown in Fig.~\ref{fig:5.8990}. The value of the parameters obtained from the fit of the low ionization lines are shown in Tab.~\ref{tab:5.8990}. 

The \OI\ absorption falls at the beginning of the \Lya\ forest.  It is still visible thanks to the presence of the QSO \Lya\ line in emission, but partially blended with some \Lya\ lines. Therefore, we consider this system in our sample of DLA analogs, but since the \OI\ column density is too uncertain, we did not use it in the subsequent analysis. For this reason, we could not derive an estimate for the \HI\ column density lower limit. 

As for the other ions, we fitted the first component by linking the redshift and the Doppler parameter of all ions;  for the the second component, the redshift of all ions except \CII\ were linked together, and the Doppler parameter was fixed at its minimum value $b=6$ \kms\ for \SiII\ and \FeII. Finally, we detected a third component in \CII\ and \MgII\ which was fitted by linking their $b$ parameters.

In the system we have also detected high ionization lines, which have a velocity structure characterized by three components of both \SiIV\ and \CIV\ (see Fig. \ref{fig:5.8990} and Tab.~ \ref{tab:5.8990}). We performed the fit of all the components by linking the redshift and the $b$ parameter of the two ions.

\begin{table}
 \centering
    \caption{Voigt parameters obtained from the fit of the metal lines in the system at $z_{\rm abs}=5.8990$ in the spectrum of DELS J1535+1943.}
    \label{tab:5.8990}
    \begin{tabular}{lccc}
        \hline
        Ion & $z_{\rm abs}$ & $\log (N_{\rm X}/{\rm cm^{-2}})$ & $b$ (\kms)\\
        \hline
        $\CII$ & $5.89758\pm 0.00003$ & $14.05\pm 0.03$ & $20\pm 2$\\
        $\SiII$ & $5.89758\pm 0.00003$ & $13.17\pm 0.13$ & $20\pm 2$\\
        $\FeII$ & $5.89758\pm 0.00003$ & $12.51\pm 0.06$ & $20\pm 2$\\
        $\MgII$ & $5.89758\pm 0.00003$ & $13.16\pm 0.09$ & $20\pm 2$\\
        \hline
        $\CII$ & $5.89933\pm 0.00002$ & $14.43\pm 0.25$ & $12\pm 3$\\
        $\SiII$ & $5.89920\pm 0.00002$ & $13.54\pm 0.08$ & $6$\\
        $\FeII$ & $5.89920\pm 0.00002$ & $13.06\pm 0.05$ & $6$\\
        $\MgII$ & $5.89920\pm 0.00002$ & $13.44\pm 0.32$ & $12\pm 4$\\
        \hline
        $\CII$ & $5.90104\pm 0.00005$ & $13.72\pm 0.05$ & $23\pm 4$\\
        $\MgII$ & $5.09106\pm 0.00011$ & $12.62\pm 0.08$ & $23\pm 4$\\
        \hline
        \hline
        $\SiIV$ & $5.89760\pm 0.00004$ & $13.05\pm 0.06$ & $19\pm 4$\\
        $\CIV$ & $5.89760\pm 0.00004$ & $13.69\pm 0.08$ & $19\pm 4$\\
        \hline
        $\SiIV$ & $5.89929\pm 0.00008$ & $12.96\pm 0.08$ & $23\pm 7$\\
        $\CIV$ & $5.89929\pm 0.00008$ & $13.56\pm 0.09$ & $23\pm 7$\\
        \hline
        $\SiIV$ & $5.90095\pm 0.00007$ & $12.96\pm 0.09$ & $22\pm 5$\\
        $\CIV$ & $5.90095\pm 0.00007$ & $13.71\pm 0.07$ & $22\pm 5$\\
        \hline
    \end{tabular}
\end{table}

\subsection*{C.24. PSO J183+05 $z_{\rm abs}=6.0642$}
\begin{figure}
\centering
	\includegraphics[width=\columnwidth]{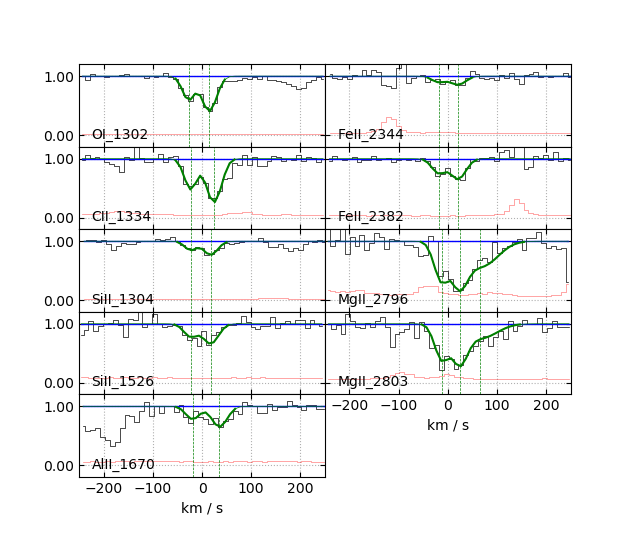}
	\includegraphics[width=\columnwidth]{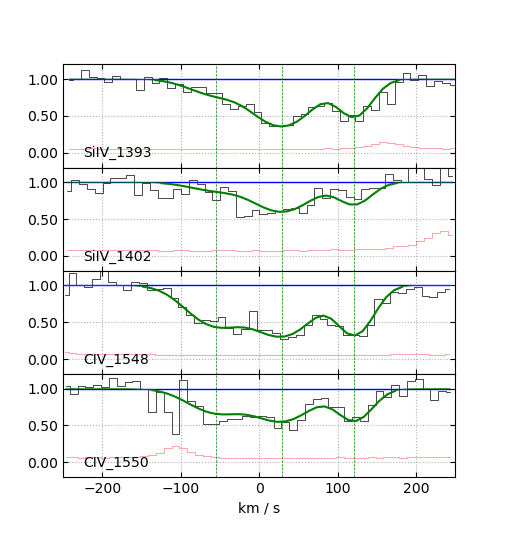}
    \caption{System at $z_{\rm abs}=6.0642$ in the spectrum of PSO J183+05. Upper panel: fit of the detected low-ionization metal lines. Lower panel: fit of the detected high-ionization lines.}
    \label{fig:6.0642}
\end{figure}

This system is an intervening DLA, with a separation from the QSO emission redshift of $\sim 15470$ \kms. We detected a velocity structure with two components for the transitions due to \OI\ $1302$, \CII\ $1334$, \SiII\ $1304$, $1526$, \AlII\ $1670$, and \FeII\ $2344$, $2382$ \AA\; while \MgII\ $2796$, $2803$ shows an additional broad component, as shown in Fig.~\ref{fig:6.0642}. The value of the parameters obtained from the fit of the low ionization lines are reported in Tab.~ \ref{tab:6.0642}. We linked the $b$ parameters of all the ions in each component, with the exception of the first component of \MgII\ for which we have fixed the value to $b=8$ \kms.
%For the first component we got $b=(8\pm 3)$ \kms, while for the second $b=(12\pm 1)$ \kms.

In the system we have also detected high ionization lines, which have a velocity structure characterized by three components of both \SiIV\ and \CIV\ covering a slightly broader velocity interval than the low ionization lines (see Fig. \ref{fig:6.0642} and Tab.~ \ref{tab:6.0642}). We performed the fit of all the components by linking the redshift and the $b$ parameter of the two ions.%; for the Doppler parameter we have obtained values of $b=(44\pm 7)$ \kms\, $b=(45\pm 5)$ \kms, and $b=(25\pm 3)$ \kms.

Finally, we estimated a lower limit for the \HI\ column density based on eq.~\ref{eq:relOH} of $\log N_{\rm HI}>19.53$.

\begin{table}
    \centering
    \caption{Voigt parameters obtained from the fit of the metal lines in the system at $z_{\rm abs}=6.0642$ in the spectrum of PSO J183+05.}
    \label{tab:6.0642}
    \begin{tabular}{lccc}
        \hline
        Ion & $z_{\rm abs}$ & $\log (N_{\rm X}/{\rm cm^{-2}})$ & $b$ (\kms)\\
        \hline
        $\OI$ & $6.06361\pm 0.00003$ & $14.06\pm 0.05$ & $8\pm 3$\\
        $\CII$ & $6.06370\pm 0.00003$ & $13.84\pm 0.15$ & $8\pm 3$\\
        $\SiII$ & $6.06370\pm 0.00006$ & $13.25\pm 0.07$ & $8\pm 3$\\
        $\AlII$ & $6.06380\pm 0.00012$ & $12.09\pm 0.13$ & $8\pm 3$\\
        $\FeII$ & $6.06382\pm 0.00007$ & $12.73\pm 0.07$ & $8\pm 3$\\
        $\MgII$ & $6.06397\pm 0.00008$ & $13.48\pm 0.32$ & $8$\\
        \hline
        $\OI$ & $6.06459\pm 0.00002$ & $14.34\pm 0.03$ & $12\pm 1$\\
        $\CII$ & $6.06482\pm 0.00003$ & $14.13\pm 0.06$ & $12\pm 1$\\
        $\SiII$ & $6.06466\pm 0.00004$ & $13.51\pm 0.04$ & $12\pm 1$\\
        $\AlII$ & $6.06508\pm 0.00008$ & $12.37\pm 0.08$ & $12\pm 1$\\
        $\FeII$ & $6.06474\pm 0.00005$ & $12.94\pm 0.05$ & $12\pm 1$\\
        $\MgII$ & $6.06482\pm 0.00008$ & $13.45\pm 0.31$ & $12\pm 8$\\
        \hline
        $\MgII$ & $6.06581\pm 0.00057$ & $13.00\pm 0.27$ & $42\pm 21$\\
        \hline
        \hline
        $\SiIV$ & $6.06296\pm 0.00016$ & $13.02\pm 0.13$ & $44\pm 7$\\
        $\CIV$ & $6.06296\pm 0.00016$ & $13.93\pm 0.07$ & $44\pm 7$\\
        \hline
        $\SiIV$ & $6.06493\pm 0.00007$ & $13.66\pm 0.04$ & $45\pm 5$\\
        $\CIV$ & $6.06493\pm 0.00007$ & $14.11\pm 0.06$ & $45\pm 5$\\
        \hline
        $\SiIV$ & $6.06711\pm 0.00005$ & $13.29\pm 0.06$ & $25\pm 3$\\
        $\CIV$ & $6.06711\pm 0.00005$ & $13.91\pm 0.04$ & $25\pm 3$\\
        \hline
    \end{tabular}
\end{table}

\subsection*{C.25. PSO J183+05 $z_{\rm abs}=6.4041$ PDLA}
\begin{figure}
\centering
    \includegraphics[width=\columnwidth]{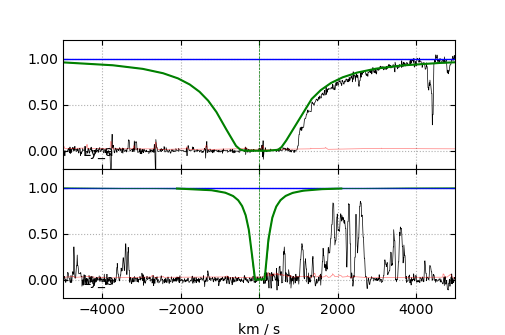}
	\includegraphics[width=\columnwidth]{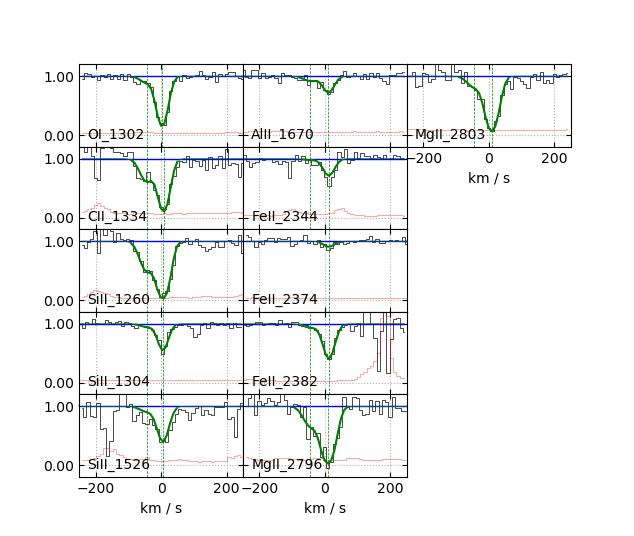}
	\includegraphics[width=\columnwidth]{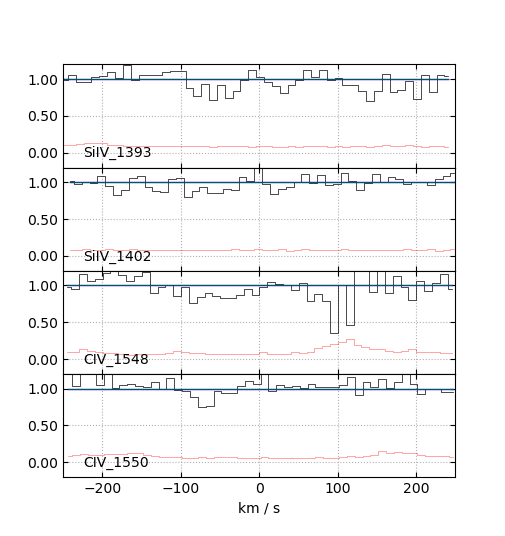}
    \caption{System at $z_{\rm abs}=6.4041$ in the spectrum of PSO J183+05. Upper panel: fit of the \HI\ \Lya\ and \Lyb\ absorption lines. Middle panel: fit of the detected low-ionization metal lines. Lower panel: region where the high ionization lines of \SiIV\ and \CIV\ would fall.}
    \label{fig:6.4041}
\end{figure}

This system is a PDLA, with a separation from the QSO emission redshift of $\sim 1390$ \kms, which presents a velocity structure characterized by two components. We detected the following ions: \OI\ $1302$, \CII\ $1334$, \SiII\ $1260$, $1304$, $1526$, \AlII\ $1670$, \FeII\ $2344$, $2374$, $2382$, and \MgII\ $2796$, $2803$, as shown in Fig.~\ref{fig:6.4041}. The value of the parameters obtained from the fit of the low ionization lines are shown in Tab.~\ref{tab:6.4041}. We performed the fit by linking the redshift and the Doppler parameter of all the ions for the weaker component, %obtaining $b=(23\pm 3)$ \kms\ for the latter,
and linking only the Doppler parameter for the stronger one. %obtaining $b=(16\pm 1)$ \kms.
The lines of \MgII\ are saturated, so we considered the column densities as lower limits.

We have also performed the fit of the \HI\ \Lya\ absorption line of this PDLA, as described in Sect.~\ref{sec:lya}, and we obtained a column density of $\log N_{\rm HI}=20.60\pm 0.10$. 
%Furthermore, this line is blended with another \Lya\ absorption line at redshift $z=6.425\pm 0.004$.

In the system, on the other hand, we did not detect high ionization lines (see Fig. \ref{fig:6.4041}) and we derived an upper limit on \SiIV\ and \CIV. %using equation (\ref{eq:s_n}) and (\ref{eq:upperlimit}). The results obtained from the calculation are shown in the Table.

%% Aggiungere confronto con Banados et al. 2019
%% differenza column densities Sodini+ vs Banados+:
%% OI 14.78       14.45    0.33
%% CII 14.52      14.30    0.22
%% SiII 13.96     13.54    0.42        
%% AlII 12.39     12.49    -0.1
%% FeII 13.36     13.19    0.17
%% MgII >14.07   13.38     0.69
This system was already identified and analyzed in \citet{Banados2019}, based on a FIRE spectrum at lower SNR than our X-SHOOTER one. The lower resolution of the FIRE spectrum did not allow to distinguish the velocity structure of the system. \citet{Banados2019} determined for each transition the equivalent width and then determined the column densities with the Apparent Optical Depth (AOD) method. 
On average, we find total column densities (see Tab.~\ref{tab:LIS}) larger by 0.3 dex (0.2 dex if we exclude the heavily saturated \MgII) which is not always consistent with the reported errors.  On the other hand, we derive an \HI\ column density slightly smaller than the one measured from the FIRE spectrum (but within uncertainties) although the adopted continuum looks quite similar. As a consequence of these discrepancies, we derive metal abundances larger than those by \citet{Banados2019} by $\sim0.2-0.4$ dex.

\begin{table}
    \centering
    \caption{Voigt parameters obtained from the fit of the metal lines in the system at $z_{\rm abs}=6.4041$ in the spectrum of PSO J183+05.}
    \label{tab:6.4041}
    \begin{tabular}{lccc}
        \hline
        Ion & $z_{\rm abs}$ & $\log (N_{\rm X}/{\rm cm^{-2}})$ & $b$ (\kms)\\
        \hline
        $\HI$ & $6.4041$ & $20.60\pm 0.10$ & $40$\\
        \hline
        \hline
        $\OI$ & $6.40301\pm 0.00006$ & $13.55\pm 0.15$ & $23\pm 3$\\
        $\CII$ & $6.40301\pm 0.00006$ & $13.73\pm 0.08$ & $23\pm 3$\\
        $\SiII$ & $6.40301\pm 0.00006$ & $12.95\pm 0.05$ & $23\pm 3$\\
        $\AlII$ & $6.40301\pm 0.00006$ & $11.76\pm 0.19$ & $23\pm 3$\\
        $\FeII$ & $6.40301\pm 0.00006$ & $12.23\pm 0.16$ & $23\pm 3$\\
        $\MgII$ & $6.40301\pm 0.00006$ & $12.68\pm 0.09$ & $23\pm 3$\\
        \hline
        $\OI$ & $6.40417\pm 0.00002$ & $14.75\pm 0.04$ & $16\pm 1$\\
        $\CII$ & $6.40429\pm 0.00003$ & $14.44\pm 0.08$ & $16\pm 1$\\
        $\SiII$ & $6.40423\pm 0.00002$ & $13.91\pm 0.03$ & $16\pm 1$\\
        $\AlII$ & $6.40436\pm 0.00008$ & $12.27\pm 0.07$ & $16\pm 1$\\
        $\FeII$ & $6.40439\pm 0.00002$ & $13.33\pm 0.03$ & $16\pm 1$\\
        $\MgII$ & $6.40433\pm 0.00003$ & $>14.05$ & $16\pm 1$\\
        \hline
        \hline
        $\SiIV$ & $6.4041$ & $<12.83$ & $26$\\
        $\CIV$ & $6.4041$ & $<13.06$ & $26$\\
        \hline
    \end{tabular}
\end{table}

\subsection*{C.26. WISEA J0439+1634 $z_{\rm abs}=6.2743$}
\begin{figure}
\centering
	\includegraphics[width=\columnwidth]{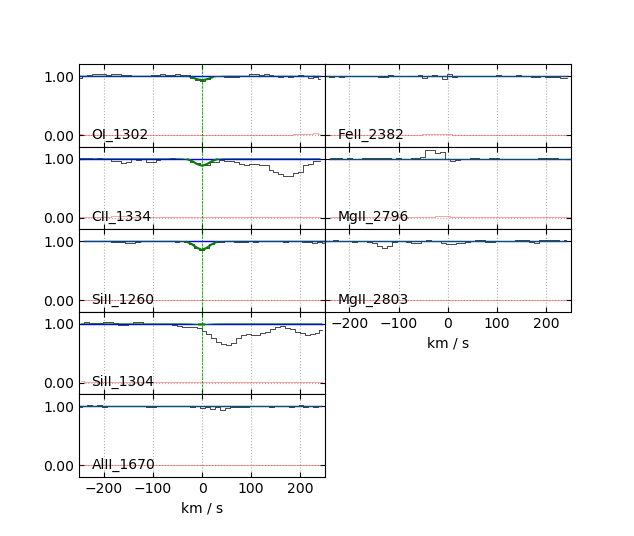}
	\includegraphics[width=\columnwidth]{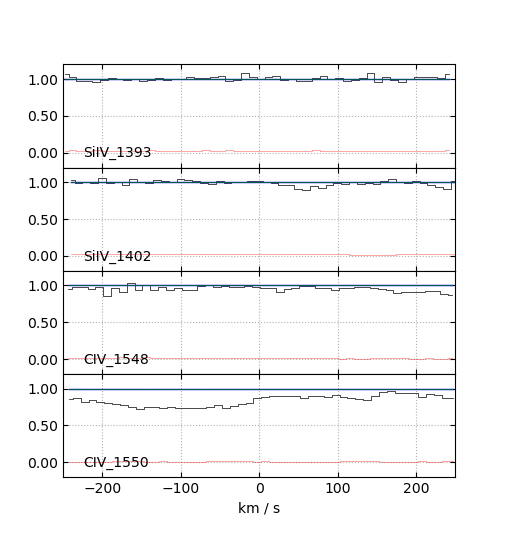}
    \caption{System at $z_{\rm abs}=6.2743$ in the spectrum of WISEA J0439+1634. Upper panel: fit of the detected low-ionization metal lines, and region where \AlII\ $1670$, \FeII\ $2382$ and \MgII\ $2796$ and $2803$ absorptions would fall. Lower panel: region where the high ionization lines of \SiIV\ and \CIV\ would fall.}
    \label{fig:6.2743}
\end{figure}

This system is a single component DLA, separated from the QSO emission redshift by $\sim 9910$ \kms. 
The following ionic transitions were detected: \OI\ $1302$, \CII\ $1334$, \SiII\ $1260$, $1304$, as shown in Fig.~\ref{fig:6.2743}. The value of the parameters obtained from the fit of the low ionization lines are shown in Tab.~ \ref{tab:6.2743}. We performed the fit by linking the redshift of these ions and freezing the parameter $b$ to its minimum value: $6$ \kms\ for \OI, \CII\ and \SiII\ which fall in the visible region of the spectrum.% We also fit the \SiII\ line at $1304$, linking the redshift and column density to the other \SiII\ transition.
We also derived an upper limit on \AlII\ $1670$, \FeII\ $2382$ and \MgII\ $2796$ by fixing the Doppler parameter to $b=7$ \kms as they are in the infrared region of the spectrum.

We did not detect high ionization lines at the redshift of the low ionization ones, and we have therefore derived an upper limit on \SiIV\ and \CIV. %, using equation (\ref{eq:s_n}) and (\ref{eq:upperlimit}). The results obtained from the calculation are shown in the Table.
On the other hand, we observe a broad CIV absorption, with $b=168\pm 5$ \kms, at $z=6.2838$ with a separation of $\sim400$ \kms\ from the low ionization lines. We assume that this \CIV\ is not related with the low ionization system and is probably part of the complex BAL system characterizing the spectrum of this QSO.

We note that this absorption system was studied also in \citet{christensen2023} based on a JWST-NIRSpec spectrum with a resolving power $R~2700$. Lines are not resolved in this spectrum, and the authors derived column densities from equivalent widths assuming they are on the linear part of the curve of growth. These column densities are in general much larger than those determined in our work, e.g., $\log N({\rm OI}) \simeq 14.22$, $\log N({\rm CII}) \simeq 13.88$ (considered as an upper limit), $\log N({\rm SiII 1304}) \simeq 13.8$ and $\log N({\rm SiII 1260}) \simeq 12.56$ (the authors did not specify how they treated these two different values). 
% OI 14.22, SiII 1260 12.56, SiII 1304 13.8, CII 1334 13.88
 What is relevant is that our determination of column densities provides relative chemical abundances which are not so extreme (see Tab.~\ref{tab:RelAbbO} and \citealt{Vanni2024}): the system falls in a region of the [Si/O] vs. [C/O] plot which is compatible with the enrichment from Pop II SNe.   

\begin{table}
    \centering
    \caption{Voigt parameters obtained from the fit of the metal lines in the system at $z_{\rm abs}=6.2743$ in the spectrum of WISEA J0439+1634.}
    \label{tab:6.2743}
    \begin{tabular}{lccc}
        \hline
        Ion & $z_{\rm abs}$ & $\log (N_{\rm X}/{\rm cm^{-2}})$ & $b$ (\kms)\\
        \hline
        $\OI$ & $6.27434\pm 0.00004$ & $13.17\pm 0.17$ & $6$\\
        $\CII$ & $6.27434\pm 0.00004$ & $13.03\pm 0.12$ & $6$\\
        $\SiII$ & $6.27434\pm 0.00004$ & $12.19\pm 0.04$ & $6$\\
        $\AlII$ & $6.2743$ & $<11.05$ & $7$\\
        $\FeII$ & $6.2743$ & $<11.60$ & $7$\\
        $\MgII$ & $6.2743$ & $<11.78$ & $7$\\
        \hline
        \hline
        $\SiIV$ & $6.2743$ & $<12.16$ & $26$\\
        $\CIV$ & $6.2743$ & $<12.50$ & $26$\\
        \hline
    \end{tabular}
\end{table}

\subsection*{C.27. VDES J0224-4711 $z_{\rm abs}=6.1228$}
\begin{figure}
\centering
	\includegraphics[width=\columnwidth]{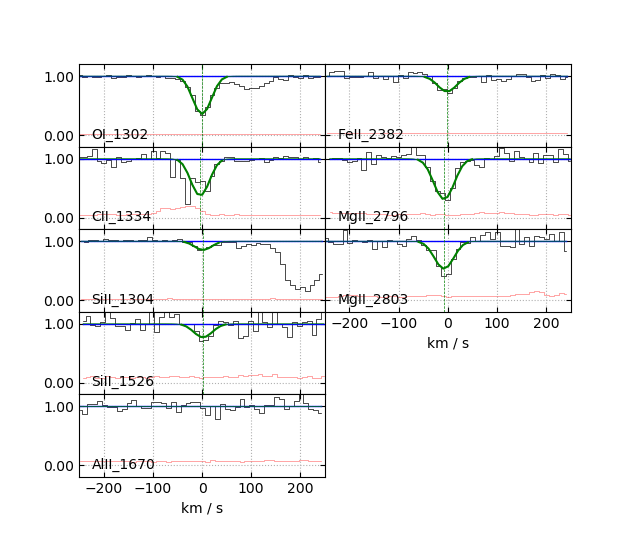}
	\includegraphics[width=\columnwidth]{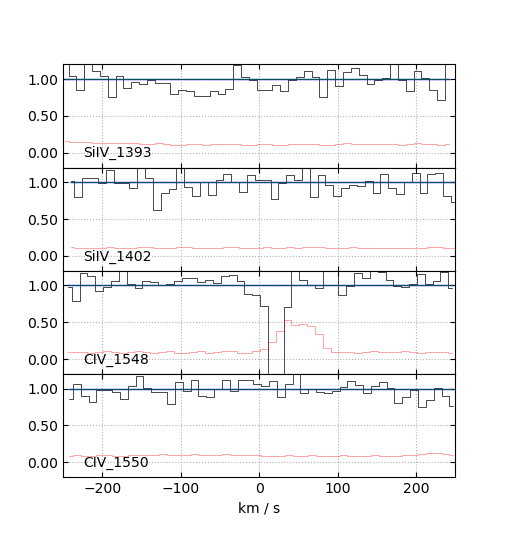}
    \caption{System at $z_{\rm abs}=6.1228$ in the spectrum of VDES J0224-4711. Upper panel: fit of the detected low-ionization metal lines, and region where \AlII\ $1670$ absorption would fall. Lower panel: region where the high ionization lines of \SiIV\ and \CIV\ would fall.}
    \label{fig:6.1228}
\end{figure}

This system is an intervening DLA, with a separation from the QSO emission redshift of $\sim 16450$ \kms.  We detected a low ionization system with a single velocity component showing the following ions: \OI\ $1302$, \CII\ $1334$, \SiII\ $1304$, $1526$, \FeII\ $2382$, and \MgII\ $2796$, $2803$, as shown in Fig.~\ref{fig:6.1228}. The value of the parameters obtained from the fit of the low ionization lines are shown in Tab.~ \ref{tab:6.1228}. We performed the fit by linking the Doppler parameters of all the ions, %, obtaining a value of $b=(17.9\pm 0.9)$ \kms.We also fit the SiII line at $1526$, linking the redshift and column density to the other SiII transition.
and we also derived an upper limit on \AlII\ $1670$. %using equation (\ref{eq:s_n}) and (\ref{eq:upperlimit}) and fixing the Doppler parameter to $b=17.9$ \kms.

\CIV\ and \SiIV\ doublets are not observed at the redshift of this system (see Fig. \ref{fig:6.1228}), therefore we derived upper limits for their column densities.% The results obtained from the calculation are shown in the Table.

Finally, we estimated a lower limit for the \HI\ column density based on eq.~\ref{eq:relOH} of $\log N_{\rm HI}>19.47$.

\begin{table}
    \centering
    \caption{Voigt parameters obtained from the fit of the metal lines in the system at $z_{\rm abs}=6.1228$ in the spectrum of VDES J0224-4711.}
    \label{tab:6.1228}
    \begin{tabular}{llccc}
        \hline
        Ion & $z_{\rm abs}$ & $\log (N_{\rm X}/{\rm cm^{-2}})$ & $b$ (\kms)\\
        \hline
        $\OI$ & $6.12283\pm 0.00001$ & $14.46\pm 0.02$ & $17.9\pm 0.9$\\
        $\CII$ & $6.12274\pm 0.00009$ & $14.00\pm 0.13$ & $17.9\pm 0.9$\\
        $\SiII$ & $6.12288\pm 0.00007$ & $13.37\pm 0.06$ & $17.9\pm 0.9$\\
        $\AlII$ & $6.1228$ & $<11.94$ & $17.9$\\
        $\FeII$ & $6.12280\pm 0.00008$ & $12.86\pm 0.06$ & $17.9\pm 0.9$\\
        $\MgII$ & $6.12265\pm 0.00004$ & $13.14\pm 0.04$ & $17.9\pm 0.9$\\
        \hline
        \hline
        $\SiIV$ & $6.1228$ & $<12.76$ & $26$\\
        $\CIV$ & $6.1228$ & $<12.76$ & $26$\\
        \hline
    \end{tabular}
\end{table}

\subsection*{C.28. PSO J036+03 $z_{\rm abs}=6.0611$}
\begin{figure}
\centering
	\includegraphics[width=\columnwidth]{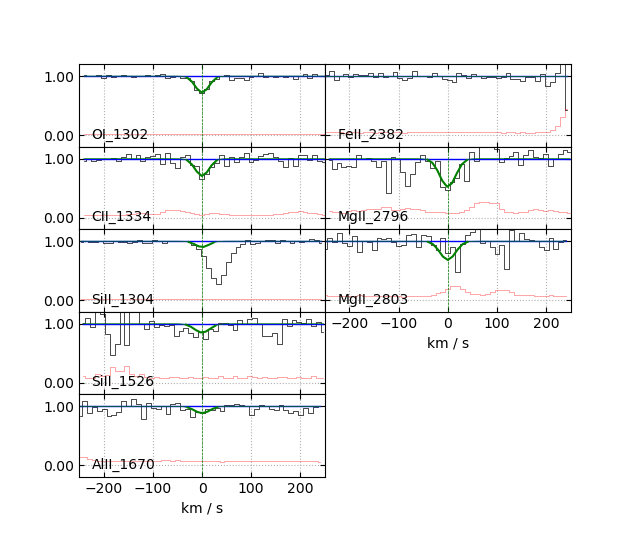}
	\includegraphics[width=\columnwidth]{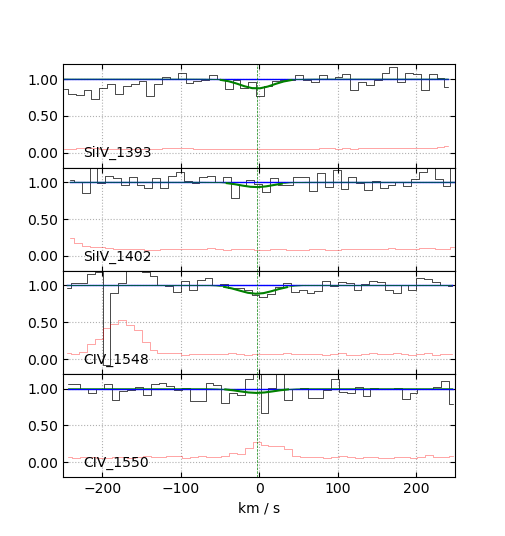}
    \caption{System at $z_{\rm abs}=6.0611$ in the spectrum of PSO J036+03.Upper panel: fit of the detected low-ionization metal lines, and region where \FeII\ $2382$ absorption would fall. Lower panel: fit of the detected high-ionization lines.}
    \label{fig:6.0611}
\end{figure}

This system is an intervening DLA, with a separation from the QSO emission redshift of $\sim 19660$ \kms, and is characterized by a simple velocity structure consisting of a single component. We detected the following ions: \OI\ $1302$, \CII\ $1334$, \SiII\ $1304$, $1526$, \AlII\ $1670$ and \MgII\ $2796$, $2803$, as shown in Fig.~\ref{fig:6.0611}. The value of the parameters obtained from the fit of the low ionization lines are shown in Tab.~ \ref{tab:6.0611}. The transition of \SiII\ $1304$ is blended with a \CII\ line at $z=5.9023$, while the \SiII\ transition at $1526$ is very weak. We then linked the redshift of \SiII\ with that of \OI\ and \CII. Also the line of \AlII\ $1670$ is very weak so, also in this case, we have linked the redshift with that of \OI\ and \CII. We performed the fit by linking the parameter $b$ of all the ions.%, obtaining a value of $b=(10\pm 3)$ \kms.
We also derived an upper limit on \FeII\ at $2382$.% using equation (\ref{eq:s_n}) and (\ref{eq:upperlimit}) and fixing the Doppler parameter to $b=10$ \kms.

High ionization lines were detected in the system, with a single component of \SiIV\ and \CIV\ at the same redshift of the low-ionization lines (see Fig. \ref{fig:6.0611} and Tab.~ \ref{tab:6.0611}). We performed the fit by linking the redshift and the parameter $b$ of the two ions.%, obtaining for the latter a value of $b=(24\pm 9)$ \kms.

Finally, we obtained a lower limit for the \HI\ column density of $\log N_{\rm HI}>19.89$, derived from eq.~\ref{eq:relOH}. 

\begin{table}
    \centering
    \caption{Voigt parameters obtained from the fit of the metal lines in the system at $z_{\rm abs}=6.0611$ in the spectrum of PSO J036+03.}
    \label{tab:6.0611}
    \begin{tabular}{lccc}
        \hline
        Ion & $z_{\rm abs}$ & $\log (N_{\rm X}/{\rm cm^{-2}})$ & $b$ (\kms)\\
        \hline
        $\OI$ & $6.06111\pm 0.00002$ & $13.88\pm 0.04$ & $10\pm 3$\\
        $\CII$ & $6.06111\pm 0.00002$ & $13.47\pm 0.09$ & $10\pm 3$\\
        $\SiII$ & $6.06111\pm 0.00002$ & $13.09\pm 0.12$ & $10\pm 3$\\
        $\AlII$ & $6.06111\pm 0.00002$ & $11.81\pm 0.23$ & $10\pm 3$\\
        $\FeII$ & $6.0611$ & $<12.13$ & $10$\\
        $\MgII$ & $6.06111\pm 0.00002$ & $12.89\pm 0.13$ & $10\pm 3$\\
        \hline
        \hline
        $\SiIV$ & $6.06103\pm 0.00013$ & $12.57\pm 0.11$ & $24\pm 9$\\
        $\CIV$ & $6.06103\pm 0.00013$ & $12.93\pm 0.15$ & $24\pm 9$\\
        \hline
    \end{tabular}
\end{table}

\subsection*{C.29. DELS J0923+0402 $z_{\rm abs}=6.3784$}
\begin{figure}
\centering
	\includegraphics[width=\columnwidth]{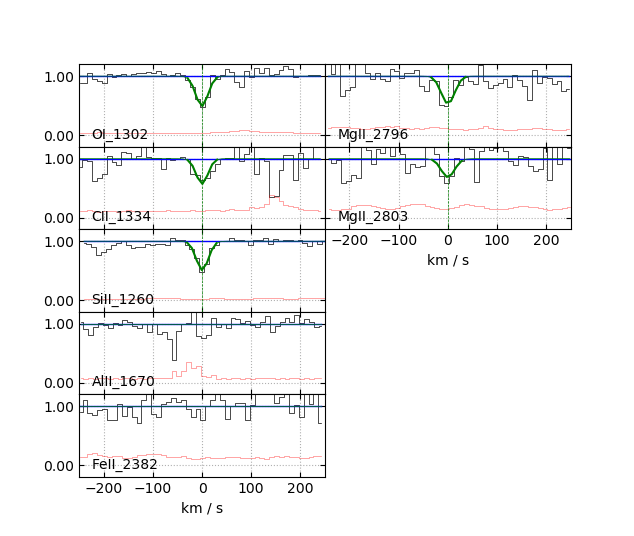}
    \caption{System at $z_{\rm abs}=6.3784$ in the spectrum of DELS J0923+0402. Fit of the detected low-ionization metal lines, and region where \AlII\ $1670$ and \FeII\ $2382$ absorptions would fall.}
    \label{fig:6.3784}
\end{figure}

This DLA falls at $\sim 10170$ \kms, and it shows a velocity structure consisting of a single component. We detected the following ions: \OI\ $1302$, \CII\ $1334$, \SiII\ $1260$ and \MgII\ $2796$, $2803$, as shown in Fig.~\ref{fig:6.3784}. The value of the parameters obtained from the fit of the low ionization lines are shown in Tab.~ \ref{tab:6.3784}. We performed the fit by linking the redshift of these ions and freezing the parameter $b$ to its minimum value: $6$ \kms\ for the transitions \OI, \CII\ and \SiII\ falling in the VIS spectrum, and $7$ \kms\ for \MgII\ which is observed in the NIR spectrum. We also derived an upper limit on \AlII\ $1670$ and \FeII\ $2382$.%, using equation (\ref{eq:s_n}) and (\ref{eq:upperlimit}) and fixing the Doppler parameter to $b=7$ \kms as they are in the infrared region of the spectrum.

Furthermore, from eq.~\ref{eq:relOH}  we derived a lower limit for the \HI\ column density of $\log N_{\rm HI}>19.57$.

The \CIV\ and \SiIV\ doublets for this system are falling in the trough of the strong BAL systems present in the spectrum of this object \citep[see][]{Bischetti2022}, therefore their presence cannot be established. Alternatively, the system could be due to a small neutral clump of gas entrained in the strong BAL outflows. As for the system at $z=5.9918$ in PSO J239-07, the intrinsic or intervening nature of this system cannot be established from the present observations, thus we assume that it is intervening and we consider it in our analysis.

\begin{table}
    \centering
    \caption{Voigt parameters obtained from the fit of the metal lines in the system at $z_{\rm abs}=6.3784$ in the spectrum of DELS J0923+0402.}
    \label{tab:6.3784}
    \begin{tabular}{lccc}
        \hline
        Ion & $z_{\rm abs}$ & $\log (N_{\rm X}/{\rm cm^{-2}})$ & $b$ (\kms)\\
        \hline
        $\OI$ & $6.37843\pm 0.00001$ & $14.56\pm 0.09$ & $6$\\
        $\CII$ & $6.37843\pm 0.00001$ & $13.84\pm 0.25$ & $6$\\
        $\SiII$ & $6.37843\pm 0.00001$ & $13.11\pm 0.05$ & $6$\\
        $\AlII$ & $6.3784$ & $<12.23$ & $7$\\
        $\FeII$ & $6.3784$ & $<12.53$ & $7$\\
        $\MgII$ & $6.37843\pm 0.00001$ & $12.86\pm 0.14$ & $7$\\
        \hline
    \end{tabular}
\end{table}

\end{appendix}
\end{document}